\documentclass[aps, 9pt, prx, twocolumn, superscriptaddress, longbibliography, nofootinbib]{revtex4-2}

\usepackage[a4paper,
            left=1.35cm,
            right=1.35cm,
            top=1.5cm,
            bottom=1.5cm,
            columnsep=0.5cm]{geometry}

\usepackage{graphicx}
\usepackage{mathtools, amssymb, amsthm, physics, bm, bbm}
\usepackage{bbold, dsfont, pifont, yfonts}
\usepackage{array}
\usepackage{tabularx}
\usepackage{booktabs}
\usepackage{hyperref}
\hypersetup{
 pdftitle = {Robust many-body quantum batteries},
 pdfauthor = {F. Schmolke, K. Hovhannisyan, M. Aguilar},
 breaklinks=true,
 pdfnewwindow=true,
 colorlinks=true,
 linkcolor=magenta,
 citecolor=cyan,
 filecolor=blue,
 urlcolor=blue
}
\usepackage[capitalize]{cleveref}

\usepackage[dvipsnames]{xcolor}
\usepackage{soul}
\usepackage{textcomp}
\usepackage{enumerate}
\usepackage{siunitx}
\usepackage{todonotes}

\usepackage[utf8]{inputenc}
\usepackage[english]{babel}
\usepackage[T1]{fontenc}
\usepackage{dcolumn}
\usepackage{csquotes}
\usetikzlibrary{calc}

\newtheorem{lemma}{Lemma}

\let\tr\relax
\DeclareMathOperator{\tr}{tr}

\DeclareMathOperator{\diam}{diam}
\DeclareMathOperator{\spec}{spec}

\DeclareMathOperator{\prob}{Prob}
\DeclareMathOperator{\supp}{supp}
\DeclareMathOperator{\spr}{spr}

\DeclareFontFamily{U}{dutchcal}{\skewchar\font=45}
\DeclareFontShape{U}{dutchcal}{m}{n}{<-> dutchcal-r}{}
\DeclareMathAlphabet{\curlyd}{U}{dutchcal}{m}{n}

\newcommand{\calh}{\mathcal{H}}
\newcommand{\calv}{\mathcal{V}}
\newcommand{\calr}{\mathcal{R}} 
\newcommand{\cald}{\mathcal{D}} 
\newcommand{\cals}{\mathcal{S}} 
\newcommand{\opt}{\mathrm{opt}}
\newcommand{\prll}{\mathrm{p}}
\newcommand{\wt}{\delta t} 
\newcommand{\fr}{\mathrm{free}}
\newcommand{\seq}{\mathrm{s}}
\newcommand{\optU}{\check{u}}
\newcommand{\ext}{\mathrm{ext}}
\newcommand{\bA}{{\hspace{2pt}\overline{\hspace{-2pt} A}}}
\newcommand{\Dtr}[2]{\curlyd{d}_{\tr}\big( #1, \, #2 \big)}
\newcommand{\opn}[1]{\big\Vert #1 \big\Vert_{\mathrm{op}}}

\newcommand{\I}{\mathrm{int}}
\newcommand{\id}{\mathbb{1}}
\newcommand{\nul}{\mathbb{0}}
\newcommand{\av}[1]{\langle #1 \rangle}
\newcommand{\Ito}{It\=o\ }
\newcommand{\m}[1]{\mathrm{#1}}
\newcommand{\erg}{\mathcal{E}}
\newcommand{\ergl}{\mathcal{E}}
\newcommand{\loc}{{\raisebox{0pt}[0pt][0pt]{\scalebox{0.6}{$\mathrm{L}$}}}}

\newcommand{\rat}{\text{\textsc{r}}}
\newcommand{\bigcircle}[4][0pt]{%
  \tikz[baseline=(X.base)]{
    \node[inner sep=0pt] (X) {\scalebox{#3}{#4}};
    \draw[draw=black] (X.center) circle [radius=#2ex];
  }%
}

\newcommand{\bigbox}[4][0pt]{%
  \tikz[baseline=(X.base)]{
    \node[
      inner sep=0pt,
      minimum width={1.1*#2ex},
      minimum height=#2ex,
      draw=black,
      fill=white,
      line width=0.7pt,
      rounded corners=1.3pt
    ] (X) {\scalebox{#3}{#4}};
  }%
}

\newcommand{\RomanNum}[1]{\uppercase\expandafter{\romannumeral #1}}

\begin{document}

\title{Robust many-body quantum batteries}

\author{Finn Schmolke}
\email{finn.schmolke@itp1.uni-stuttgart.de}
\affiliation{Institute for Theoretical Physics I, University of Stuttgart, D-70550 Stuttgart, Germany}

\author{Karen Hovhannisyan}
\email{karen.hovhannisyan@uni-potsdam.de}
\affiliation{University of Potsdam, Institute of Physics and Astronomy, Karl-Liebknecht-Str. 24-25, 14476 Potsdam, Germany}

\author{Milton Aguilar}
\email{maguilar@itp1.uni-stuttgart.de}
\affiliation{Institute for Theoretical Physics I, University of Stuttgart, D-70550 Stuttgart, Germany}

\begin{abstract}
    Realistic work extraction from many-body quantum batteries must be local. However, only a small fraction of energy eigenstates of a generic many-body system, called scars, can support local extraction. The remaining bulk is useless for the task due to the eigenstate thermalization hypothesis.
    Here we devise a universal low-complexity protocol that steers any initial state towards exactly one scar---representing a charged state of the battery---from which a macroscopic amount of work can be extracted using local unitary operations.
    This is achieved by leveraging the nontrivial interplay of engineered dissipation and continuous indirect measurement that, in addition, leads to enhanced stability and charging speed compared to any other strategy using these processes independently. Moreover, the protocol works directly on the hardware level, in that it requires no simulation or suppression of interactions between subsystems.
    Our construction thereby enables macroscopic charge storage in steady states of generic nonintegrable many-body systems indefinitely, from which a reliable stream of work can be extracted via purely local means.
\end{abstract}

\maketitle

Energy storage and processing in out-of-equilibrium many-body systems has been the main topic in thermodynamics since its inception \cite{Callen}. Work is considered to be the most versatile form of energy, and the devices that reliably store and give on-demand access to it are called batteries. As quantum technologies mature, quantum batteries are emerging as work storing devices with the potential to significantly outperform their classical counterparts \cite{Campaioli_2024}. In particular, uniquely quantum resources such as coherence and entanglement can be leveraged to achieve faster charging and work extraction \cite{Alicki_2013, Hovhannisyan_2013, Andolina_2019, Gyhm_2022, Campaioli_2024}. 

Equally appealing is the prospect that working with individual constituents could enable higher energy densities, whose increase remains a major challenge for classical batteries \cite{Ye_2024}. In practice, achieving that goal requires tightly packed microscopic cells, inevitably causing nonnegligible interactions via tunneling, Coulomb blockade, crosstalk, etc. Thus, large quantum batteries should be regarded as strongly interacting and, unless extreme fine tuning enforcing symmetries is implemented, as nonintegrable many-body systems. However, quantumness comes with severe maintenance costs in this regime. The ``charge'' a quantum battery holds is determined by coherences and population inversions, which are prone to decay via decoherence \cite{Zurek_2003} and spontaneous emission \cite{Loudon_2000}. The complexity of preserving and harnessing these resources gets exacerbated in many-body systems, as it grows exponentially with their size \cite{Preskill_2018}.

Scaling batteries to larger sizes comes with an additional fundamental challenge. Extracting the full amount of stored work requires global many-body operations and implementing them is exponentially complex in general. Therefore, realistic operations providing robust extractions must be local. However, we prove here that no work can be extracted by local operations from generic stable states of nonintegrable many-body systems. This occurs because the majority of the energy eigenstates obey the eigenstate thermalization hypothesis (ETH) \cite{Popescu_2006, Rigol_2008}, rendering them locally thermal.

We demonstrate in this paper that quantum many-body scars provide a solution to these problems. Scars are atypical energy eigenstates occurring in a wide variety of systems, that comprise a vanishingly small fraction of their energy spectra \cite{Shiraishi_2017, Turner_2018, Bernien_2017, Turner_2018, Serbyn_2021, Moudgalya_2022}. By leveraging the fact that scars are locally nonthermal, we show that they enable robust storage of locally-extractable macroscopic work in a nonintegrable quantum many-body system.

\begin{figure*}[t!]
    \centering
    \begin{tikzpicture}
        \node (a) at (0,0) {\includegraphics[width=1\linewidth]{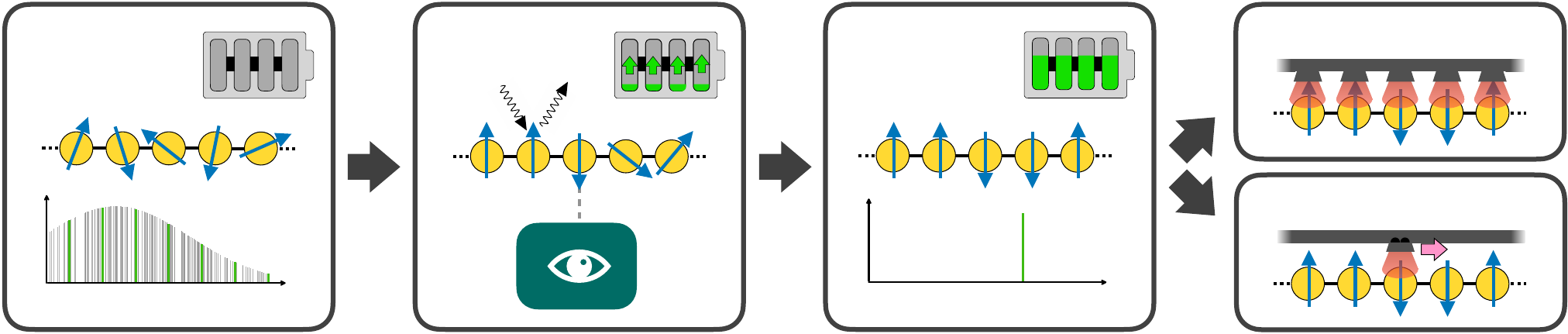}};
        \draw[line width = 0.9, rounded corners=2] (-7.5, 2.5) rectangle +(0.5, 0.5) node[pos=0.5] {\RomanNum{1}};
        \draw[line width = 0.9, rounded corners=2] (-7.2, 2.2) node {\small \textbf{Generic initial state}};
        \draw (-6.2, 1.735) node {\scriptsize Discharged};
        \draw (-7.2, -1.6) node {\scriptsize Energy spectrum};
        %
        \draw[line width = 0.9, rounded corners=2] (-2.62, 2.5) rectangle +(0.5, 0.5) node[pos=0.5] {\RomanNum{2}};
        \draw[line width = 0.9, rounded corners=2] (-2.45, 2.15) node {\small \textbf{Charging protocol}};
        \draw (-1.4, 1.735) node {\scriptsize Charging};
        
        \draw[line width = 0.9, rounded corners=2] (2.1, 2.5)  rectangle +(0.5, 0.5) node[pos=0.5] {\RomanNum{3}};
        \draw[line width = 0.9, rounded corners=2] (2.35, 2.2) node {\small \textbf{Scar reached}};
        \draw (3.4, 1.735) node {\scriptsize Charged};
        \draw (2.5, -1.6) node {\scriptsize Energy spectrum};

        \draw[line width = 0.9, rounded corners=2] (6.95, 2.5) rectangle +(0.5,0.5) node[pos=0.5] {\RomanNum{4}};
        \draw[line width = 0.9, rounded corners=2] (7.15, 2.2) node {\small \textbf{Local work extraction}};
        
        \draw[color=darkgray, fill=white, line width = 0.7] (5.57,  1.58) circle (0.2cm) node {a};
        \draw (7.2, 1.5) node {\scriptsize Parallel};

        \draw[color=darkgray, fill=white, line width = 0.7] (5.57, -0.41) circle (0.2cm) node {b};
        \draw (7.2, -0.49) node {\scriptsize Sequential};
    \end{tikzpicture}
    
    \caption{\textbf{Schematic representation of the battery charging and discharging protocols.} The charge--discharge cycle proceeds in four stages.
    \bigbox{2.5}{0.8}{\RomanNum{1}} The battery starts in some arbitrary state represented in terms of a distribution over its energy spectrum. The ETH bulk (gray lines) is interspersed by rare equidistantly spaced scars (green lines). This is considered a discharged state of the battery, as it is useless for local work extraction.
    \bigbox{2.5}{0.8}{\RomanNum{2}} The universal charging protocol sets in, where the concurrent action of local engineered dissipation and continuous measurement of a local operator filters out the ETH bulk while driving the system to one of the scars.
    \bigbox{2.5}{0.8}{\RomanNum{3}} Each scar represents a charged state that allows for local work extraction from the battery, with the total amount depending on the scar energy. Since scars are steady states of both the dissipative and internal dynamics, their maintenance has zero operational cost.
    \bigbox{2.5}{0.8}{\RomanNum{4}} When work is demanded from the battery, there are two generic scenarios for extraction: \bigcircle{1.28}{0.9}{a} all cells of the battery are drained in parallel by the simultaneous action of local unitaries, and \bigcircle{1.28}{0.9}{b} the locally extracting ``tip'' moves sequentially from cell to cell, outputting a steady stream of work while gradually depleting the battery. Afterwards, the now empty state of the battery can be recharged by simply feeding it into step \bigbox{2.5}{0.8}{\RomanNum{1}} and repeating the protocol.
    }
    \label{fig:protocol}
\end{figure*}
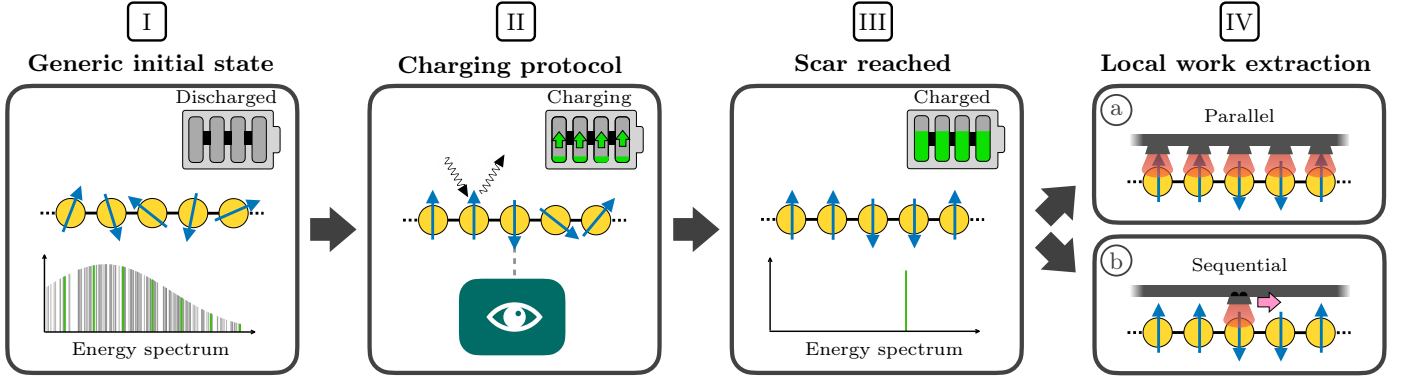

The final obstacle is to actually stabilize the system in a scar. We overcome it by introducing a universal battery charging protocol, illustrated in Fig.~\ref{fig:protocol}. Starting from an arbitrary initial configuration of the system, the protocol reliably singles out exactly one scar as a nonequilibrium steady state of the dynamics. Afterwards, work can be extracted from each individual cell in parallel (all at the same time) or sequentially (one by one). A key feature of our protocol is that it has an exceptionally low complexity. Namely, it achieves the complete filtration of the unwanted typical eigenstates by combining engineered \textit{local} dissipation and continuous monitoring of a \textit{local} many-body operator. We exemplify the protocol on a paradigmatic spin-$1$ $XY$ chain. Our findings thus identify quantum scars as an essential resource in building large many-body batteries.

\textbf{Local ergotropy as battery charge.}---The most important characteristic of a battery is its charge level, which is standardly measured by the ergotropy \cite{Hatsopoulos_1976, Pusz_1978, Allahverdyan_2004, Campaioli_2024}. It is defined as the maximal amount of work extractable from the system, $\erg (\rho) \coloneq \tr (\rho H) - \min_{U} \tr (U \rho U^{\dagger} H)$, with $\rho$ its state, $H$ the Hamiltonian, and the minimization is carried over all possible unitary operators $U$. Constructing the optimal $U$ requires the full set of eigenvectors of both $\rho$ and $H$ \cite{Allahverdyan_2004}, which quickly becomes unmanageable for quantum many-body systems. For large systems, these eigenvectors are typically nonlocal and computing them is extremely challenging numerically. Even with perfect knowledge of the eigenvectors, implementing such a highly nonlocal $U$ would require complete control of the system and its environment, which is well beyond the capabilities of current or near-future platforms. Moreover, generic many-body spectra become exponentially dense with the system size \cite{Hartmann_2005, Keating_2015}, so addressing individual eigenvectors comes with an additional metrological overhead. Hence, in many-body scenarios, the full, ``global'' ergotropy is a somewhat unrealistic upper bound on the feasibly achievable performance of a battery.

Instead, a more realistic figure of merit is the work extractable via \textit{local} operations.
Since the cells interact, any manipulation of the local degrees of freedom of a cell also affects the neighboring ones \cite{Frey_2014, Salvia_2023, Castellano_2024}. In the limit of vanishing extraction time, the unitary acts locally, yielding the so-called ``local ergotropy'' \cite{Salvia_2023}. We introduce a slight generalization of this notion where such local operations act in parallel:
\begin{align} \label{parallel_erg_def}
    \ergl_\prll (\rho) \coloneq \tr(\rho H ) - \min_{\{ u_j\}} \tr\Big(\bigotimes_j u_j^{\phantom{\dagger}} \, \rho \, \bigotimes_j u_j^\dagger H\Big).
\end{align}
Each unitary $u_j$ acts only on cell $j$, and we denote its optimal value by $\check{u}_j$. This quantity provides a benchmark for total capacity of the battery under local extracting operations, and we refer to it as \textit{parallel charge}.

In most meaningful use cases, the battery will need to provide work in small chunks, gradually discharging in the process. We model this scenario by introducing the notion of sequential discharging with the time interval $\wt$ between subsequent instantaneous extractions. There, one applies $\optU_1$ to cell $1$, then lets the system evolve by itself under $U_\fr \coloneq \exp(- \m{i} \wt H)$, then applies $\optU_2$ to cell $2$, etc. Thus, by the $\ell$'th iteration, the extracted work will be
\begin{align} \label{sequential_erg}
    \ergl_{\seq,\ell}(\rho, \wt) \coloneq \tr(H \rho) - \tr\big( H \, U_{[1 \cdots \ell]}^{\phantom{\dagger}} \, \rho \, U_{[1 \cdots \ell]}^\dagger \big),
\end{align}
where $U_{[1 \cdots \ell]} \coloneq \check{u}_\ell \, U_\fr \, \cdots \, U_\fr \, \check{u}_1$. We refer to it as \textit{sequential work}, and by construction, $\ergl_{\seq,N} (\rho, 0) = \ergl_\prll(\rho)$. 

A major obstacle on the way towards large practicable quantum batteries is that, for typical states of generic many-body systems, these two key functional characteristics [Eqs.~\eqref{parallel_erg_def} and~\eqref{sequential_erg}] are essentially zero. Indeed, almost all energy eigenstates of such systems satisfy the ETH \cite{Berry_1977, Deutsch_1991, Srednicki_1992, Popescu_2006, Rigol_2008} and are hence locally similar to (mean force \cite{Trushechkin_2022}) Gibbs states. In the SI, we prove that no macroscopic amount of work can be extracted locally from any convex combinations of ETH states: $\ergl_\prll(\rho_{\mathrm{ETH}}) = o(N)$. This leaves scars as the only remaining candidates for storing macroscopic locally extractable work.

\textbf{Properties of scars.}---Scars usually come in two flavors: as an unstructured set of isolated eigenstates, or as a structured collection emerging from an effective Lie algebra realized within a subspace of the Hilbert space $\calh$ \cite{Moudgalya_2022, Matsui_2025}. 
If the resulting algebra is $\mathfrak{su} (2)$, then the scars appear as (one or many) towers of equidistant energy eigenstates, while if it is $\mathfrak{su} (3)$, then richer structures are possible \cite{Choi_2019, ODea_2020, Mark_2020, Ren_2021, Matsui_2025}. Here we focus on structured scars, because their properties are conveniently exploitable for reliable battery charging. For simplicity of exposition, we limit ourselves to the $\mathfrak{su} (2)$ case, but the results can be readily extended to $\mathfrak{su} (3)$ and beyond. In models presenting towers of scars, there exist raising and lowering ladder operators, $Q^{+}$ and $Q^{-}$, and a subspace $\calv$ of $\calh$ such that  
\begin{equation} \label{eq:sga}
    \big( [H, Q^{\pm}] \mp \omega Q^{\pm} \big) \big|_{\mathcal{V}} = 0,
\end{equation}
for some $\omega \in \mathbb{R}$. Thus, $H$ and $Q^\pm$ define a spectrum-generating algebra that guarantees the existence of equally spaced eigenstates with increasing energy in the spectrum of $H$---the scars (see Fig.~\ref{fig:protocol}). Repeated application of $Q^{+}$ on the ground scar $\ket{s_0}$ then yields all higher energy scars $\ket{s_n} \propto (Q^{+})^n \ket{s_0}$. 
Additionally, the Hamiltonian decomposes as 
\begin{equation}\label{eq:H}
    H = H_{0} + H_{\mathrm{ZE}},
\end{equation}
where the Zeeman-like term $H_{\mathrm{ZE}}$ forms a closed $\mathfrak{su} (2)$ algebra along with $H$ and $Q^\pm$ and breaks the degeneracy of the scars,
\begin{equation} \label{eq:deg}
    \big( [H_{\mathrm{ZE}}, Q^{\pm}] \mp \zeta Q^{\pm} \big) \big|_{\calv} = 0,
\end{equation}
for some $\zeta \in \mathbb{R}$. That is, $\ket{s_n} $ are simultaneous eigenstates of $H$ and $H_{\mathrm{ZE}}$ with distinct eigenvalues.

\textbf{Universal charging protocol.}---The following protocol prepares the many-body system in a charged steady state, priming it for local work extraction. This is achieved by steering the system towards one of the structured scars. Crucially, these states are stationary under both the protocol-driven as well as intrinsic dynamics of the system, and can therefore be maintained indefinitely at no extra cost.

The procedure consists of two tasks that could be performed consecutively, but, as explained below, concurrent implementation leads to faster charging. First is the dissipation engineering, where dissipative processes are tailored for specific tasks by manipulating the system--environment interface \cite{Diehl_2008, Verstraete_2009, Buca_2019, Harrington_2022}. Here, we use it to establish the scar subspace $\cals \coloneq \mathrm{span} \{ \ket{s_n} \} \subseteq \calv$ as the unique invariant and attractive manifold in $\calh$, where each scar is decoherence-free. Second, continuous indirect measurement that breaks the degeneracy inside $\cals$ and dynamically stabilizes the system in one and only one of the scars. Attractiveness and uniqueness ensure the asymptotic dynamics takes place only inside $\cals$, while being decoherence-free is necessary for individual scars to be stationary states and thus get singled out by the measurement.

We stress that any of the two tasks by itself does not suffice---it is their combined action that enables steering the state of the system towards a single scar. Indeed, only implementing engineered dissipation produces as asymptotic state a (generically coherent and time-evolving) mixture of scars that depends on the initial state, while only performing continuous monitoring would map the many-body system outside of $\cals$ almost surely (see SI).

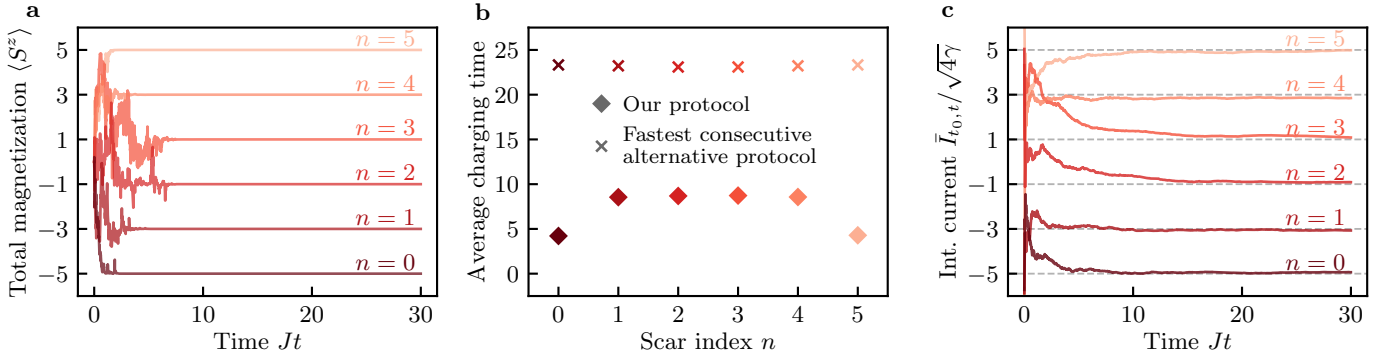
\begin{figure*}[t!]
    \centering
    \begin{tikzpicture}    
        \node (a) at (0,0) {\includegraphics{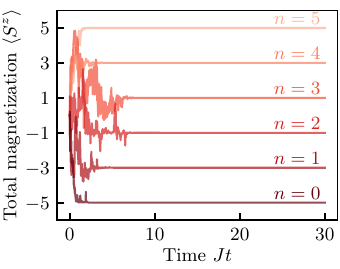}};
        \node (b) at (6.15,0) {\includegraphics{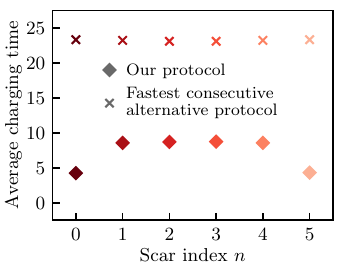}};
        \node (c) at (12.3,0) {\includegraphics{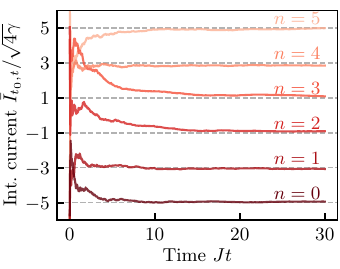}};

        \draw (-2.5,2.3) node {\bfseries a};
        \draw (3.45,2.3) node {\bfseries b};
        \draw (9.6, 2.3) node  {\bfseries c};
    \end{tikzpicture}
    \caption{\textbf{Battery charging and certification for the spin-$1$ $XY$ chain.} The universal charging protocol [Eq.~\eqref{eq:quantumTrajectory}] drives every initial state into a scar. Here it is realized by choosing $M = S^z$, where $S_z$ is the total magnetization, and the jump operators as in Eq.~\eqref{the_choice}. For $N=5$ the model hosts six scars $\ket{s_{n}}$ with $n = 0 , \dots, 5$, represented by different colors indicated on top of the figure.
    \textbf{a}, Six typical realizations of the protocol for the same initial state $\rho(0) = \mathds{1} / 3^N$. The convergence of trajectories to scars is witnessed by tracking $\av{S^z} = \tr(\rho_t^\m{c} S^z)$---by construction, each scar is an eigenvector of $S^z$ with a distinct eigenvalue.
    \textbf{b}, Average charging time of our protocol, quantified by the mean value of the scar-resolved first-passage time $\mathbb{E} [\tau^\m{c}_n(\epsilon)]$ [Eq.~\eqref{eq:first-passage}] (diamonds), compared to the charging times for the fastest possible alternative consecutive protocol, $\tau_{\m{ED},n}(\epsilon)$ (crosses). That is, engineered dissipation followed by an instantaneous projective measurement. The average charging time is improved at least by a factor of $\approx 2.6$. 
    \textbf{c}, Time-integrated measurement current [Eq.~\eqref{eq:current}] for the same trajectories as \textbf{a}, sharply certifying a scar has been reached and thus the battery charged.
    Parameters are $h/J = 1$ and $D/J = 1.3$, $\gamma/J = 2$, $\Gamma_j/J = 2$.
    }
    \label{fig:relaxation}
\end{figure*}

Each realization of the protocol gives rise to a quantum trajectory, i.e., a stochastic path taken by the system state conditioned upon the measurement results of the continuously monitored observable \cite{Wiseman_2009}. Provided the above prerequisites are satisfied, every quantum trajectory will single out and stabilize exactly one of the scars. The dynamics of the conditional state $\rho_{t}^{\m{c}}$, including the engineered dissipation and measurement backaction, is described by the \Ito stochastic differential equation,
\begin{equation} \label{eq:quantumTrajectory}
\begin{split}
  \dd{\rho_{t}^{\m{c}}} 
   & = \Big[ \! - \! \m{i} [H,\rho_{t}^{\m{c}}] 
  \! + \! \sum_{j} \Gamma_j \textsf{D}[L_j](\rho_{t}^{\m{c}}) \! + \!  \gamma \textsf{D}[M](\rho_{t}^{\m{c}}) \Big] \dd{t}
  \\
  & + \sqrt{\gamma} \big[M \rho_{t}^{\m{c}} \! + \! \rho_{t}^{\m{c}} M^{\dagger} \! - \! \tr (M \rho_{t}^{\m{c}} \! + \! \rho_{t}^{\m{c}} M^{\dagger}) \rho_{t}^{\m{c}} \big] \dd W_t,
\end{split}
\end{equation}
where, $\textsf{D}[X] (\bullet) = X \bullet X^{\dag} - \{X^{\dag} X, \bullet \}/2$, $\Gamma_j$ are the dissipation rates, $\gamma$ is the measurement strength, and $\dd W_t$ is a Wiener increment \cite{Wiseman_2009, Barchielli_2009, Albarelli_2024, Jordan_2024}. Here, $\textsf{D}[L_{j}]$ correspond to the engineered dissipation, whereas the terms containing $M$ represent the effective action of the continuous indirect measurement on the system. Note that $M$ does not have to be Hermitian. 

As we prove in the SI, Eq.~\eqref{eq:quantumTrajectory} has exactly the desired properties. Namely, in conjunction with the engineered dissipation, the continuous measurement drives irreversible localization transitions that asymptotically map any initial state $\rho_{0}$ into an individual scar,
\begin{equation} \label{eq:charging}
    \rho_{0} \longrightarrow \rho_{\infty}^{\m{c}} = \dyad{s_n} \;\;\; \text{with probability} \;\;\; p_n (\rho_0).
\end{equation}
Here, $p_n(\rho_0)$ are determined from a generalized Born rule (see SI). To achieve this remarkable behavior, the protocol exploits nonlinear quantum measurement backaction.

We now describe the individual components of the protocol separately. Regarding the engineered dissipation, the necessary and sufficient conditions the operators $L_{j}$ must satisfy for $\cals$ to be an invariant and attractive subspace of the dynamics can be informally stated as follows (see the SI for precise statements). First, to guarantee invariance, there cannot be outflow from $\cals$ to its orthogonal complement $\cals^{\perp}$.
Second, attractiveness requires states getting trapped only inside $\cals$, meaning there are no other invariant subspaces besides $\cals$, and inflow takes place only from $\cals^{\perp}$ to $\cals$. With these met, the support of any initial state on $\cals^{\perp}$ will asymptotically vanish, confining the remanent evolution to $\cals$.

Forcing the tiny scar manifold to be a globally attractive subspace of the total Hilbert space $\calh$ appears to be very challenging. However, notably, it lends itself to a natural and universal construction by leveraging the fact that most known scar Hamiltonians can be put into a generalized Shiraishi--Mori form \cite{Shiraishi_2017, Moudgalya_2024, Omiya_2026}. There, the $H_{0}$ term in Eq.~\eqref{eq:H} is decomposed as $H_{0} = \sum_{j} P_{[j]}^{\dagger} h_{[j]} P_{[j]}$, where $P_{[j]}$ and $h_{[j]}$ are strictly local operators acting only in a neighborhood of site $j$, and their range does not scale with the system size. Importantly, all scars are annihilated by all $P_{[j]}$, $P_{[j]} \ket{s_n} = 0$, and $H_{\m{ZE}}$ leaves invariant the common kernel of the $P_{[j]}$, so that $\cals$ is an invariant subspace of $H$. Thus, by implementing local operators of the form $L_{j} = O_{[j]} P_{[j]}$, $\cals$ is guaranteed to be an invariant subspace of the entire dynamics \cite{Wang_2024}. Attractivity demanding inflow from $\cals^{\perp}$ to $\cals$, necessitates the $O_{[j]}$ to be chosen such that the scars are not left eigenvectors of $L_{j}$. Note that the number of operators $L_j$ is not fixed and could even be one. The recipe for an appropriate form of engineered dissipation is therefore already built into the framework of quantum-many body scars. The locality of $L_{j}$ and the significant freedom in the choice of $O_{[j]}$ may facilitate practical implementations of the protocol.

Once inside $\cals$, the continuous monitoring realized by $M$ discriminates between the individual scars. The associated measurement backaction then collapses the state in one and only one of the scars. This is achieved by any operator $M$, provided it has the scars as simultaneous left and right eigenvectors with distinct eigenvalues. Furthermore, the continuous measurement suppresses leakage out of the individual scar subspaces and therefore provides additional protection and sustained stabilization that would not be present were projective measurements performed instead \cite{Zanardi_2014}. Out of the many possible choices for $M$, a straightforward one could be the Zeeman-like term $H_{\mathrm{ZE}}$ already contained in the Hamiltonian [see Eqs.~\eqref{eq:H} and~\eqref{eq:deg}]. In the subsequent analysis, we however keep $M$ general.

\textbf{Charging speed.}---We define the charging time $\tau_n^\m{c}(\epsilon)$ as the first-passage time of the system state to get $\epsilon$-close to a scar [Eq.~\eqref{eq:charging}]. That is, for a given tolerance $\epsilon > 0$ and scar $\ket{s_n}$,
\begin{equation} \label{eq:first-passage}
    \tau_n^\m{c}(\epsilon) \coloneq \inf\{ t \; \colon \bra{s_n} \rho^\m{c}_{t} \ket{s_n} > 1-\epsilon\}.
\end{equation}
Crucially, since the trajectories are stochastic, $\tau^\m{c}_n (\epsilon)$ is a random variable. We analyze the speed at which individual trajectories approach the scars in the SI. We conclude that the protocol-induced convergence to an individual scar is exponential in time with an asymptotic rate $\rat^\m{c}$.
Thus, the charging time is $\tau^\m{c}_n(\epsilon) = \tau^\m{c}_{\m{T}} + O(1/\rat^\m{c})$, where $\tau^\m{c}_{\m{T}}$ is the transient time.

The fastest alternative consecutive protocol to ours consists of first reaching $\cals$ by using engineered dissipation, and then performing an instantaneous projective measurement of $M$ to select a scar. Its charging time thus equals the relaxation time of the engineered dissipation, $\tau_{\m{ED},n}(\epsilon)$ [defined analogously to Eq.~\eqref{eq:first-passage}]. We find in the example below that our protocol, on average, charges the battery faster: $\mathbb{E} [\tau^\m{c}_n(\epsilon)] \leq \tau_{\m{ED},n}(\epsilon)$.

This boost in charging speed results from the mutual reinforcement of engineered dissipation and continuous monitoring. As the engineered dissipation drags the state towards $\cals$, the backaction of the continuous monitoring constantly pushes it to an eigenspace of $M$. Since both processes take place at the same time, their combined action steers the state to a simultaneous invariant subspace (a scar) in a time shorter than $\tau_{\m{ED},n}(\epsilon)$. We expect the discrepancy between $\mathbb{E} [\tau^\m{c}_n(\epsilon)]$ and $\tau_{\m{ED},n}(\epsilon)$ to be present in any system and become even more pronounced in the many-body regime. While $\tau_{\m{ED},n}(\epsilon)$ typically increases with the particle number $N$ \cite{Znidaric_2015}, $\mathbb{E} [\tau^\m{c}_n(\epsilon)]$ is much less affected.

Importantly, this measurement-powered speedup mechanism is fundamentally different from the usual approaches leveraging collective effects \cite{Campaioli_2017, Gyhm_2022}. In the standard unitary setting, speedup is achieved by increasing the degree of nonlocality of the driving Hamiltonian. Here instead, the degrees of locality of $H$, $L_j$, and $M$ are fixed, while the asymptotic charging speed is proportional to the measurement strength $\gamma$.

\textbf{Scar certification.}---How can one be sure the battery was successfully charged?

Fortunately, our protocol is naturally supplied with state-detection capabilities and thus avoids costly state tomography that would otherwise be required to verify the presence and order $n$ of a scar $\ket{s_n}$. Indeed, continuous monitoring yields a real-time measurement current $\dd{Y}_{t} = \sqrt{\gamma} \tr[(M + M^{\dag}) \rho_{t}^{\m{c}}] \dd{t} + \dd W_{t}$ that can be used to infer the steady state. Although, in general, it only provides limited information about the instantaneous state $\rho_{t}^{\m{c}}$ \cite{Wiseman_2009}, once the system enters a one-dimensional decoherence-free subspace, $Y_{t}$ determines the state uniquely. Convergence to a scar can then be directly inferred from the time-integrated measurement current $\bar{I}_{t, t_{0}} = t^{-1}\int_{t_{0}}^{t_{0} + t} \dd Y_{t^{\prime}}$ as
\begin{equation} \label{eq:current}
    \bar{I}_{t, t_{0}} \longrightarrow \sqrt{4 \gamma} \, \text{Re} (m_n),
\end{equation}
where $M \ket{s_n} = m_n \ket{s_n}$, and the limit is for $t \gg \tau^\m{c}_n(\epsilon)$. Furthermore, because the distribution of $\dd Y_t$ is asymptotically Gaussian, the standard deviation is $\sigma[\bar{I}_{t, t_{0}}] = t^{-1/2}$ for $t_0 \gtrsim \tau_n^{\m{c}}(\epsilon)$, leading to the signal-to-noise ratio $\bar{I}_{t, t_{0}} / \sigma[\bar{I}_{t, t_{0}}] \propto t^{1/2}$. Thus, tracking $\bar{I}_{t, t_{0}}$ makes the different outcomes clearly distinguishable, yielding a precise and robust scar certification method.
 
\begin{figure*}[t!]
    \centering
    \begin{tikzpicture}
        \node (a) at (6.4,2.9) {\includegraphics{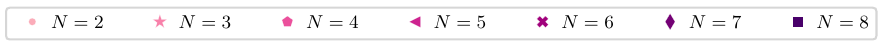}};
        
        \node (a) at (0,0) {\includegraphics{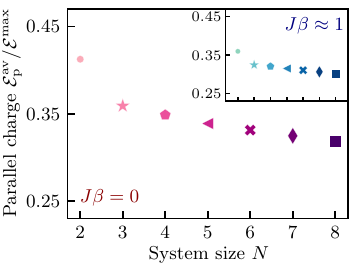}};
        \node (b) at (6.2,0) {\includegraphics{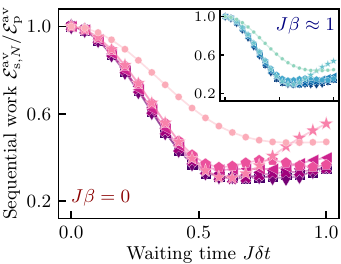}};
        \node (c) at (12.32,0) {\includegraphics{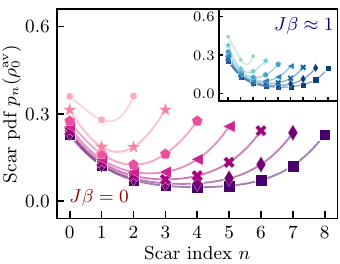}};

        \draw (-2.45,2.3) node {\bfseries a};
        \draw (3.85,2.3) node {\bfseries b};
        \draw (9.9, 2.3) node  {\bfseries c};
    \end{tikzpicture}
    \caption{\textbf{Battery characterization of the spin-$1$ $XY$ chain.} The initial state $\rho_0$ is sampled from the Scrooge ensemble \cite{McGinley_2025}, so that $\rho_0^{\mathrm{av}} = \exp(-\beta H) / Z$, with $Z$ the partition function. In the main panels, $\beta = 0$, whereas in the insets, $J \beta \approx 1$ (see SI for details on this choice).
    \textbf{a,} Average charge of the battery that can be retrieved by locally discharging all cells simultaneously, $\ergl_{\prll}^{\mathrm{av}}$, scaled by its maximal capacity $\erg^{\max}$, as a function of the system size. Recall that $\erg^{\max} = E_{\max} - E_{\min}$, with $E_{\max}$ and $E_{\min}$ the largest and smallest eigenvalues of $H$, and thus it can only be attained by initializing the system in its highest energy eigenstate and then rotating it to the ground state via highly complex global operations. It is therefore quite remarkable that, despite $\ergl_{\prll}^{\mathrm{av}}$ being obtained by arguably minimal-complexity operations [Eq.~\eqref{parallel_erg_def}] and $\rho_0$ being sampled from positive-temperature states, it remains above $30\%$ of $\mathcal{E}^{\max}$.
    \textbf{b,} Average work delivered by sequential local discharging of all cells, $\ergl_{\seq , N}^{\mathrm{av}} (\wt)$, compared to being discharged simultaneously, as a function of the waiting time $J \delta t$ between subsequent extractions. The point of transition between two qualitatively different discharging behaviors, $J \delta t \approx 0.5$, is almost independent of $N$.
    \textbf{c,} Asymptotic probability distribution $p_n (\rho_{0}^{\mathrm{av}})$ of converging to different scars $\dyad{s_n}$. This population inversion pattern persists even when the system is initially prepared in a low-temperature state, indicating the protocol creates pathways from low to high energy regions of the Hilbert space. All parameters are the same as in Fig.~\ref{fig:relaxation}.
    }
    \label{fig:battery}
\end{figure*}
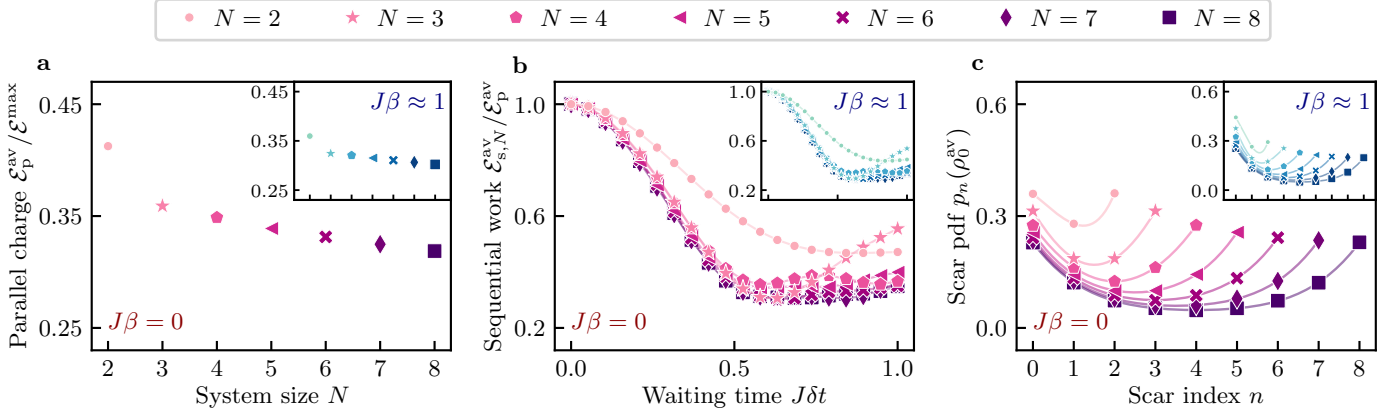

\textbf{Example.}---Let us illustrate our general construction with the paradigmatic example of a scarred many-body system: a spin-$1$ $XY$ chain with Hamiltonian
\begin{equation} \label{eq:exampleH}
    H = J \! \sum_{j = 1}^{N-1} \! (S_{j}^{x} S_{j+1}^{x} \! + \! S_{j}^{y} S_{j+1}^{y}) + \sum_{j=1}^{N} \! \left[h S_{j}^{z} \! + \! D (S_{j}^{z})^{2} \right] \! ,
\end{equation}
where $S_{j}^{\alpha}$ is the spin-$1$ operator in the $\alpha = x, y, z$ direction on site $j$ and $J, h , D > 0$ \cite{Schecter_2019}. This model hosts a tower of $N+1$ scars generated by successive application of the ladder operator $Q^{+} = \sum_{j = 1}^{N} (-1)^{j} (S_{j}^{x} + i S_{j}^{y})^{2} / 2$ onto the lowest energy scar, $\ket{s_{0}} = \ket{-1, \dots , -1}$. The Hamiltonian is consistent with the Shiraishi--Mori form \cite{Wang_2024, Omiya_2026} (see SI), and a possible choice for engineered dissipation and measurement operators is
\begin{equation} \label{the_choice}
    L_j = S_{j}^{x} (S_{j}^{x} S_{j+1}^{x} \! + \! S_{j}^{y} S_{j+1}^{y}) \quad\, \mathrm{and} \quad\, 
    M = \sum_{j = 1}^{N} S_{j}^{z}.
\end{equation}
Note that the $L_j$ are $2$-local operators, and $M$ is simply the total magnetization, different from $H_{\mathrm{ZE}} = \sum_{j} \left[h S_{j}^{z} + D (S_{j}^{z})^{2} \right]$ but sharing the same eigenvectors. The scars $\ket{s_n}$ are simultaneous eigenstates of $H$ and $M$ with eigenvalues $E_n = N (D-h) + 2 n h$ and $m_n = - (N-2n)$, respectively [so, here, $\omega = 2h$ and $\zeta = 2$; cf.~Eqs.~\eqref{eq:sga} and~\eqref{eq:deg}]. 

The $XY$ chain plays the role of a many-body battery. The protocol [Eq.~\eqref{eq:quantumTrajectory}] leads any initial state $\rho_{0}$ to a charged state $\rho_{\infty}^{\mathrm{c}} = \dyad{s_n}$ [Eq.~\eqref{eq:charging}]. Figure~\ref{fig:relaxation}(a) shows typical realizations for a given $\rho_{0}$ converging to different scars, as witnessed by the total magnetization $M$.
Figure~\ref{fig:relaxation}(b) presents the average charging time $\mathbb{E} [ \tau_n^\m{c}(\epsilon)]$ [Eq.~\eqref{eq:first-passage}], demonstrating that the concurrent implementation charges the battery at least $2.6$ times as fast compared to consecutive alternatives of the protocol (see SI for the full histograms).
The charged state of the battery is now easily certified by inspecting the time-integrated measurement current $\bar{I}_{t, t_{0}}$, displayed in Fig.~\ref{fig:relaxation}(c), which approaches an eigenvalue of $M$ [Eq.~\eqref{eq:current}].

Each spin of the chain may be considered as a single cell of the battery, which can be individually depleted by local operations. We show numerically in the SI that the optimal single-site unitaries $\check{u}_j$ extracting maximal parallel ergotropy [Eq.~\eqref{parallel_erg_def}] are all equal and given by a permutation in the local $S^z_j$ basis. Explicitly, we obtain
\begin{equation}\label{eq:XY-local-ergotropy}
    \begin{aligned}
        \ergl_\prll (\dyad{s_n }) = & \, n \left[ h + D + 4 J \left( 1 - n/N \right) \right]
        \\
        & + \theta (2 n - N) (h - D) (2 n - N),
    \end{aligned}
\end{equation}
where $\theta$ is the Heaviside step function. When $n \propto N$, the parallel charge of the corresponding scar becomes macroscopic: $\ergl_\prll (\dyad{s_n}) \propto N$.

As the scar selection process itself is stochastic, besides the per-scar ergotropy [Eq.~\eqref{eq:XY-local-ergotropy}], it is meaningful to consider the average asymptotic ergotropy obtained by sampling over an ensemble of random initial states, $\ergl_{\prll}^{\mathrm{av}} \coloneq \mathbb{E}_{\rho_0}[\ergl_\prll(\rho^c_\infty(\rho_0))]$. We show in the SI that this equals the mean ergotropy of the ensemble-averaged state $\rho_0^{\mathrm{av}} \coloneq \mathbb{E}_{\rho_0}[\rho_{0}]$:
\begin{equation} \label{av_prll_erg}
    \ergl_{\prll}^{\mathrm{av}} = \sum_{n = 0}^{N} p_n (\rho_{0}^{\mathrm{av}}) \, \ergl_\prll ( \dyad{s_n }).
\end{equation}
Figure~\ref{fig:battery}(a) shows $\ergl_{\prll}^{\mathrm{av}}$ scaled by the maximal capacity $\mathcal{E}^{\max}$, which equals the largest eigenvalue difference of $H$, with $\rho_{0}^{\mathrm{av}} \propto \exp (- \beta H)$, for different values of $\beta$.
As argued above, a more realistic usage scenario for a many-body battery is the sequential cell-by-cell discharging, characterized by $\ergl_{\seq,\ell}(\rho, \wt)$ [Eq.~\eqref{sequential_erg}]. Figure~\ref{fig:battery}(b) shows that its ensemble average, $\ergl_{\seq, N}^\mathrm{av}(\wt)$, is appreciably robust with respect to variations of $\wt$. Furthermore, we show in the SI that certain scars support stable, almost linear in $\ell$, discharging for waiting times as large as $J \wt \approx 0.42$.

The apparent macroscopicity of $\ergl_{\prll}^{\mathrm{av}}$ in Fig.~\ref{fig:battery}(a) results from the counterintuitive behavior of the asymptotic scar distribution $p_n(\rho_0^\m{av})$ displayed in Fig.~\ref{fig:battery}(c). Indeed, the large occupation of the negative-temperature half of the spectrum combined with $\ergl_\prll ( \dyad{s_n }) \propto N$ for large $n$ implies the scaling $\ergl_{\prll}^{\mathrm{av}} \propto N$.

\textbf{Discussion.}---The functionally pertinent measure of energy density for many-body batteries is the local ergotropy density, $\varepsilon \coloneq \ergl_\prll / N$ (as opposed to, e.g., full ergotropy per particle). Our platform-independent demonstration that scars can be isolated as steady states and that they hold a finite $\varepsilon$ [Fig.~\ref{fig:battery}] thus lays the groundwork for robust macroscopic quantum batteries. The analysis of sequential extraction additionally shows that such batteries are capable of providing stable discharge rates.

Beyond requiring only local control, the experimental feasibility of the protocol is furthered by the fact that, even though scars make up a vanishing fraction of all energy eigenstates, accessing them is well within the reach of current platforms. They have been observed in Rydberg atoms \cite{Bernien_2017}, ultracold Fermi gases \cite{Scherg_2021}, superconducting qubits \cite{Zhang_2023}, and graphene quantum dots \cite{Ge_2024}.

Several routes remain open for investigation. First, the engineered dissipation operators $L_{j}$ were provided by the Hamiltonian and chosen due to being local and realizable in quantum simulators \cite{Wang_2024}. However, depending on platform constraints, better choices may exist.
Second, while continuous measurements are routinely implemented in quantum optical setups \cite{Jordan_2024}, performing them in many-body systems still remains challenging.
Third, our protocol relies on exact Hilbert space structures which may be distorted in actual experimental realizations, causing leakage into non-scar subspaces and thus requiring additional stabilization strategies. Although our charging and extraction protocols are stable with respect to small imperfections, concrete error bounds are still needed.

\let\oldaddtocontents\addtocontents
\renewcommand{\addtocontents}[2]{}

\begin{acknowledgments}

F.S. and M.A. are grateful for support from the University of Stuttgart. K.H. is grateful for support from the University of Potsdam. The authors further thank Sofia Sevitz for helpful discussions and advice.

\medskip

During the final stages of this project, we found out of Refs.~\cite{Zhi_2025, Liu_2026}, where certain ergotropic features of scar states in spin-chain models are studied.

\end{acknowledgments}

\bibliography{references}

@book{Callen,
  author = {Callen, Herbert B.},
  title = {Thermodynamics and an Introduction to Thermostatistics},
  publisher = {Wiley, New York},
  edition = {2},
  year = {1985},
}

@article{Ye_2024,
  author = {Ye, Yusheng and Xu, Rong and Huang, Wenxiao and Ai, Huayue and Zhang, Wenbo and Affeld, Jordan Otto and Cui, Andy and Liu, Fang and Gao, Xin and Chen, Zhouyi and Li, Tony and Xiao, Xin and Zhang, Zewen and Peng, Yucan and Vila, Rafael A. and Wu, Yecun and Oyakhire, Solomon T. and Kuwajima, Hideaki and Suzuki, Yoshiaki and Matsumoto, Ryuhei and Masuda, Yasuyuki and Yuuki, Takahiro and Nakayama, Yuri and Cui, Yi},
  title = {Quadruple the rate capability of high-energy batteries through a porous current collector design},
  journal = {Nat. Energy},
  volume = {9},
  number = {6},
  pages = {643--653},
  year = {2024},
  publisher = {Springer Science and Business Media LLC},
  doi = {10.1038/s41560-024-01473-2},
}

@article{Hatsopoulos_1976,
  author = {Hatsopoulos, George N. and Gyftopoulos, Elias P.},
  title = {A unified quantum theory of mechanics and thermodynamics. Part IIa. Available energy},
  journal = {Found. Phys.},
  volume = {6},
  pages = {127--141},
  year = {1976},
  doi = {10.1007/BF00708955},
}

@article{Pusz_1978,
  author = {Pusz, Wies\l{}aw and Woronowicz, Stanis\l{}aw Lech},
  title = {Passive states and {KMS} states for general quantum systems},
  journal = {Commun. Math. Phys.},
  volume = {58},
  number = {3},
  pages = {273--290},
  issn = {1432-0916},
  year = {1978},
  doi = {10.1007/BF01614224},
}

@article{Lenard_1978,
  author = {Lenard, Andrew},
  title = {Thermodynamical proof of the {Gibbs} formula for elementary quantum systems},
  journal = {J. Stat. Phys.},
  volume = {19},
  issue = {6},
  pages = {575--586},
  year = {1978},
  doi = {10.1007/BF01011769},
}

@article{Allahverdyan_2004,
  title = {Maximal work extraction from finite quantum systems},
  author = {Allahverdyan, Armen E. and Balian, Roger and Nieuwenhuizen, Theodorus M.},
  journal = {Europhys. Lett},
  volume = {67},
  pages = {565--571},
  year = {2004},
  doi = {10.1209/epl/i2004-10101-2}
}

@article{Tirone_2021,
  title = {Quantum Energy Lines and the Optimal Output Ergotropy Problem},
  author = {Tirone, Salvatore and Salvia, Raffaele and Giovannetti, Vittorio},
  journal = {Phys. Rev. Lett.},
  volume = {127},
  issue = {21},
  pages = {210601},
  year = {2021},
  publisher = {American Physical Society},
  doi = {10.1103/PhysRevLett.127.210601},
}

@article{Safranek_2023,
  title = {Work Extraction from Unknown Quantum Sources},
  author = {\v{S}afr\'{a}nek, Dominik and Rosa, Dario and Binder, Felix C.},
  journal = {Phys. Rev. Lett.},
  volume = {130},
  issue = {21},
  pages = {210401},
  year = {2023},
  publisher = {American Physical Society},
  doi = {10.1103/PhysRevLett.130.210401},
}

@misc{Hovhannisyan_2024,
  author = {Hovhannisyan, Karen and Simon, Rick P. A. and Anders, Janet},
  title = {Concentration of ergotropy in many-body systems},
  year = {2024},
  eprint = {2412.19801},
  archivePrefix = {arXiv},
  primaryClass = {quant-ph},
}

@article{Chakraborty_2025,
  author = {Chakraborty, Shantanav and Das, Siddhartha and Ghorui, Arnab and Hazra, Soumyabrata and Singh, Uttam},
  title = {Sample Complexity of Black Box Work Extraction},
  journal = {Quantum Sci. Technol.},
  volume = {10},
  pages = {045070},
  year = {2025},
  doi = {10.1088/2058-9565/ae0e4d},
}

@article{Frey_2014,
  title = {Strong local passivity in finite quantum systems},
  author = {Frey, Michael and Funo, Ken and Hotta, Masahiro},
  journal = {Phys. Rev. E},
  volume = {90},
  issue = {1},
  pages = {012127},
  year = {2014},
  publisher = {American Physical Society},
  doi = {10.1103/PhysRevE.90.012127},
}

@article{Kaneko_2019,
  title = {Work extraction from a single energy eigenstate},
  author = {Kaneko, Kazuya and Iyoda, Eiki and Sagawa, Takahiro},
  journal = {Phys. Rev. E},
  volume = {99},
  issue = {3},
  pages = {032128},
  year = {2019},
  publisher = {American Physical Society},
  doi = {10.1103/PhysRevE.99.032128},
}

@misc{Baba_2023,
  author = {Baba, Shotaro Z. and Yoshioka, Nobuyuki and Sagawa, Takahiro},
  title = {Work extractability from energy eigenstates under optimized local operations},
  year = {2023},
  eprint = {2308.03537},
  archivePrefix = {arXiv},
  primaryClass = {quant-ph},
}

@article{Salvia_2023,
  title = {Optimal local work extraction from bipartite quantum systems in the presence of Hamiltonian couplings},
  author = {Salvia, Raffaele and De Palma, Giacomo and Giovannetti, Vittorio},
  journal = {Phys. Rev. A},
  volume = {107},
  issue = {1},
  pages = {012405},
  year = {2023},
  publisher = {American Physical Society},
  doi = {10.1103/PhysRevA.107.012405},
}

@article{Castellano_2024,
  title = {Extended Local Ergotropy},
  author = {Castellano, Riccardo and Farina, Donato and Giovannetti, Vittorio and Acin, Antonio},
  journal = {Phys. Rev. Lett.},
  volume = {133},
  issue = {15},
  pages = {150402},
  year = {2024},
  publisher = {American Physical Society},
  doi = {10.1103/PhysRevLett.133.150402},
}

@misc{Zhi_2025,
  author = {Zhi, Zhaohui and Qian, Qingyun and Liu, Jin-Guo and Zhu, Guo-Yi},
  title = {Ergotropy of quantum many-body scars},
  year = {2025},
  eprint = {2512.19801},
  archivePrefix = {arXiv},
  primaryClass = {quant-ph},
}

@misc{Liu_2026,
  author = {Liu, Jinglin and Liu, Yiming and Xu, Mingdi and Pan, Lei},
  title = {Dissipative Quantum Battery from Many-Body Scars},
  year = {2026},
  eprint = {2609.11595},
  archivePrefix = {arXiv},
  primaryClass = {quant-ph},
}

@article{Hokkyo_2025,
  title = {Universal Upper Bound on Ergotropy and No-Go Theorem by the Eigenstate Thermalization Hypothesis},
  author = {Hokkyo, Akihiro and Ueda, Masahito},
  journal = {Phys. Rev. Lett.},
  volume = {134},
  issue = {1},
  pages = {010406},
  year = {2025},
  publisher = {American Physical Society},
  doi = {10.1103/PhysRevLett.134.010406},
}

@misc{Chiba_2026,
  author = {Chiba, Yuuya and Yoneta, Yasushi and Hamazaki, Ryusuke and Shimizu, Akira},
  title = {Second law of thermodynamics in closed quantum many-body systems},
  year = {2026},
  eprint = {2602.06657},
  archivePrefix = {arXiv},
  primaryClass = {quant-ph},
}

@article{Campaioli_2017,
  title = {Enhancing the Charging Power of Quantum Batteries},
  author = {Campaioli, Francesco and Pollock, Felix A. and Binder, Felix C. and C\'eleri, Lucas and Goold, John and Vinjanampathy, Sai and Modi, Kavan},
  journal = {Phys. Rev. Lett.},
  volume = {118},
  issue = {15},
  pages = {150601},
  year = {2017},
  publisher = {American Physical Society},
  doi = {10.1103/PhysRevLett.118.150601},
}

@article{Campaioli_2024,
  title = {Colloquium: Quantum batteries},
  author = {Campaioli, Francesco and Gherardini, Stefano and Quach, James Q. and Polini, Marco and Andolina, Gian Marcello},
  journal = {Rev. Mod. Phys.},
  volume = {96},
  issue = {3},
  pages = {031001},
  year = {2024},
  publisher = {American Physical Society},
  doi = {10.1103/RevModPhys.96.031001},
}

@article{Alicki_2013,
  title = {Entanglement boost for extractable work from ensembles of quantum batteries},
  author = {Alicki, Robert and Fannes, Mark},
  journal = {Phys. Rev. E},
  volume = {87},
  issue = {4},
  pages = {042123},
  year = {2013},
  publisher = {American Physical Society},
  doi = {10.1103/PhysRevE.87.042123},
}

@article{Hovhannisyan_2013,
  title = {Entanglement Generation is Not Necessary for Optimal Work Extraction},
  author = {Hovhannisyan, Karen V. and Perarnau-Llobet, Mart\'{i} and Huber, Marcus and Ac\'{i}n, Antonio},
  journal = {Phys. Rev. Lett.},
  volume = {111},
  issue = {24},
  pages = {240401},
  year = {2013},
  publisher = {American Physical Society},
  doi = {10.1103/PhysRevLett.111.240401},
}

@article{Perarnau_2015,
  title = {Extractable Work from Correlations},
  author = {Perarnau-Llobet, Martí and Hovhannisyan, Karen V. and Huber, Marcus and Skrzypczyk, Paul and Brunner, Nicolas and Acín, Antonio},
  journal = {Phys. Rev. X},
  volume = {5},
  issue = {4},
  pages = {041011},
  year = {2015},
  publisher = {American Physical Society},
  doi = {10.1103/PhysRevX.5.041011},
}

@article{Andolina_2019,
  title = {Extractable Work, the Role of Correlations, and Asymptotic Freedom in Quantum Batteries},
  author = {Andolina, Gian Marcello and Keck, Maximilian and Mari, Andrea and Campisi, Michele and Giovannetti, Vittorio and Polini, Marco},
  journal = {Phys. Rev. Lett.},
  volume = {122},
  issue = {4},
  pages = {047702},
  year = {2019},
  publisher = {American Physical Society},
  doi = {10.1103/PhysRevLett.122.047702},
}

@article{Bernards_2019,
  author = {Bernards, Fabian and Kleinmann, Matthias and Gühne, Otfried and Paternostro, Mauro},
  title = {Daemonic Ergotropy: Generalised Measurements and Multipartite Settings},
  journal = {Entropy},
  volume = {21},
  number = {8},
  pages = {771},
  year = {2019},
  doi = {10.3390/e21080771},
}

@article{Gyhm_2022,
  title = {Quantum Charging Advantage Cannot Be Extensive without Global Operations},
  author = {Gyhm, Ju-Yeon and \v{S}afr\'{a}nek, Dominik and Rosa, Dario},
  journal = {Phys. Rev. Lett.},
  volume = {128},
  issue = {14},
  pages = {140501},
  year = {2022},
  publisher = {American Physical Society},
  doi = {10.1103/PhysRevLett.128.140501},
}

@book{Simon,
  author = {Simon, Barry},
  title = {The Statistical Mechanics of Lattice Gases},
  volume = {1},
  publisher = {Princeton University Press, Princeton},
  year = {1993},
}

@book{HornJohnson,
  author = {Horn, Roger A. and Johnson, Charles R.},
  title = {Matrix analysis},
  edition = {2},
  publisher = {Cambridge University Press, New York},
  year = {2013},
}

@article{DAlessio_2016,
  author = {D'Alessio, Luca and Kafri, Yariv and Polkovnikov, Anatoli and Rigol, Marcos},
  title = {From quantum chaos and eigenstate thermalization to statistical mechanics and thermodynamics},
  journal = {Ad. Phys.},
  volume = {65},
  number = {3},
  pages = {239--362},
  year = {2016},
  publisher = {Taylor and Francis},
  doi = {10.1080/00018732.2016.1198134},
}

@article{Gogolin_2016,
  author = {Gogolin, Christian and Eisert, Jens},
  title = {Equilibration, thermalisation, and the emergence of statistical mechanics in closed quantum systems},
  journal = {Rep. Prog. Phys.},
  volume = {79},
  number = {5},
  pages = {056001},
  year = {2016},
  publisher = {{IOP} Publishing},
  doi = {10.1088/0034-4885/79/5/056001},
}

@article{Araki_1969,
  author = {Araki, Huzihiro},
  title = {Gibbs states of a one dimensional quantum lattice},
  journal = {Commun. Math. Phys.},
  volume = {14},
  pages = {120--157},
  year = {1969},
  doi = {10.1007/BF01645134},
}

@article{Lieb_1972,
  author = {Lieb, Elliott H. and Robinson, Derek W.},
  title = {The finite group velocity of quantum spin systems},
  journal = {Commun.Math. Phys.},
  volume = {28},
  pages = {251--257},
  year = {1972},
  doi = {10.1007/BF01645779},
}

@article{Araki_1974,
  author = {Araki, Huzihiro},
  title = {On the equivalence of the {KMS} condition and the variational principle for quantum lattice systems},
  journal = {Commun. Math. Phys.},
  volume = {38},
  pages = {1--10},
  year = {1974},
  doi = {10.1007/BF01651545},
}

@article{Berry_1977,
  title = {Regular and irregular semiclassical wavefunctions},
  author = {Berry, Michael V.},
  year = {1977},
  publisher = {IOP Publishing},
  journal = {J. Phys. A},
  volume = {10},
  pages = {2083--2091},
  doi = {10.1088/0305-4470/10/12/016},
}

@article{Deutsch_1991,
  title = {Quantum statistical mechanics in a closed system},
  author = {Deutsch, J. M.},
  journal = {Phys. Rev. A},
  volume = {43},
  issue = {4},
  pages = {2046--2049},
  year = {1991},
  publisher = {American Physical Society},
  doi = {10.1103/PhysRevA.43.2046},
}

@article{Srednicki_1992,
  title = {Chaos and quantum thermalization},
  author = {Srednicki, Mark},
  journal = {Phys. Rev. E},
  volume = {50},
  issue = {2},
  pages = {888--901},
  year = {1994},
  publisher = {American Physical Society},
  doi = {10.1103/PhysRevE.50.888},
}

@article{Tasaki_1998,
  title = {From Quantum Dynamics to the Canonical Distribution: General Picture and a Rigorous Example},
  author = {Tasaki, Hal},
  journal = {Phys. Rev. Lett.},
  volume = {80},
  issue = {7},
  pages = {1373--1376},
  year = {1998},
  publisher = {American Physical Society},
  doi = {10.1103/PhysRevLett.80.1373},
}

@article{Hartmann_2005,
  author = {Hartmann, Michael and Mahler, Gunther and Hess, Ortwin},
  title = {Spectral Densities and Partition Functions of Modular Quantum ystems as Derived from a Central Limit Theorem},
  journal = {J. Stat. Phys.},
  volume = {119},
  pages = {1139--1151},
  year = {2005},
  doi = {10.1007/s10955-004-4298-5},
}

@article{Popescu_2006,
  title = {Entanglement and the foundations of statistical mechanics},
  author = {Sandu Popescu and Anthony J. Short and Andreas Winter},
  year = {2006},
  publisher = {Springer Science and Business Media LLC},
  journal = {Nat. Phys.},
  volume = {2},
  pages = {754--758},
  doi = {10.1038/nphys444},
}

@article{Rigol_2008,
  title = {Thermalization and its mechanism for generic isolated quantum systems},
  author = {Marcos Rigol and Vanja Dunjko and Maxim Olshanii},
  year = {2008},
  publisher = {Springer Science and Business Media LLC},
  journal = {Nature},
  volume = {452},
  pages = {854-858},
  doi = {10.1038/nature06838},
}

@article{Muller_2015,
  author = {Müller, Markus P. and Adlam, Emily and Masanes, Lluís and Wiebe, Nathan},
  title = {Thermalization and Canonical Typicality in Translation-Invariant Quantum Lattice Systems},
  journal = {Commun. Math. Phys.},
  volume = {340},
  pages = {499–561},
  year = {2015},
  doi = {10.1007/s00220-015-2473-y},
}

@article{Keating_2015,
  author = {Keating, Jonathan P. and Linden, Noah and Wells, Huw J.},
  title = {Spectra and Eigenstates of Spin Chain Hamiltonians},
  journal = {Commun. Math. Phys.},
  volume = {338},
  pages = {81--102},
  year = {2015},
  doi = {10.1007/s00220-015-2366-0},
}

@misc{Brandao_2015,
  author = {Brandão, Fernando G. S. L. and Cramer, Marcus},
  title = {Equivalence of Statistical Mechanical Ensembles for Non-Critical Quantum Systems},
  year = {2015},
  eprint = {1502.03263},
  archivePrefix = {arXiv},
  primaryClass = {quant-ph},
}

@article{Goldstein_2015,
  title = {Thermal Equilibrium of a Macroscopic Quantum System in a Pure State},
  author = {Goldstein, Sheldon and Huse, David A. and Lebowitz, Joel L. and Tumulka, Roderich},
  journal = {Phys. Rev. Lett.},
  volume = {115},
  issue = {10},
  pages = {100402},
  year = {2015},
  publisher = {American Physical Society},
  doi = {10.1103/PhysRevLett.115.100402},
}

@article{Shiraishi_2017,
  title = {Systematic Construction of Counterexamples to the Eigenstate Thermalization Hypothesis},
  author = {Shiraishi, Naoto and Mori, Takashi},
  journal = {Phys. Rev. Lett.},
  volume = {119},
  issue = {3},
  pages = {030601},
  year = {2017},
  publisher = {American Physical Society},
  doi = {10.1103/PhysRevLett.119.030601},
}

@article{Bernien_2017,
  title = {Probing many-body dynamics on a 51-atom quantum simulator},
  author = {Hannes Bernien and Sylvain Schwartz and Alexander Keesling and Harry Levine and Ahmed Omran and Hannes Pichler and Soonwon Choi and Alexander S. Zibrov and Manuel Endres and Markus Greiner and Vladan Vuletić and Mikhail D. Lukin},
  year = {2017},
  publisher = {Springer Science and Business Media LLC},
  journal = {Nature},
  volume = {551},
  pages = {579-584},
  doi = {10.1038/nature24622},
}

@article{Turner_2018,
  author = {C. J. Turner and A. A. Michailidis and D. A. Abanin and M. Serbyn and Z. Papić},
  title = {Weak ergodicity breaking from quantum many-body scars},
  journal = {Nat. Phys.},
  volume = {14},
  pages = {745-749},
  year = {2018},
  publisher = {Springer Science and Business Media LLC},
  doi = {10.1038/s41567-018-0137-5},
}

@article{Moudgalya_2018,
  author = {Moudgalya, Sanjay and Rachel, Stephan and Bernevig, B. Andrei and Regnault, Nicolas},
  title = {Exact excited states of nonintegrable models},
  journal = {Phys. Rev. B},
  volume = {98},
  issue = {23},
  pages = {235155},
  year = {2018},
  publisher = {American Physical Society},
  doi = {10.1103/PhysRevB.98.235155},
}

@article{Tasaki_2018,
  author = {Tasaki, Hal},
  title = {On the Local Equivalence Between the Canonical and the Microcanonical Ensembles for Quantum Spin Systems},
  journal = {J. Stat. Phys.},
  volume = {172},
  issue = {4},
  pages = {905--926},
  year = {2018},
  doi = {10.1007/s10955-018-2077-y},
}

@article{Dymarsky_2018,
  title = {Subsystem eigenstate thermalization hypothesis},
  author = {Dymarsky, Anatoly and Lashkari, Nima and Liu, Hong},
  journal = {Phys. Rev. E},
  volume = {97},
  issue = {1},
  pages = {012140},
  numpages = {7},
  year = {2018},
  publisher = {American Physical Society},
  doi = {10.1103/PhysRevE.97.012140},
}

@article{Nachtergaele_2019,
  author = {Nachtergaele, Bruno and Sims, Robert and Young, Amanda},
  title = {Quasi-locality bounds for quantum lattice systems. I. {Lieb}--{Robinson} bounds, quasi-local maps, and spectral flow automorphisms},
  journal = {J. Math. Phys.},
  volume = {60},
  number = {6},
  pages = {061101},
  year = {2019},
  doi = {10.1063/1.5095769},
}

@article{Choi_2019,
  author = {Choi, Soonwon and Turner, Christopher J. and Pichler, Hannes and Ho, Wen Wei and Michailidis, Alexios A. and Papić, Zlatko and Serbyn, Maksym and Lukin, Mikhail D. and Abanin, Dmitry A.},
  title = {Emergent $\mathrm{SU}(2)$ Dynamics and Perfect Quantum Many-Body Scars},
  journal = {Phys. Rev. Lett.},
  volume = {122},
  issue = {22},
  pages = {220603},
  year = {2019},
  publisher = {American Physical Society},
  doi = {10.1103/PhysRevLett.122.220603},
}

@article{Schecter_2019,
  title = {Weak Ergodicity Breaking and Quantum Many-Body Scars in Spin-1 $XY$ Magnets},
  author = {Schecter, Michael and Iadecola, Thomas},
  journal = {Phys. Rev. Lett.},
  volume = {123},
  issue = {14},
  pages = {147201},
  year = {2019},
  publisher = {American Physical Society},
  doi = {10.1103/PhysRevLett.123.147201},
}

@article{Kuwahara_2020ETH,
  title = {Eigenstate Thermalization from the Clustering Property of Correlation},
  author = {Kuwahara, Tomotaka and Saito, Keiji},
  journal = {Phys. Rev. Lett.},
  volume = {124},
  issue = {20},
  pages = {200604},
  year = {2020},
  publisher = {American Physical Society},
  doi = {10.1103/PhysRevLett.124.200604},
}

@article{Kuwahara_2020LiebRobi,
  title = {Strictly Linear Light Cones in Long-Range Interacting Systems of Arbitrary Dimensions},
  author = {Kuwahara, Tomotaka and Saito, Keiji},
  journal = {Phys. Rev. X},
  volume = {10},
  issue = {3},
  pages = {031010},
  year = {2020},
  publisher = {American Physical Society},
  doi = {10.1103/PhysRevX.10.031010},
}

@article{ODea_2020,
  title = {From tunnels to towers: Quantum scars from {Lie} algebras and $q$-deformed {Lie} algebras},
  author = {O'Dea, Nicholas and Burnell, Fiona and Chandran, Anushya and Khemani, Vedika},
  journal = {Phys. Rev. Res.},
  volume = {2},
  issue = {4},
  pages = {043305},
  year = {2020},
  publisher = {American Physical Society},
  doi = {10.1103/PhysRevResearch.2.043305},
}

@article{Mark_2020,
  title = {Unified structure for exact towers of scar states in the {Affleck}--{Kennedy}--{Lieb}--{Tasaki} and other models},
  author = {Mark, Daniel K. and Lin, Cheng-Ju and Motrunich, Olexei I.},
  journal = {Phys. Rev. B},
  volume = {101},
  issue = {19},
  pages = {195131},
  year = {2020},
  publisher = {American Physical Society},
  doi = {10.1103/PhysRevB.101.195131},
}

@article{Ren_2021,
  title = {Quasisymmetry Groups and Many-Body Scar Dynamics},
  author = {Ren, Jie and Liang, Chenguang and Fang, Chen},
  journal = {Phys. Rev. Lett.},
  volume = {126},
  issue = {12},
  pages = {120604},
  year = {2021},
  publisher = {American Physical Society},
  doi = {10.1103/PhysRevLett.126.120604},
}

@article{Scherg_2021,
  title = {Observing non-ergodicity due to kinetic constraints in tilted {Fermi}--{Hubbard} chains},
  author = {Scherg, Sebastian and Kohlert, Thomas and Sala, Pablo and Pollmann, Frank and Madhusudhana, Bharath Hebbe and Bloch, Immanuel and Aidelsburger, Monika},
  journal = {Nat. Commun.},
  volume = {12},
  pages = {4490},
  year = {2021},
  publisher = {Springer Science and Business Media LLC},
  doi = {10.1038/s41467-021-24726-0},
}

@article{Serbyn_2021,
  title = {Quantum many-body scars and weak breaking of ergodicity},
  author = {Maksym Serbyn and Dmitry A. Abanin and Zlatko Papić},
  journal = {Nat. Phys.},
  volume = {17},
  pages = {675--685},
  year = {2021},
  publisher = {Springer Science and Business Media LLC},
  doi = {10.1038/s41567-021-01230-2},
}

@article{Cresser_2021,
  title = {Weak and Ultrastrong Coupling Limits of the Quantum Mean Force {Gibbs} State},
  author = {Cresser, James D. and Anders, Janet},
  journal = {Phys. Rev. Lett.},
  volume = {127},
  issue = {25},
  pages = {250601},
  year = {2021},
  publisher = {American Physical Society},
  doi = {10.1103/PhysRevLett.127.250601},
}

@article{Trushechkin_2022,
  author = {Trushechkin, Anton S. and Merkli, Marco and Cresser, James D. and Anders, Janet},
  title = {Open quantum system dynamics and the mean force {Gibbs} state},
  journal = {AVS Quantum Sci.},
  volume = {4},
  number = {1},
  pages = {012301},
  year = {2022},
  doi = {10.1116/5.0073853},
}

@article{Moudgalya_2022,
  author = {Moudgalya, Sanjay and Bernevig, B Andrei and Regnault, Nicolas},
  title = {Quantum many-body scars and {Hilbert} space fragmentation: a review of exact results},
  journal = {Rep. Prog. Phys.},
  volume = {85},
  number = {8},
  pages = {086501},
  year = {2022},
  publisher = {IOP Publishing},
  doi = {10.1088/1361-6633/ac73a0},
}

@article{Gotta_2023,
  title = {Asymptotic Quantum Many-Body Scars},
  author = {Gotta, Lorenzo and Moudgalya, Sanjay and Mazza, Leonardo},
  journal = {Phys. Rev. Lett.},
  volume = {131},
  issue = {19},
  pages = {190401},
  year = {2023},
  publisher = {American Physical Society},
  doi = {10.1103/PhysRevLett.131.190401},
}

@article{Zhang_2023,
  author = {Pengfei Zhang and Hang Dong and Yu Gao and Liangtian Zhao and Jie Hao and Jean-Yves Desaules and Qiujiang Guo and Jiachen Chen and Jinfeng Deng and Bobo Liu and Wenhui Ren and Yunyan Yao and Xu Zhang and Shibo Xu and Ke Wang and Feitong Jin and Xuhao Zhu and Bing Zhang and Hekang Li and Chao Song and Zhen Wang and Fangli Liu and Zlatko Papić and Lei Ying and H. Wang and Ying-Cheng Lai},
  title = {Many-body {Hilbert} space scarring on a superconducting processor},
  journal = {Nat. Phys.},
  volume = {19},
  pages = {120--125},
  year = {2022},
  publisher = {Springer Science and Business Media LLC},
  doi = {10.1038/s41567-022-01784-9},
}

@article{Chen_2023,
  author = {Chen, Chi-Fang (Anthony) and Lucas, Andrew and Yin, Chao},
  title = {Speed limits and locality in many-body quantum dynamics},
  journal = {Rep. Progr. Phys.},
  volume = {86},
  number = {11},
  pages = {116001},
  year = {2023},
  publisher = {IOP Publishing},
  doi = {10.1088/1361-6633/acfaae},
}

@article{Ge_2024,
  author = {Zhehao Ge and Anton M. Graf and Joonas Keski-Rahkonen and Sergey Slizovskiy and Peter Polizogopoulos and Takashi Taniguchi and Kenji Watanabe and Ryan Van Haren and David Lederman and Vladimir I. Fal’ko and Eric J. Heller and Jairo Velasco},
  title = {Direct visualization of relativistic quantum scars in graphene quantum dots},
  journal = {Nature},
  volume = {635},
  pages = {841--846},
  year = {2024},
  publisher = {Springer Science and Business Media LLC},
  doi = {10.1038/s41586-024-08190-6},
}

@article{Wang_2024,
  title = {Embedding Quantum Many-Body Scars into Decoherence-Free Subspaces},
  author = {Wang, He-Ran and Yuan, Dong and Zhang, Shun-Yao and Wang, Zhong and Deng, Dong-Ling and Duan, L.-M.},
  journal = {Phys. Rev. Lett.},
  volume = {132},
  issue = {15},
  pages = {150401},
  numpages = {8},
  year = {2024},
  publisher = {American Physical Society},
  doi = {10.1103/PhysRevLett.132.150401},
}

@article{Moudgalya_2024,
  title = {Exhaustive Characterization of Quantum Many-Body Scars Using Commutant Algebras},
  author = {Moudgalya, Sanjay and Motrunich, Olexei I.},
  journal = {Phys. Rev. X},
  volume = {14},
  issue = {4},
  pages = {041069},
  year = {2024},
  publisher = {American Physical Society},
  doi = {10.1103/PhysRevX.14.041069},
}

@misc{McGinley_2025,
  author = {McGinley, Max and Schuster, Thomas},
  title = {The {Scrooge} ensemble in many-body quantum systems},
  year = {2025},
  eprint = {2511.17172},
  archivePrefix = {arXiv},
  primaryClass = {quant-ph},
}

@misc{Matsui_2025,
  author = {Matsui, Chihiro},
  title = {Scar subspaces stabilized by algebraic closure: Beyond equally-spaced spectra and exact solvability},
  year = {2026},
  eprint = {2604.11015},
  archivePrefix = {arXiv},
  primaryClass = {quant-ph},
}

@misc{Omiya_2026,
  author = {Omiya, Keita},
  title = {Any local Hamiltonian with ferromagnetic quantum many-body scars has a generalized Shiraishi-Mori form},
  year = {2026},
  eprint = {2601.11806},
  archivePrefix = {arXiv},
  primaryClass = {quant-ph},
}

@misc{Schmolke_2025,
  author = {Schmolke, Finn},
  title = {Asymptotic Fate of Continuously Monitored Quantum Systems},
  year = {2025},
  eprint = {2506.10873},
  archivePrefix = {arXiv},
  primaryClass = {quant-ph},
}

@article{Albarelli_2024,
  author = {Albarelli, Francesco and Genoni, Marco G.},
  title = {A pedagogical introduction to continuously monitored quantum systems and measurement-based feedback},
  journal = {Phys. Lett. A}, 
  volume = {494},
  pages = {129260},
  publisher = {Elsevier BV},  
  year = {2024}, 
  doi = {10.1016/j.physleta.2023.129260}, 
}

@book{Wiseman_2009, 
  author = {Wiseman, Howard M. and Milburn, Gerard J.},
  place = {Cambridge},
  title = {Quantum Measurement and Control},
  publisher = {Cambridge University Press},
  year = {2009},
  doi = {10.1017/CBO9780511813948},
}

@book{Jordan_2024,
  author = {Jordan, Andrew N. and Siddiqi, Irfan A.},
  title = {Quantum Measurement: Theory and Practice},
  year = {2024},
  publisher={Cambridge University Press},
  isbn = {9781009100069},
  doi = {10.1017/9781009103909},
}

@book{Barchielli_2009,
  author = {Barchielli, Alberto and Gregoratti, Matteo},
  title = {Quantum Trajectories and Measurements in Continuous Time},
  subtitle = {The Diffusive Case},
  publisher = {Springer Berlin, Heidelberg},
  year = {2009},
  isbn = {978-3-642-01298-3},
  doi = {10.1007/978-3-642-01298-3},
}

@article{Benoist_2014,
  author={Benoist, Tristan and Pellegrini, Clément},
  title = {Large Time Behavior and Convergence Rate for Quantum Filters Under Standard Non Demolition Conditions},
  journal = {Commun. Math. Phys.},
  volume = {331},
  number = {2},
  pages={703--723},
  year = {2014},
  publisher = {Springer Science and Business Media LLC},
  doi = {10.1007/s00220-014-2029-6},
}

@article{Benoist_2017,
  author = {Benoist, Tristan and Pellegrini, Clément and Ticozzi, Francesco},
  title = {Exponential Stability of Subspaces for Quantum Stochastic Master Equations},
  journal = {Ann. Henri Poincaré},
  volume = {18},
  number = {6},
  pages = {2045-–2074},
  year = {2017},
  publisher = {Springer Science and Business Media LLC},
  doi = {10.1007/s00023-017-0556-3},
}

@misc{Benoist_2025,
  author = {Tristan Benoist and Linda Greggio and Clément Pellegrini},
  title = {Exponentially fast selection of sectors for quantum trajectories beyond non demolition measurements},
  year = {2025},
  eprint = {2407.18864},
  archivePrefix = {arXiv},
  primaryClass = {math-ph},
}

@article{Roldan_2023,
  author = {Roldán, \'{E}dgar and Neri, Izaak and Chetrite, Raphael and Gupta, Shamik and Pigolotti, Simone and Jülicher, Frank and Sekimoto, Ken},
  title = {Martingales for physicists: a treatise on stochastic thermodynamics and beyond},
  journal = {Adv. Phys.},
  volume = {72},
  number = {1-2},
  pages = {1--258},
  year = {2023},
  publisher = {Taylor and Francis},
  doi = {10.1080/00018732.2024.2317494},
}

@article{Baumgartner_2008,
  author = {Baumgartner, Bernhard and Narnhofer, Heide},
  title = {Analysis of quantum semigroups with {GKS}–{Lindblad} generators: {II}. General},
  journal = {J. Phys. A},
  volume = {41},
  number = {39},
  pages = {395303},
  year = {2008}, 
  doi = {10.1088/1751-8113/41/39/395303},
}

@article{Baumgartner_2012,
  author = {Baumgartner, Bernhard and Narnhofer, Heide},
  title = {The structures of state space concerning quantum dynamical semigroups},
  journal = {Rev. Math. Phys.},
  volume = {24},
  number = {2},
  pages = {1250001},
  year = {2012},
  doi = {10.1142/S0129055X12500018},
}

@article{Ticozzi_2012,
  author = {Ticozzi, Francesco and Lucchese, Riccardo and Cappellaro, Paola and Viola, Lorenza},
  title = {Hamiltonian Control of Quantum Dynamical Semigroups: Stabilization and Convergence Speed},
  journal = {IEEE Trans. Automatic Control},
  volume = {57},
  number = {8},
  pages = {1931--1944},
  year={2012},
  doi = {10.1109/TAC.2012.2195858}
}

@article{Albert_2016,
  author = {Albert, Victor V. and Bradlyn, Barry and Fraas, Martin and Jiang, Liang},
  title = {Geometry and Response of {Lindbladians}},
  journal = {Phys. Rev. X},
  volume = {6},
  issue = {4},
  pages = {041031},
  year = {2016},
  publisher = {American Physical Society},
  doi = {10.1103/PhysRevX.6.041031},
}

@misc{Ladenburger_2025,
  title = {Universal first-passage time statistics for quantum diffusion}, 
  author = {Ladenburger, Guido and Schmolke, Finn and Lutz, Eric},
  year = {2025},
  eprint = {2511.03455},
  archivePrefix = {arXiv},
  primaryClass = {quant-ph},
}

@article{Zanardi_2014,
  title = {Coherent Quantum Dynamics in Steady-State Manifolds of Strongly Dissipative Systems},
  author = {Zanardi, Paolo and Campos Venuti, Lorenzo},
  journal = {Phys. Rev. Lett.},
  volume = {113},
  issue = {24},
  pages = {240406},
  year = {2014},
  publisher = {American Physical Society},
  doi = {10.1103/PhysRevLett.113.240406},
}

@article{Diehl_2008,
  author = {Diehl, Sebastian and Micheli, Andrea and Kantian, Adrian and Kraus, Barbara and Büchler, Hans Peter and Zoller, Peter},
  title = {Quantum states and phases in driven open quantum systems with cold atoms},
  journal = {Nat. Phys.},
  volume = {4},
  number = {11},
  pages = {878–-883},
  year = {2008},
  publisher = {Springer Science and Business Media LLC},
  doi = {10.1038/nphys1073},
}

@article{Verstraete_2009,
  author = {Verstraete, Frank and Wolf, Michael M. and Ignacio Cirac, J.},
  title = {Quantum computation and quantum-state engineering driven by dissipation},
  journal = {Nat. Phys.},
  volume = {5},
  number = {9}, 
  pages = {633–636}, 
  year = {2009},
  publisher = {Springer Science and Business Media LLC},
  doi = {10.1038/nphys1342},
}

@article{Buca_2019,
  author = {Buča, Berislav and Tindall, Joseph and Jaksch, Dieter},
  title = {Non-stationary coherent quantum many-body dynamics through dissipation},
  journal = {Nat. Commun.},
  volume = {10},
  number = {1},
  pages = {1730},
  year = {2019},
  publisher = {Springer Science and Business Media LLC}, 
  doi = {10.1038/s41467-019-09757-y},
}

@article{Harrington_2022,
  author = {Harrington, Patrick M. and Mueller, Erich J. and Murch, Kater W.},
  title = {Engineered dissipation for quantum information science},
  journal = {Nat. Rev. Phys.},
  volume = {4},
  number = {10},
  pages = {660-–671},
  year = {2022},
  publisher = {Springer Science and Business Media LLC}, 
  doi = {10.1038/s42254-022-00494-8},
}

@article{Lidar_1998,
   author = {D. A. Lidar and I. L. Chuang and K. B. Whaley},
   doi = {10.1103/PhysRevLett.81.2594},
   issn = {0031-9007},
   issue = {12},
   journal = {Physical Review Letters},
   month = {9},
   pages = {2594-2597},
   title = {Decoherence-Free Subspaces for Quantum Computation},
   volume = {81},
   year = {1998},
}

@article{Znidaric_2015,
  title = {Relaxation times of dissipative many-body quantum systems},
  author = {\v{Z}nidari\v{c}, Marko},
  journal = {Phys. Rev. E},
  volume = {92},
  issue = {4},
  pages = {042143},
  numpages = {17},
  year = {2015},
  publisher = {American Physical Society},
  doi = {10.1103/PhysRevE.92.042143},
}

@article{Song_2019,
  title = {Non-{Hermitian} Skin Effect and Chiral Damping in Open Quantum Systems},
  author = {Song, Fei and Yao, Shunyu and Wang, Zhong},
  journal = {Phys. Rev. Lett.},
  volume = {123},
  issue = {17},
  pages = {170401},
  year = {2019},
  publisher = {American Physical Society},
  doi = {10.1103/PhysRevLett.123.170401},
}

@article{Mori_2020,
  title = {Resolving a Discrepancy between {Liouvillian} Gap and Relaxation Time in Boundary-Dissipated Quantum Many-Body Systems},
  author = {Mori, Takashi and Shirai, Tatsuhiko},
  journal = {Phys. Rev. Lett.},
  volume = {125},
  issue = {23},
  pages = {230604},
  year = {2020},
  publisher = {American Physical Society},
  doi = {10.1103/PhysRevLett.125.230604},
}

@article{Lee_2023,
  title = {Anomalously large relaxation times in dissipative lattice models beyond the non-{Hermitian} skin effect},
  author = {Lee, Gideon and McDonald, Alexander and Clerk, Aashish},
  journal = {Phys. Rev. B},
  volume = {108},
  issue = {6},
  pages = {064311},
  year = {2023},
  publisher = {American Physical Society},
  doi = {10.1103/PhysRevB.108.064311},
}

@article{Haga_2021,
  title = {Liouvillian Skin Effect: Slowing Down of Relaxation Processes without Gap Closing},
  author = {Haga, Taiki and Nakagawa, Masaya and Hamazaki, Ryusuke and Ueda, Masahito},
  journal = {Phys. Rev. Lett.},
  volume = {127},
  issue = {7},
  pages = {070402},
  year = {2021},
  publisher = {American Physical Society},
  doi = {10.1103/PhysRevLett.127.070402},
}

@article{Znidaric_2023,
  title = {Phantom relaxation rate of the average purity evolution in random circuits due to {Jordan} non-{Hermitian} skin effect and magic sums},
  author = {\v{Z}nidari\v{c}, Marko},
  journal = {Phys. Rev. Res.},
  volume = {5},
  issue = {3},
  pages = {033145},
  year = {2023},
  publisher = {American Physical Society},
  doi = {10.1103/PhysRevResearch.5.033145},
}

@book{Loudon_2000,
  author = {Loudon, Rodney},
  title = {The Quantum Theory of Light},
  publisher = {Oxford University Press, New York},
  year = {2000},
}

@article{Zurek_2003,
  title = {Decoherence, einselection, and the quantum origins of the classical},
  author = {Zurek, Wojciech Hubert},
  journal = {Rev. Mod. Phys.},
  volume = {75},
  issue = {3},
  pages = {715--775},
  year = {2003},
  publisher = {American Physical Society},
  doi = {10.1103/RevModPhys.75.715},
}

@article{Preskill_2018,
  title = {Quantum computing in the {NISQ} era and beyond},
  author = {Preskill, John},
  journal = {Quantum},
  volume = {2},
  pages = {79},
  year = {2018},
  publisher = {Verein zur F\"{o}rderung des Open Access Publizierens in den Quantenwissenschaften},
  doi = {10.22331/q-2018-08-06-79},
}

@article{Bronzan_1988,
  title = {Parametrization of {SU}$(3)$},
  author = {Bronzan, J. B.},
  journal = {Phys. Rev. D},
  volume = {38},
  issue = {6},
  pages = {1994--1999},
  year = {1988},
  publisher = {American Physical Society},
  doi = {10.1103/PhysRevD.38.1994},
}

@article{Wales_1997,
    author = {Wales, David J. and Doye, Jonathan P. K.},
    title = {Global Optimization by Basin-Hopping and the Lowest Energy Structures of {Lennard}--{Jones}
Clusters Containing up to 110 Atoms},
    journal = {J. Phys. Chem. A},
    volume = {101},
    number = {28},
    pages = {5111--5116},
    year = {1997},
    doi = {10.1021/jp970984n},
}

@software{Cerisola_2026,
  author = {Cerisola, Federico},
  title = {LocalErgotropy.jl},
  year = {2026},
  publisher = {Zenodo},
  version = {v1.0.0},
  doi = {10.5281/zenodo.20660782},
}

@article{qutip5,
  title = {QuTiP 5: The Quantum Toolbox in {Python}},
  author = {Lambert, Neill and Gigu{`e}re, Eric and Menczel, Paul and Li, Boxi and
    Hopf, Patrick and Su{'a}rez, Gerardo and Gali, Marc and Lishman, Jake and
    Gadhvi, Rushiraj and Agarwal, Rochisha and Galicia, Asier and Shammah, Nathan and
    Nation, Paul and Johansson, J. R. and Ahmed, Shahnawaz and Cross, Simon and
    Pitchford, Alexander and Nori, Franco
  },
  journal = {Physics Reports},
  volume = {1153},
  pages = {1-62},
  year = {2026},
  doi = {10.1016/j.physrep.2025.10.001},
}



\clearpage
\widetext
\begin{center}
\textbf{\large Supplemental Information: Robust many-body quantum batteries}
\end{center}
\setcounter{equation}{0}
\setcounter{figure}{0}
\setcounter{table}{0} 
\setcounter{page}{1}
\makeatletter
\renewcommand{\theequation}{S\arabic{equation}}
\renewcommand{\thefigure}{S\arabic{figure}}


\renewcommand{\figurename}{Supplementary Figure} 
\renewcommand{\theequation}{S\arabic{equation}}
\renewcommand{\thefigure}{S\arabic{figure}}
\setcounter{secnumdepth}{2}
\makeatletter
\def\@seccntformat#1{\csname the#1\endcsname.\quad}
\makeatother
\renewcommand{\thesection}{\Roman{section}}
\thispagestyle{empty}

\let\addtocontents\oldaddtocontents

\tableofcontents

\newpage

\section{Local work extraction from typical energy eigenstates}
\label{app:ethergotropy}

Here we prove that any convex combination of positive-temperature ETH states of a generic many-body system is effectively passive with respect to local work extraction. A generic many-body system is here understood to be a lattice system composed of $N$ components and featuring $\mathbb{k}$-local interactions\footnote{Not to be confused with $\mathbb{k}$-body interactions. While $\mathbb{k}$-local interactions are a special case of $\mathbb{k}$-body interactions, the latter spans also long-range systems where $h_a$ may act on components that are far apart.}, with $\mathbb{k}$ a finite, $N$-independent number. Each component lives in a Hilbert space of dimension upper bounded by some finite, $N$-independent number $\mathfrak{d}$. The interaction structure of the lattice is described by a graph with $N$ vertices, representing the components, and cliques of size at most $\mathbb{k}$. Thus, the Hamiltonian of the system can be written as
\begin{align} \label{genham}
    H = \sum_{a=1}^{\aleph} h_a,
\end{align}
where each $h_a$ acts on at most $\mathbb{k}$ vertices. The number of terms $\aleph \propto N$, and we enforce extensivity by requiring that $0 < \curlyd{h} \leq \Vert h_a \Vert_\mathrm{op} \leq \curlyd{H}$ $\forall a$, where $\curlyd{h}$ and $\curlyd{H}$ are $N$-independent constants. This general construction covers, among others, all possible spin lattice models with finite-range interactions.

Provided there is no strong disorder in the system and it is non-integrable (chaotic), it is expected to violate the eigenstate thermalization hypothesis (ETH) \cite{Deutsch_1991, Srednicki_1992, Tasaki_1998, Rigol_2008, DAlessio_2016} only weakly \cite{Serbyn_2021, Moudgalya_2022}. Namely, all of its energy eigenstates satisfy the ETH except a small, typically (but not necessarily \cite{Shiraishi_2017}) polynomial in $N$, subset of them---called quantum many-body scars \cite{Bernien_2017, Turner_2018, Moudgalya_2018}---that violates it. Furthermore, systems with such generic local Hamiltonians typically satisfy the so-called equivalence of ensembles \cite{Simon, Muller_2015, Brandao_2015, Gogolin_2016, Tasaki_2018}.

A high-level definition of the ETH \cite{Rigol_2008, Shiraishi_2017} we use here is as follows. An energy eigenstate $\ket{E_k}$ is said to satisfy the ETH if, for \textit{any} local many-body operator $Q = \sum_b q_b$,
\begin{align} \label{ETH_1}
    \bra{E_k} Q \ket{E_k} = \tr(\rho_\mathrm{mc} Q) + c_1(N),
\end{align}
where $\rho_\mathrm{mc}$ is a microcanonical state such that $\tr(\rho_\mathrm{mc} H) = E_k$, and $c_1(N)$ denotes the finite-size correction such that $c_1(N) = o(N)$ in the thermodynamic limit $N \to \infty$. Note that, in that limit, $\bra{E_k} Q \ket{E_k}$ and $\tr(\rho_\mathrm{mc} Q)$ are generically $\propto N$ (see Refs.~\cite{Araki_1969, Araki_1974, Simon, Muller_2015} for rigorous results on this in translationally invariant systems). Whenever the ``detailed'' ETH in the sense of Refs.~\cite{Deutsch_1991, Srednicki_1992} is satisfied, it can be shown that $c_1(N) = e^{-\Theta(N)}$ \cite{DAlessio_2016}, where $\Theta$ is standard asymptotic notation for ``of the same order.'' However, in general, exponential convergence in Eq.~\eqref{ETH_1} is not necessary for the ETH to hold. Note that our definition of the ETH \eqref{ETH_1} is sometimes referred to as microscopic thermal equilibrium (MITE) \cite{Goldstein_2015}, and all the results in this section apply to arbitrary MITE states, even those that are not ETH energy eigenstates.

In turn, the equivalence of ensembles states that, for any local many-body operator $Q$,
\begin{align}
    \tr(\rho_\mathrm{mc} Q) = \tr(\tau_\beta Q) + c_2(N).
\end{align}
Here, $\tau_\beta \propto \exp(-\beta H)$ is the Gibbs state of the lattice at the inverse temperature $\beta$ uniquely determined from the condition $\tr(\tau_\beta H) = \tr(\rho_\mathrm{mc} H)$. The finite-size correction $c_2(N) = o(N)$ (it generically, but not necessarily, scales as $N^{1/2}$ \cite{Muller_2015, Brandao_2015, Gogolin_2016}).

Thus, by a generic non-disordered and non-integrable $\mathbb{k}$-local Hamiltonian we mean a Hamiltonian of the form \eqref{genham} the non-scar states of which satisfy, for any local many-body operator $Q$,
\begin{align} \label{ETH_2}
    \bra{E_k} Q \ket{E_k} = \tr(\tau_{\beta_k} Q) + c(N),
\end{align}
where $\beta_k$ is the unique solution of $\tr(\tau_\beta H) = E_k$ for $\beta$, and the finite-size correction $c(N) = c_1(N) + c_2(N) = o(N)$. Sometimes, the ETH is introduced directly in this manner \cite{Kuwahara_2020ETH}.

Next, let us consider a partition of the lattice into $\beth = \Theta(N)$ spatially localized subsystems $\{ A_b\}_{b=1}^\beth$ each containing at most $\mathbb{l}$ vertices. Let $Q = \sum_b q_b$ be a local many-body operator such that $q_b$ acts only on $A_b$. Since Eq.~\eqref{ETH_2} holds for any such $Q$, i.e., as each $q_b$ spans all Hermitian operators acting on the Hilbert space of $A_b$, it is straightforward to show that
\begin{align} \label{subsysETH}
    \sum_{b=1}^\beth \big\Vert \rho_{A_b} - \tr_{\bA_b} [\tau_{\beta_k}] \big\Vert_1 = o(N),
\end{align}
where $\rho_{A_b}$ is the reduced state of subsystem $A_{b}$ and $\bA_b$ denotes the complement of $A_b$. In this local-state form, the ETH is referred to as the ``subsystem ETH'' \cite{Dymarsky_2018}. We also note that the local states of a subsystem $S$ of the form $\tr_{\bA} [\tau_{\beta}]$ are called ``mean-force Gibbs states'' \cite{Trushechkin_2022}. Thus, the ETH can be phrased as follows: when the system is in the state $\ket{E_k}$, the local states of finite subsystems are given by the corresponding mean-force Gibbs states, with a correction $o(1)$ in the thermodynamic limit.

\medskip

We now address local work extraction from an ETH state $\ket{E_k}$. Namely, we take a partition $\{ A_b \}_{b=1}^\beth$ of the graph into subsystems consisting of at most $\mathbb{l}$ adjacent vertices and apply
\begin{align}
    U_\loc = \bigotimes_{b=1}^\beth u_b,
\end{align}
where each $u_b$ is unitary and acts on $A_b$ only. Like $\mathbb{k}$ above, $\mathbb{l}$ is a finite, $N$- and $b$-independent integer. The work extracted by this operation is then
\begin{align}
    W_\loc(\dyad{E_k}, H) = E_k - \bra{E_k} U_\loc^\dagger H U_\loc^{\phantom{\dagger}} \ket{E_k}.
\end{align}
First, by the very definition of $\beta_k$, we have that $E_k = \tr(\tau_{\beta_k} H)$. Then, we note that
\begin{align*}
    H' \coloneq U_\loc^\dagger H U_\loc^{\phantom{\dagger}} = \sum_a \underbrace{U_\loc^\dagger h_a U_\loc^{\phantom{\dagger}}}_{h_a'}
\end{align*}
is a local many-body operator as each $h_a'$ acts on at most $\mathbb{k} + \mathbb{l} - 1$ vertices. Thus, Eq.~\eqref{ETH_2} applies to $H'$, giving us $\bra{E_k} H' \ket{E_k} = \tr(\tau_{\beta_k} H') + o(N)$. Altogether,
\begin{align} \nonumber
    W_\loc(\dyad{E_k}, H) &= \tr(\tau_{\beta_k} H) - \tr(\tau_{\beta_k} U_\loc^\dagger H U_\loc^{\phantom{\dagger}}) + o(N)
    \\ \label{locallyextracted}
    &= - \beta_k^{-1} S\big(U_\loc^{\phantom{\dagger}} \tau_{\beta_k} U_\loc^\dagger \, \big\Vert \,  \tau_{\beta_k} \big) + o(N),
\end{align}
where $S(\bullet \Vert \bullet)$ is the quantum relative entropy, and $\tr(\tau_\beta H) - \tr(U \tau_\beta U^\dagger H) = - \beta^{-1} S(U \tau_\beta U^\dagger  \Vert \tau_\beta) \leq 0$ is simply the reflection of the fact that thermal states are passive \cite{Hatsopoulos_1976, Pusz_1978, Lenard_1978}. In view of Eq.~\eqref{locallyextracted}, it is thus obvious that, for any $\ket{E_k}$ satisfying the ETH, the ``parallel'' local ergotropy (or simply parallel charge) is
\begin{align} \label{par_erg_pass}
    \ergl_\prll(\dyad{E_k}, \, H) \coloneq \max_{\{u_b\}_{b=1}^\beth} W_\loc(\dyad{E_k}, \, H) = o(N),
\end{align}
where we have made the dependence of $\ergl_\prll$ on the Hamiltonian explicit for later convenience. For the local ergotropy (see Ref.~\cite{Salvia_2023} and Eq.~\eqref{locergdef} below for its general definition) of an individual subsystem $A_b$,
\begin{align} \label{locerg_E_k}
    \ergl^{(b)}(\dyad{E_k}, \, H) \coloneq \max_{\substack{u_b \\ u_{b' \neq b} = \id_{b'}}} \!\! W_\loc(\dyad{E_k}, \, H),
\end{align}
Eq.~\eqref{locallyextracted} yields the same $o(N)$ scaling, which is a somewhat crude estimate. A more refined bound follows from the subsystem ETH expressed in Eq.~\eqref{subsysETH}. Indeed, keeping in mind that $\beth = \Theta(N)$, we can rewrite Eq.~\eqref{subsysETH} as
\begin{align*}
    \frac{1}{\beth} \sum_{b=1}^\beth \Dtr{\rho_{A_b}}{\tr_{\bA_b} [\tau_{\beta_k}]} = o(1),
\end{align*}
where $\Dtr{\rho_1}{\rho_2} \coloneq \tfrac{1}{2} \, \Vert \rho_1 - \rho_2 \Vert_1$ is the trace distance. In other words, the arithmetic mean of all local deviations from the mean-force Gibbs state is $o(1)$.

Now, let us choose a ``typical'' $A_b$ in the sense that
\begin{align} \label{subsysETH_1}
    \Dtr{\rho_{A_b}}{\tr_{\bA_b} [\tau_{\beta_k}]} = o(1),
\end{align}
and consider its local ergotropy \eqref{locerg_E_k}. Let us furthermore assume that $A_b^\ext$---the extension of $A_b$ that includes an ``interaction belt'' around $A_b$---is also typical in the same sense:
\begin{align} \label{subsysETH_2}
    \Dtr{\rho_{A_b^\ext}}{\tr_{\bA_b^\ext} [\tau_{\beta_k}]} = o(1).
\end{align}
More precisely, we define $A_b^\ext$ as
\begin{align}
    A_b^\ext \coloneq \bigcup_{\supp(h_a) \bigcap A_b \neq \emptyset} \supp(h_a),
\end{align}
where $\supp(\bullet)$ gives the vertices on which its argument acts nontrivially. Note that $A_b^\ext$ contains at most $\mathbb{l} (\mathbb{k} - 1)$ vertices.

Thus, introducing the operators
\begin{align}
    u_b^\ext \coloneq u_b \otimes \id_{A_b^\ext \backslash A_b} \qquad \mathrm{and} \qquad h_b^\ext \coloneq \sum_{\supp(h_a) \bigcap A_b \neq \emptyset} h_a,
\end{align}
both supported only on $A_b^\ext$, we can write the work extracted by the local action of $u_b$ as
\begin{align}
    W_\loc(\dyad{E_k}, H) &= \tr \Big[ \dyad{E_k} \Big( \sum_a h_a \Big) \Big] - \tr \Big[ \dyad{E_k} \big(u_b^\dagger \otimes \id_{\bA_b} \big) \Big( \sum_a h_a \Big) \big(u_b \otimes \id_{\bA_b}\big) \Big]
    \\
    &= \tr \big[ \rho_{A_b^\ext} \,\, h_b^\ext \big] - \tr\big[ \rho_{A_b^\ext} \,\, u_b^{\ext \, \dagger} \, h_b^\ext \, u_b^\ext \big],
\end{align}
where we have explicitly written $\sum_a h_a$ instead of $H$ simply to make the calculation leading from the first line to the second more transparent. Hence, for the local ergotropy of $A_b$, we have
\begin{align} \label{locerg_to_Uerg_1}
    \ergl^{(b)}(\dyad{E_k} , H) &= \tr \big[ \rho_{A_b^\ext} \,\, h_b^\ext \big] - \min_{u_b^\ext} \tr\big[ \rho_{A_b^\ext} \,\, u_b^{\ext \, \dagger} \, h_b^\ext \, u_b^\ext \big].
\end{align}
Here, the right-hand side is nothing but the local ergotropy of a much smaller system ($A_b^\ext$) than the full lattice. Namely, Eq.~\eqref{locerg_to_Uerg_1} simply means
\begin{align} \label{locerg_to_Uerg_2}
    \ergl^{(b)}(\dyad{E_k}, \, H) = \ergl^{(b)} \big( \rho_{A_b^\ext}, h_b^\ext \big).
\end{align}
Obviously, the above analysis applies to any state of the lattice, so we also have that
\begin{align} \label{locerg_to_Uerg_3}
    \ergl^{(b)}(\tau_{\beta_k}, \, H) = \ergl^{(b)}\big( \tr_{\bA_b^\ext} [\tau_{\beta_k}], h_b^\ext \big).
\end{align}
On the other hand, any positive-temperature Gibbs state is a passive state, so no work can be extracted from it by unitary operations (local or not); namely, $\ergl^{(b)}(\tau_{\beta_k}, \, H) = 0$, and hence,
\begin{align} \label{loc_pass_therm}
    \ergl^{(b)}\big( \tr_{\bA_b^\ext} [\tau_{\beta_k}], h_b^\ext \big) = 0.
\end{align}
Finally, accounting for Eqs.~\eqref{locerg_to_Uerg_2}, ~\eqref{locerg_to_Uerg_3}, and~\eqref{loc_pass_therm}, and next applying Lemma~\ref{thm:convexity_continuity} or, more explicitly, Eq.~\eqref{contbound_loc_1} in Sec.~\ref{app:U-ergotropy}, we have that
\begin{align} \label{almostthere}
    \ergl^{(b)}(\dyad{E_k}, \, H) = \big\vert \ergl^{(b)}(\rho_{A_b^\ext}, \, h_b^\ext) - \ergl^{(b)}\big( \tr_{\bA_b^\ext} [\tau_{\beta_k}], h_b^\ext \big) \big\vert \leq 2 \, \spr(h_b^\ext) \, \Dtr{\rho_{A_b^\ext}}{\tr_{\bA_b^\ext} [\tau_{\beta_k}]}.
\end{align}
Furthermore, since $A_b^\ext$ contains at most $\mathbb{l} (\mathbb{k} - 1)$ vertices and $\opn{h_a} \leq \curlyd{H} \, \forall a$, we have that $\spr(h_b^\ext) = O(1)$. Taking this fact along with Eq.~\eqref{subsysETH_2} into account in Eq~\eqref{almostthere} leads us to
\begin{align} \label{loc_erg_pass}
    \ergl^{(b)}(\dyad{E_k}, \, H) = o(1),
\end{align}
where, we remind, the asymptotic scaling $o(1)$ is in the $N \to \infty$ limit.

\medskip

Let us, in passing, comment on the implications of the above argumentation on extended local ergotropy \cite{Castellano_2024}, even though we do not use this quantity in this paper. There, the extraction time $t$ is finite, therefore, even if only $A_b$'s degrees of freedom are manipulated, the intra-lattice interactions cause the affected region to extend beyond $A_b$. Despite this complication, when $t$ is finite, the analysis here proceeds exactly as in the previous strictly local case, the only modification being that the size of the affected region $A_b^\ext(t)$ will depend on time. To account for that, we will assume that the Lieb--Robinson bound \cite{Lieb_1972, Nachtergaele_2019, Kuwahara_2020LiebRobi, Chen_2023} applies to the system, as is generically expected even for systems with sufficiently rapidly decaying polynomial interactions \cite{Kuwahara_2020LiebRobi}. With this assumption, we have that $\diam(A_b^\ext(t)) \propto t$, so the number of vertices in $A_b^\ext(t)$, and therefore the number of $h_a$'s overlapping with it, is $\propto t^{\curlyd{D}}$, where $\curlyd{D}$ is the spatial dimension of the lattice. Thus, by the triangle inequality, we have that $\spr(h_b^\ext(t)) \propto t^{\curlyd{D}}$. Reading from Eq.~\eqref{almostthere}, we conclude that the extended local ergotropy
\begin{align} \label{loc-ext_erg_pass}
    \mathcal{E}_{\text{ext loc}}^{(b)}(\dyad{E_k}, \, H) = t^{\curlyd{D}} \, o(1).
\end{align}
So, for finite $t$, ETH states are asymptotically passive also with respect to extended local ergotropy.

\medskip

Now, invoking the convexity property of ergotropy-like quantities [Lemma~\ref{thm:convexity_continuity}], it is straightforward to see that any probabilistic mixture of positive-temperature ETH states is asymptotically passive with respect to parallel, local, and extended local ergotropies. Indeed, taking the lattice in the state
\begin{align}
    \rho_\mathrm{ETH} = \sum_k p_k \dyad{E_k},
\end{align}
where $p_k \geq 0$ and $\sum_k p_k = 1$, and $\ket{E_k}$ are ETH states with $\beta_k \geq 0$, we have [accounting for Eqs.~\eqref{par_erg_pass},~\eqref{loc_erg_pass}, and~\eqref{loc-ext_erg_pass}]
\begin{align}
\begin{split}
    \ergl_\prll(\rho_\mathrm{ETH}, \, H) &\leq \sum_k p_k \ergl_\prll(\dyad{E_k}, \, H) = \sum_k p_k \, o(N)_k = o(N).
    \\
    \ergl^{(b)}(\rho_\mathrm{ETH}, \, H) &\leq \sum_k p_k \ergl^{(b)}(\dyad{E_k}, \, H) = \sum_k p_k \, o(1)_k = o(1).
    \\
    \ergl_{\text{ext loc}}^{(b)}(\rho_\mathrm{ETH}, \, H) &\leq \sum_k p_k \ergl_{\text{ext loc}}^{(b)}(\dyad{E_k}, \, H) = \sum_k p_k \, t^{\curlyd{D}} \, o(1)_k = t^{\curlyd{D}} o(1).
\end{split}
\end{align}
Note that we do not expect local passivity of this kind to hold for arbitrary coherent superpositions of ETH states of the form $\sum_k \sqrt{p_k} \ket{E_k}$. This is due partly to the so-called asymptotic scarring phenomenon \cite{Gotta_2023} where such superpositions are able to replicate certain properties of scar states.

\medskip

Lastly, a few comments on the previous literature on the matter. A notion of local work extraction from energy eigenstates of nonintegrable many-body systems has been investigated in Refs.~\cite{Kaneko_2019, Baba_2023}, but only numerically and on specific models. In Ref.~\cite{Hokkyo_2025}, the problem is considered in the limit where the subsystems are themselves macroscopic and their local states in thermal equilibrium are almost Gibbs states. Here, none of these assumptions are made---the subsystems are small and their equilibrium mean-force Gibbs states can be far from being Gibbsian \cite{Cresser_2021}.

While finalizing this manuscript, we became aware of Ref.~\cite{Chiba_2026}, which independently considers related aspects of local work extraction from systems in microscopic thermal equilibrium.

\subsection{Continuity and convexity of ergotropy-like quantities}
\label{app:U-ergotropy}

Consider a general quantum system with a finite, $d$-dimensional Hilbert space and described by a Hamiltonian $H$ (no additional structural assumptions are made here). Suppose the allowed unitary operations one can perform on the system belong to a subset $\mathfrak{U}$ of $\mathrm{U}(d)$---the set of all unitary operators. Then, the maximal extractable work from the system under such unitaries is
\begin{align}
    \erg_\mathfrak{U}(\rho, H) \coloneq \tr(\rho H) - \min_{U \in \mathfrak{U}} \tr(\rho U^\dagger H U).
\end{align}
We will call this ergotropy-like quantity $\mathfrak{U}$-ergotropy. This notion subsumes all known ergotropy-like quantities such as local and extended local ergotropies \cite{Salvia_2023, Castellano_2024} and extractable work in ``black-box'' scenarios \cite{Safranek_2023, Chakraborty_2025}. Here, we will prove the following lemma that extends the properties of convexity \cite{Perarnau_2015, Bernards_2019, Tirone_2021} and Lipschitz-continuity \cite{Hovhannisyan_2024} of the standard, ``full'' ergotropy to general $\mathfrak{U}$-ergotropy.

\begin{lemma} \label{thm:convexity_continuity}
    For any $\mathfrak{U} \subseteq \mathrm{U}(d)$, $\erg_\mathfrak{U}(\rho, H)$ is convex and Lipschitz-continuous with respect to the trace distance as a function of $\rho$. The continuity bound is
    \begin{align} \label{contbound}
        \big\vert \erg_\mathfrak{U}(\rho_1, H) -  \erg_\mathfrak{U}(\rho_2, H) \big\vert \leq 2 \spr(H) \, \Dtr{\rho_1}{\rho_2},
    \end{align}
    where $\spr(H)$ is the spread of $H$.
\end{lemma}

The (spectral) spread of a matrix is the largest distance between its eigenvalues \cite{HornJohnson}: $\spr(H) \coloneq \max\limits_{k,j} | E_k - E_j |$, where $E_k$ the eigenvalues of $H$. Note that, when $H$ is the Hamiltonian (as is the case here) and its ground state is chosen to have zero energy (i.e., $\min\limits_k E_k = 0$), then $\spr(H)$ is simply equal to the operator norm $\opn{H}$.

The first claim of Lemma~\ref{thm:convexity_continuity}---convexity of $\erg_\mathfrak{U}(\rho, H)$---follows straightforwardly from the elementary superadditivity property of minimization. Namely, for any two abstract functions $f(x)$ and $g(x)$ defined on the same domain $\mathcal{X}$,
\begin{align*}
    \min_{x \in \mathcal{X}} \, [f(x) + g(x)] \geq \min_{x \in \mathcal{X}} f(x) + \min_{x \in \mathcal{X}} g(x).
\end{align*}
Indeed, keeping this in mind, we have, for $\forall \, \rho_1, \rho_2$ and $\forall \, \lambda_1, \lambda_2 \geq 0$ such that $\lambda_1 + \lambda_2 = 1$,
\begin{align}
\begin{split}
     \erg_\mathfrak{U}(\lambda_1 \rho_1 + \lambda_2 \rho_2, H) &= \tr[(\lambda_1 \rho_1 + \lambda_2 \rho_2) H] - \min_{U \in \mathfrak{U}} \tr[(\lambda_1 \rho_1 + \lambda_2 \rho_2) U^\dagger H U]
     \\
     &= \lambda_1 \tr(\rho_1 H) + \lambda_2 \tr(\rho_2 H) - \min_{U \in \mathfrak{U}} \big[\lambda_1 \tr(\rho_1 U^\dagger H U) + \lambda_2 \tr(\rho_2 U^\dagger H U) \big]
     \\
     &\leq \lambda_1 \tr(\rho_1 H) + \lambda_2 \tr(\rho_2 H) - \lambda_1 \min_{U \in \mathfrak{U}} \tr(\rho_1 U^\dagger H U) - \lambda_2 \min_{U \in \mathfrak{U}} \tr(\rho_2 U^\dagger H U)
     \\
     &= \lambda_1 \erg_\mathfrak{U}(\rho_1, H) + \lambda_2 \erg_\mathfrak{U}(\rho_2, H).
\end{split}
\end{align}
Note that, arguing identically, one can show that $\erg_\mathfrak{U}(\rho, H)$ is convex also as a function of $H$.

\medskip

Next, let us prove the second claim of Lemma~\ref{thm:convexity_continuity}; namely, the continuity bound $\eqref{contbound}$. To that end, we introduce the operator
\begin{align}
    \Upsilon_U \coloneq H - U^\dagger H U,
\end{align}
so that the definition of $\erg_\mathfrak{U}$ rewrites as
\begin{align} \label{ergconstr}
    \erg_\mathfrak{U}(\rho, H) = \max_{U \in \mathfrak{U}} \tr(\rho \Upsilon_U).
\end{align}
Then, introducing the $\mathfrak{U}$-ergotropy-extracting unitaries
\begin{align*}
    U_1^\opt = \arg \max_{U \in \mathfrak{U}} \tr(\rho_1 \Upsilon_U) \qquad \mathrm{and} \qquad U_2^\opt = \arg \max_{U \in \mathfrak{U}} \tr(\rho_2 \Upsilon_U),
\end{align*}
we can write
\begin{align*}
    \erg_\mathfrak{U}(\rho_1, H) = \tr\big(\rho_1 \Upsilon_{U_1^\opt}\big) \qquad \mathrm{and} \qquad \erg_\mathfrak{U}(\rho_2, H) = \tr\big(\rho_2 \Upsilon_{U_2^\opt}\big).
\end{align*}
Thus, noting that, by definition,
\begin{align}
    \erg_\mathfrak{U}(\rho_2, H) \geq \tr\big(\rho_2 \Upsilon_{U_1^\opt}\big),
\end{align}
we have
\begin{align} \label{zaal}
    \erg_\mathfrak{U}(\rho_1, H) - \erg_\mathfrak{U}(\rho_2, H) \leq \tr\big[\big(\rho_1 - \rho_2\big) \Upsilon_{U_1^\opt} \big].
\end{align}
Now, recalling that, for any two operators $R_1$ and $R_2$, $\tr(R_1 R_2) \leq \Vert R_1 \Vert_\mathrm{op} \, \Vert R_2 \Vert_1$ \cite{HornJohnson}, and applying that to the last term in Eq.~\eqref{zaal}, we get
\begin{align} \label{contbound1}
    \erg_\mathfrak{U}(\rho_1, H) - \erg_\mathfrak{U}(\rho_2, H) \leq 2 \opn{\Upsilon_{U_1^\opt}} \, \Dtr{\rho_1}{\rho_2} \leq 2 \Big(\max_{U \in \mathfrak{U}} \opn{\Upsilon_U}\Big) \, \Dtr{\rho_1}{\rho_2}.
\end{align}
Performing an identical analysis for $\erg_\mathfrak{U}(\rho_2, H) - \erg_\mathfrak{U}(\rho_1, H)$, we similarly obtain that 
\begin{align} \label{contbound2}
    \erg_\mathfrak{U}(\rho_2, H) - \erg_\mathfrak{U}(\rho_1, H) \leq 2 \Big(\max_{U \in \mathfrak{U}} \opn{\Upsilon_U}\Big) \, \Dtr{\rho_1}{\rho_2},
\end{align}
which is also obvious from the symmetry of the problem with respect to $\rho_1$ and $\rho_2$. Together, Eqs.~\eqref{contbound1} and~\eqref{contbound2} simply mean that
\begin{align} \label{contbound3}
    \big\vert \erg_\mathfrak{U}(\rho_2, H) - \erg_\mathfrak{U}(\rho_1, H) \big\vert \leq 2 \Big(\max_{U \in \mathfrak{U}} \opn{\Upsilon_U}\Big) \, \Dtr{\rho_1}{\rho_2}
\end{align}
To finalize the proof of Eq.~\eqref{contbound}, it remains to show that
\begin{align} \label{spreadups}
    \max_{U \in \mathrm{U}(d)} \opn{\Upsilon_U} = \spr(H).
\end{align}
To do so, we note that $\Upsilon_U$ is invariant under the constant-shift transformation $H \to H - \upsilon \id_d$ of the Hamiltonian. Thus, by choosing $\upsilon_0 = \min\limits_k E_k$ and denoting $\overline{H} \coloneq H - \upsilon_0 \id_d \geq \nul$, we have that
\begin{align}
    \Upsilon_U = \overline{H} - U^\dagger \overline{H} U \qquad \Longrightarrow \qquad \bra{\psi} \Upsilon_U \ket{\psi} \overset{(\spadesuit)}{\leq} \bra{\psi} \overline{H} \ket{\psi} \;\; \forall \ket{\psi}, \, U \qquad \Longleftrightarrow \qquad \opn{\Upsilon_U} \leq \opn{\overline{H}} \;\; \forall U,
\end{align}
where the inequality ($\spadesuit$) is due to the fact that $U^\dagger \overline{H} U \geq \nul$ because $\overline{H} \geq \nul$. Also, from the definition of $\overline{H}$ it immediately follows that
\begin{align} \label{spreadopn}
    \opn{\overline{H}} = \spr(H).
\end{align}
Furthermore, note that when $U$ is a permutation of the eigenbasis of $\overline{H}$ which swaps the smallest and largest eigenvalues of $\overline{H}$, then $\opn{\Upsilon_U} = \opn{\overline{H}}$. This simply means that $\max\limits_{U \in \mathrm{U}(d)} \opn{\Upsilon_U} = \opn{\overline{H}}$ and, along with Eq.~\eqref{spreadopn}, establishes Eq.~\eqref{spreadups}, which, combined with Eq.~\eqref{contbound3}, proves Eq.~\eqref{contbound}.

\medskip

Finally, let us apply the continuity bound \eqref{contbound} to the case of local ergotropy. There, one is given a system composed of a subsystem $A$ and its complement $\bA$ and described by the total interacting Hamiltonian $H_{A\bA}$. Starting in some state $\rho_{A\bA}$, the work extracted from the total system by the unitary $U_A$ acting only on the subsystem $A$ will be
\begin{align}
    W_\loc \big(\rho_{A\bA}, H_{A\bA}, U_A \big) \coloneq& \, \tr \big( \rho_{A\bA} H_{A\bA} \big) -  \tr \big( \rho_{A\bA} \, U_A^\dagger \otimes \id_\bA \, H_{A\bA} \, U_A \otimes \id_\bA \big).
\end{align}
The maximal amount of such locally extractable work is called ``local ergotropy'' \cite{Salvia_2023}:
\begin{align} \label{locergdef}
    \ergl^{(A)} \big(\rho_{A\bA}, H_{A\bA}\big) \coloneq \max_{U_A} W_\loc \big(\rho_{A\bA}, H_{A\bA}, U_A\big).
\end{align}
Obviously, $\ergl^{(A)}$ is a $\mathfrak{U}$-ergotropy with
\begin{align*}
    \mathfrak{U} = \{ U_A \otimes \id_\bA \; \colon \; U_A \in \mathrm{U}(d_A) \} \subset \mathrm{U}(d_{A \bA}),
\end{align*}
where $d_A$ and $d_{A \bA}$ are the Hilbert-space dimensions of, respectively, $A$ and $A \bA$. Accordingly, as a $\mathfrak{U}$-ergotropy, $\ergl^{(A)} \big( \rho_{A\bA}, H_{A\bA} \big)$ obeys Lemma~\ref{thm:convexity_continuity} and satisfies the continuity bound
\begin{align} \label{contbound_loc_1}
    \big\vert \ergl^{(A)}(\rho_{A\bA}, H_{A\bA}) - \ergl^{(A)}(\vartheta_{A\bA}, H_{A\bA}) \big\vert \, \leq \, 2 \, \spr\!\big(H_{A\bA} \big) \, \Dtr{\rho_{A\bA}}{\vartheta_{A\bA}}.
\end{align}
We also note that, whenever the Hamiltonian has the structure
\begin{align}
    H_{A\bA} = H_A \otimes \id_\bA + H_\I + \id_A \otimes H_\bA,
\end{align}
the locally extracted work is insensitive to $H_\bA$. Namely, denoting
\begin{align}
    H_A^\ext \coloneq H_A \otimes \id_\bA + H_\I,
\end{align}
we have that
\begin{align}
    W_\loc \big(\rho_{A\bA}, H_{A\bA}, U_A \big) = W_\loc \big(\rho_{A\bA}, H_A^\ext, U_A\big) \qquad \Longrightarrow \qquad \ergl^{(A)}(\rho_{A\bA}, H_{A\bA}) = \ergl^{(A)}(\rho_{A\bA}, H_A^\ext),
\end{align}
and therefore,
\begin{align} \label{contbound_loc_2}
    \big\vert \ergl^{(A)}(\rho_{A\bA}, H_{A\bA}) - \ergl^{(A)}(\vartheta_{A\bA}, H_{A\bA}) \big\vert \, \leq \, 2 \, \spr\!\big(H_A^\ext \big) \, \Dtr{\rho_{A\bA}}{\vartheta_{A\bA}}.
\end{align}
This form of the continuity bound for $\erg_\loc^{(A)}$ can be useful when $\bA$ is a large bath with a macroscopic $\opn{H_\bA}$. Indeed, despite the large $\opn{H_\bA}$ necessitating a large $\opn{H_{A\bA}}$, the continuity bound is determined by $\opn{H_A^\ext} \leq \opn{H_A} + \opn{H_\I}$, which will be $O\big(\opn{H_A}\big)$ whenever $\opn{H_\I} = O\big(\opn{H_A}\big)$. The latter is satisfied not only when the interaction of $A$ with $\bA$ is of finite range [like in our analysis between Eqs.~\eqref{locerg_E_k} and~\eqref{locerg_to_Uerg_2} above], but also when it is decaying polynomially with a sufficiently large exponent \cite{Kuwahara_2020ETH}.

\section{Universal charging protocol}

We here provide more details on the general charging procedure that maps any initial state onto one of the many-body scars. The protocol consists of two processes. On the one hand, engineered dissipation implemented via Lindbladian dynamics 
\begin{equation}
    \dot{\rho}_t 
    = - \m{i} [H,\rho_{t}] + \sum_{j} \Gamma_j \textsf{D}[L_j](\rho_{t})
    \coloneq \textsf{L}_\m{ED} (\rho_t)
\end{equation}
where $\textsf{D}[X] (\bullet) = X \bullet X^{\dag} - \{X^{\dag} X,\bullet \}/2$ denotes the dissipator with yet unspecified Lindblad operators $L_j$ and rates $\Gamma_j$, and we have defined the associated generator $\textsf{L}_\m{ED}$. And, on the other, continuous monitoring described by a quantum trajectory
\begin{align}
    \dd \rho^\m{c}_t = 
    - \m{i} [H,\rho_{t}] \dd t + \gamma \textsf{D}[M](\rho^\m{c}_{t}) \dd t + \sqrt{\gamma} [\rho^\m{c}_t M + M^{\dag} \rho^\m{c}_t - \tr(\rho^\m{c}_t M + M^{\dag} \rho^\m{c}_t) \rho^\m{c}_t] \dd W_t,
    \label{eq:full-protocol}
\end{align}
where $\rho^\m{c}_t$ is the state of the system conditioned on the measurement outcomes, the operator $M$ mediates the effective action of the measurement on $\rho^\m{c}_t$, and $\dd W_t$ is a standard Wiener process.
In short, the engineered dissipation drives any initial state into the scar manifold $\cals$, whereas the continuous monitoring then probabilistically selects one of the scars. 
Performing both the engineered dissipation and the continuous monitoring concurrently, the complete protocol is described by the \Ito stochastic differential equation for the evolution of the state
\begin{equation}
  \dd{\rho_{t}^{\m{c}}} 
  = \Big( -\m{i} [H,\rho_{t}^{\m{c}}] 
  + \sum_{j} \Gamma_j \textsf{D}[L_j](\rho_{t}^{\m{c}}) + \gamma \textsf{D}[M](\rho_{t}^{\m{c}}) \Big) \dd{t} 
  + \sqrt{\gamma} [M \rho_{t}^{\m{c}} + \rho_{t}^{\m{c}} M^{\dag} - \tr(\rho^\m{c}_t M + M^{\dag} \rho^\m{c}_t) \rho^\m{c}_t] \dd{W_t}.
  \label{Seq:quantum-trajectory}
\end{equation}
It should be noted that, in principle, one may alternatively use the engineered dissipation first and then perform a projective measurement to obtain one of the scars.
However, it turns out that concurrent implementation of dissipation and continuous monitoring usually leads to enhanced charging times as discussed in \cref{sec:charging-speed}.
Denote by $\rho_t = \mathbb{E}[\rho_{t}^{\m{c}}]$ the state averaged over all quantum trajectories. 
Its evolution is given by the average of \cref{Seq:quantum-trajectory} which, given that $\mathbb{E}[ \dd W_t] = 0$, reduces to the Lindblad equation
\begin{equation}
  \begin{split}
    \dot{\rho_t} 
    = -\m{i} [H,\rho_t] 
    + \sum_{j} \Gamma_j \textsf{D}[L_j](\rho_t) + \gamma \textsf{D}[M](\rho_t)
    \coloneq \textsf{L}(\rho_t),
  \end{split}
  \label{eq:full-Lindblad}
\end{equation}
where we have defined the corresponding generator $\textsf{L}$ of the average protocol.

In order to stabilize an individual scar, it must remain invariant under the dynamics induced by \cref{Seq:quantum-trajectory} and hence be stationary.
Since this requirement must moreover hold for all scars, each corresponds to a stationary state of the full dynamics.
Other steady states are not admissible since then there would always exist initial states that eventually result in non-scar asymptotic states.
Thus, the scar subspace $\cals$ must correspond to the unique invariant and attractive manifold in Hilbert space.
We now detail the necessary and sufficient conditions for this to be the case. 
These were originally derived for Lindbladian dynamics, but the same statements hold true for quantum trajectories \cite{Ticozzi_2012,Benoist_2017,Schmolke_2025}.
In particular, a state is stationary on the trajectory level [\cref{Seq:quantum-trajectory}] only if it is stationary on the ensemble level [Eq.~\eqref{eq:full-Lindblad}] \cite{Schmolke_2025}.
The operators $L_j$ and $M$ must therefore satisfy the same structural conditions.
For clarity of notation, let us hence refer to them collectively by $\{ F_{k} \} = \{ L_{j} , M \}$. 
Define now $[F_{k}]_{\mathcal{P} , \mathcal{Q}}$ as the only nonzero block of the matrix $P_{\mathcal{P}} F_{k} P_{\mathcal{Q}}$, where $P_{\mathcal{P} (\mathcal{Q})}$ is the orthogonal projector over the subspace $\mathcal{P}$ ($\mathcal{Q}$). 
A linear operator $X$ can thus be represented accordingly in a block decomposition based on the structure of the Hilbert space (see also Eq.~\eqref{eq:coarse-decomposition})
\begin{equation}
  X = 
  \begin{pmatrix}
    [X]_{\cals, \cals} & [X]_{\cals, \cals^\perp} \\
    [X]_{\cals^\perp, \cals} & [X]_{\cals^\perp, \cals^\perp}
  \end{pmatrix}.
\end{equation}
Then, $\cals$ is an \emph{invariant} subspace under the evolution generated by \cref{Seq:quantum-trajectory} and Eq.~\eqref{eq:full-Lindblad} if and only if the operators $F_{k}$ and the Hamiltonian $H$ satisfy \cite{Baumgartner_2008,Ticozzi_2012,Albert_2016,Benoist_2017}
\begin{equation}
    F_k = 
    \begin{pmatrix}
        [F_k]_{\cals, \cals} & [F_k]_{\cals, \cals^\perp} \\
        0                    & [F_k]_{\cals^\perp, \cals^\perp}
    \end{pmatrix},
    \qquad 
    \m{i} [H]_{\cals,\cals^\perp} - \frac{1}{2} \sum_{k} [F_k]^\dag_{\cals, \cals} [F_k]_{\cals, \cals^\perp} = 0.
    \label{eq:invariance}
\end{equation} 
These conditions guarantee that there can be inflow but there is no outflow out of $\cals$. 
Invariance by itself does however not suffice to ensure global asymptotic stability since there might still exist additional invariant subspaces residing inside of $\cals^{\perp}$. 
These are excluded and $\cals$ is furthermore \emph{attractive} if and only if the subspace
\begin{equation}
    \mathcal{K} = \bigcap_{k} \ker ([F_{k}]_{\cals , \cals^{\perp}})
    \label{eq:attractivity}
\end{equation}
does not contain any invariant subspaces \cite{Ticozzi_2012}.
This guarantees that there are no other invariant subspaces that could prematurely collect probability flowing out of $\cals^\perp$ before it could ever reach $\cals$. 

In our charging protocol the primary purpose of the engineered dissipation is to establish $\cals$ as invariant and attractive while $M$ only needs to be able to discriminate between scars and not necessarily control the inflow into $\cals$.
It is indeed sufficient that the scars are simultaneous \emph{left and right} eigenvectors of $M$ with distinct eigenvalues. 
In this way, it takes on a block diagonal form that respects the structure of the Hilbert space imposed by the operators $L_{j}$ and the Hamiltonian $H$, that is, $[M]_{\cals, \cals^\perp} = [M]_{\cals^\perp, \cals} = 0$, and hence does not break invariance and attractivity of $\cals$.
Any pure stationary state of a Lindblad equation is necessarily a decoherence-free state, a simultaneous eigenstate of both the Hamiltonian and all Lindblad operators \cite{Lidar_1998}.
The same statement is true for quantum trajectories \cite{Schmolke_2025}.
Taken together, it must hence hold that
\begin{equation} \label{eq:dfs-conditions}
    L_{j} \ket{s_n} = c_{j,n} \ket{s_n}, \quad 
    M \ket{s_n}     = m_n \ket{s_n}, \ \forall j,n,
\end{equation}
with arbitrary complex numbers $c_{j,n}$ and where $M$ restricted to $\cals$ has only simple eigenvalues $m_n \neq m_k$, $\forall n \neq k$.
If this were not the case, some initial states would end up in a mixture or superposition of multiple scars with generally unknown coefficients therefore precluding optimal local unitary work extraction.

In the spin-$1$ example of the main text, $M = S^z$ is Hermitian and since $[L_j]_{\cals, \cals} = 0$, we have the following structure
\begin{equation}
  H = 
  \begin{bmatrix}
    [H]_{\cals, \cals} & 0 \\
    0         & [H]_{\cals^\perp, \cals^\perp}
  \end{bmatrix}, \quad
  L_j =
  \begin{bmatrix}
    0 & [L_j]_{\cals, \cals^\perp} \\
    0 & [L_j]_{\cals^\perp, \cals^\perp}
  \end{bmatrix}, \quad 
  M = 
  \begin{bmatrix}
    [M]_{\cals, \cals} & 0 \\
    0           & [M]_{\cals^\perp, \cals^\perp}
  \end{bmatrix}.
  \label{eq:block-decomposition}
\end{equation}
Invariance is thus straightforwardly satisfied.
Attractiveness is harder to show in general but can, in principle, be verified numerically with the algorithm presented in Sec.~III of Ref.~\cite{Ticozzi_2012}.
However, for a sufficiently chaotic Hamiltonian, one would typically not expect $\mathcal{K}$ to host an invariant subspace satisfying \eqref{eq:invariance}.
It is therefore typically guaranteed that any initial state of the many body system will eventually get trapped inside $\cals$ \cite{Wang_2024}.

In principle, one could also allow for several operators $M_{j}$.
In this case 
\begin{align}
    M_{j} \ket{s_n} = m^{(j)}_n \ket{s_n}, \ \forall n,j,
\end{align}
and the condition becomes $\forall n \neq k, \exists j \ \m{s.t.} \ m^{(j)}_n \neq m^{(j)}_k$.

\subsection{Scar selection}
\label{sec:stabilization}

\begin{figure*}[t]
  \centering
  \includegraphics{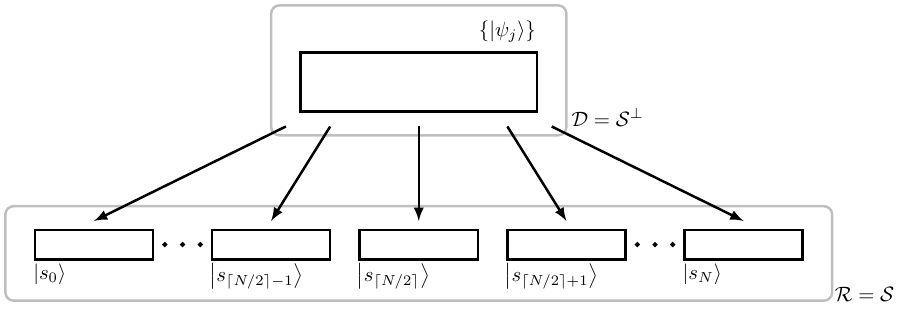}
  \caption{Hilbert space structure of the universal charging protocol, induced by the Lindblad equation Eq.~\eqref{eq:full-Lindblad}. The Hilbert space decomposes into the decaying subspace $\cald = \cals^{\perp}$ with states $\ket{\psi_j}$ and the the asymptotic subspace $\calr = \cals$. The engineered dissipation drives probability into $\cals$, while the continuous monitoring selects exactly one of the scars $\ket{s_n}$ to which the state converges in the long-time limit with probability $p_n$. Each scar state has a its own decay channel, indicated by arrows, cf.~Eq.~\eqref{eq:S_n(t)}.
  }
  \label{fig:scar-structure}
\end{figure*}

In this section we show that every realization of the general charging protocol, described by Eq.~\eqref{eq:quantumTrajectory} in the main text or \cref{Seq:quantum-trajectory} here, asymptotically converges to exactly one of the scars $\ket{s_n}$, effectively yielding Eq.~\eqref{eq:charging}. 

In general, continuously monitored quantum systems are susceptible to the Hilbert space structure of the corresponding average Lindblad dynamics and asymptotically converge towards the minimal invariant subspaces \cite{Schmolke_2025}.
To prove scar convergence we essentially follow the steps outlined in Ref.~\cite{Schmolke_2025} (see also Ref.~\cite{Benoist_2025} for a formal proof in the discrete time scenario).
Any Lindblad generator $\textsf{L}$ induces a unique decomposition of the Hilbert space according to \cite{Baumgartner_2008,Baumgartner_2012}
\begin{equation}
  \calh = \cald \oplus \calr,
  \label{eq:coarse-decomposition}
\end{equation}
where $\cald$ is the decaying subspace 
\begin{align}
  \cald = \left\{\ket{\psi} \in \calh \, \big\vert \, \forall \rho, \lim_{t \to \infty}\bra{\psi} \rho_t \ket{\psi} = 0\right\},
  \label{eq:decaying-subspace}
\end{align}
and its complement, $\calr$, is the recurrent subspace
\begin{align}
  \calr = \left\{\ket{\psi} \in \calh \, \big\vert \, \exists \rho \text{ s.t. } \lim_{t \to \infty}\bra{\psi} \rho_t \ket{\psi} > 0\right\}.
  \label{eq:asymptotic-subspace}
\end{align}
$\cald$ corresponds to the largest subspace in which any initial state asymptotically loses all support, while $\calr$ is the largest subspace in which any initial state has asymptotically full support, $\lim_{t \to \infty} \tr[P_\calr \rho_t] = 1 \, \forall \rho_0$.
In our charging protocol, the scar subspace corresponds to the entire recurrent subspace $\calr = \cals$, and the decaying subspace is $\cald = \cals^{\perp}$.
See \cref{fig:scar-structure} for a pictorial representation of the structure of the Hilbert space.

To prove that individual quantum realizations of the protocol [\cref{Seq:quantum-trajectory}] converge to a single scar state it is now useful to consider the bipartition of the total Hilbert space into a single scar subspace which we denote $\cals_1$ and its complement, denoted accordingly by $\cals_1^\perp$,
\begin{equation}
  \calh = P_1 \calh + P^\perp_1 \calh =  \cals_1 \oplus \cals_1^{\perp},
\end{equation}
where $P_1 = \dyad{s_1}$ and $P^\perp_1 = \id - P_1$ are the corresponding orthogonal projectors.
The probability to find the conditional state $\rho^\m{c}_t$ inside $\cals_1$ at time $t$ is given by the overlap
\begin{equation}
    |\cals_1(t)|^2 \coloneq \tr[P_1 \rho^\m{c}_t].
\end{equation}
Its evolution follows straightforwardly by inserting the stochastic master equation \cref{Seq:quantum-trajectory}
\begin{equation}
  \dd (|\cals_1(t)|^2)
  = \tr[P_1 \textsf{L}(\rho^\m{c}_t)] \dd{t} +  \sqrt{\gamma} |\cals_1(t)|^2  \left[ 2\m{Re}(m_1) (1 - |\cals_1(t)|^2) - \tr[\rho^\m{c}_t (P^\perp_1 M + M^\dag P^\perp_1)] \right] \dd{W_t},
  \label{eq:dS1-full}
\end{equation}
where $m_{1}$ is the eigenvalue of $M$ corresponding to the scar being considered.
Using the fact that $\cals_1$ is an invariant subspace and operators thus must adhere to the general structure Eq.~\eqref{eq:invariance}, we thus obtain a diffusive process with non-negative drift 
\begin{equation}
    \tr[P_1 \textsf{L}(\rho^\m{c}_t)]
    = \sum_j \Gamma_j \tr([L_j]^\dag_{\cals_1,\cals_1^\perp} [L_j]_{\cals_1,\cals^\perp_1} [\rho^\m{c}_t]_{\cals_1^\perp,\cals_1^\perp}) \ge 0.
\end{equation}
which drives the conditional state into $\cals_1$ and a state-dependent diffusion coefficient.

Since the recurrent subspace is invariant and attractive, asymptotic dynamics takes place only inside of $\cals$ \cite{Benoist_2017}.
The Hilbert space thus effectively contracts according to (see \cref{sec:charging-speed} for details on the time scale)
\begin{equation}
  \calh \xrightarrow{t \to \infty} \cals,
\end{equation}
as any trajectory will be eventually supported only on $\cals$. 
Then, inside $\cals$ Eq.~\eqref{eq:dS1-full} reduces to
\begin{equation}
  \dd{(|\cals_1(t)|^2)} 
  = 2 \sqrt{\gamma} |\cals_1(t)|^2 \Big[\m{Re}(m_1) - \sum_i \m{Re}(m_i) |\cals_i(t)|^2 \Big] \dd{W_t},
  \label{eq:dS1-R}
\end{equation}
where $m_{i}$ are the distinct eigenvalues of $M$ corresponding to the scars.
Since total probability must be conserved, $\sum_i |\cals_i(t)|^2 = 1$, the evolution corresponds to a driftless state-dependent Brownian motion on a simplex.
The diffusion coefficient vanishes at the boundaries $|\cals_1|^2 \in \{0, 1\}$, which correspond to stable fixed points of the dynamics (cf.~Ref.~\cite{Ladenburger_2025} for a more detailed discussion of the boundaries).
Moreover, one expects the diffusive dynamics to lead the trajectory to explore the whole interval $[0,1]$ and thus eventually hit one of the boundaries where the evolution stops.
This intuition can be made more rigorous by using the fact that (for the asymptotic evolution in $\cals$) the quantity $|\cals_1(t)|^2$ is a bounded martingale, which means it has no drift, and remains on average equal to its asymptotic value $\mathbb{E}[|\cals_1(t)|^2] = p_1$ given by $p_1 = \lim_{t \to \infty} \tr[P_1 \rho_t]$.
By the martingale convergence theorem, $|\cals_1(t)|^2$ must eventually converge to a random variable \cite{Roldan_2023, Schmolke_2025, Benoist_2025}.
Since we have assumed that all eigenvalues $m_i$ are different, there do not exist stable fixed points other than the boundaries, all scar subspaces are distinguishable by the operator $M$ and the state reaches either full support and converges to the scar ($|\cals_1(t)|^2 \to 1$) or completely looses support and converges to the scar complement ($|\cals_1(t)|^2 \to 0$).
In the former case we obtain a charged battery ready for work extraction and we are done.
In the latter case, one may again partition $\cals_1^{\perp} = \cals_2 \oplus (\cals_2^{\perp} \ominus \cals_1)$.
Applying the same reasoning repeatedly, eventually shows that any quantum trajectory selects exactly one of the scars.

To obtain the probability for these irreversible localization events, we investigate the behavior of the averaged support in a scar, $\mathbb{E}[|\cals_1(t)|^2] = \tr[P_1 \rho_t]$.
The support in each scar subspace is not conserved on average.
There is however a complete set of conserved observables corresponding to the infinite time projectors
\begin{equation}
    \textsf{P}_n^\infty \coloneq \lim_{t \to \infty} e^{\textsf{L}^\dag t} \, \textsf{P}_n,
\end{equation}
where $\textsf{P}_n (\bullet) = P_n \bullet P_n$ and $\textsf{L}^\dag$ is the adjoint Lindbladian of \eqref{eq:full-Lindblad} governing the Heisenberg evolution of observables. 
The $P_n^\infty$ collect all the probability already residing in $\cals$ and the one that flows down from $\cals^{\perp}$ into $\cals_n$, and thus determine the asymptotic support of any initial state $\rho_{0}$ in the scar subspaces
\begin{equation}
    p_n(\rho_0) = \lim_{t \to \infty} \tr[P_n \rho_t] = \tr[P^\infty_n \rho_{0}].
\end{equation}
Completeness of the $P^\infty_n$ ensures that $\sum_n p_n(\rho_0) = 1$.
Since any realization of the measurement process will converge to exactly one scar, the probability to converge to a particular state $\ket{s_n}$ is exactly given by $p_n(\rho_0)$.
See \cref{fig:battery}c for the $p_n(\rho_0)$ distribution in the spin-$1$ $XY$ model.
In general, the stationary probability distribution of the density matrix, for a given initial state, is hence
\begin{equation}
  \m{P}^\m{s}(\rho_0) = \sum_n p_n(\rho_0) \, \delta(|\cals_n|^2 - 1) \dyad{s_n},
\end{equation}
where the sum runs over all scars.
The charging protocol Eq.~\eqref{eq:quantumTrajectory} thus realizes a generalized asymptotic projection that effectively maps the initial state onto one of the scars according to the update rule
\begin{equation}
  \rho_{0} \to 
  \rho^\m{c}_\infty 
  = \frac{\textsf{P}^{\infty}_n \rho_{0}}{\tr[\textsf{P}^\infty_n \rho_{0}]} 
  = \dyad{s_n},
\end{equation}
with probability given by the generalized Born rule $p_n(\rho_0) = \tr[\textsf{P}^\infty_n \rho_{0}]$ [cf.~Eq.~\eqref{eq:charging}].

\section{Charging speed}
\label{sec:charging-speed}

We here provide an in-depth analysis on the charging speed given by the asymptotic convergence rate of our protocol towards a scar and the associated charging time of the battery.
The charging speed results from a nontrivial interplay of concurrently implemented engineered dissipation and continuous monitoring, and it is typically enhanced compared to first implementing engineered dissipation followed by any type of measurement scheme.
Before analyzing the total charging time, it is however instructive to first discuss the characteristic relaxation time scale of the engineered dissipation and the stochastic scar selection time of the continuous monitoring separately.

\subsection{Relaxation time of the engineered dissipation}

In the absence of measurement, the engineered dissipation drains the decaying subspace $\cald = \cals^\perp$ and drives any initial state into the scar subspace $\calr = \cals$.
The relaxation time thus corresponds to the time of the effective Hilbert space contraction $\calh \to \cals$, set by the generator $\textsf{L}_\m{ED}$ which we repeat here for convenience
\begin{equation}
  \dot{\rho}_t
  = \textsf{L}_\m{ED}(\rho_t)
  = -\m{i} [H,\rho_t] + \sum_{j} \Gamma_j \textsf{D}[L_j](\rho_t).
  \label{eq:engineered-dissipation}
\end{equation}
Since the engineered dissipation induces decay, i.e., $\mathcal{D} \neq \emptyset$, the generator $\textsf{L}_\m{ED}$ is not normal and can hence not be diagonalized by a unitary transformation.
Denote by $\Lambda_k$ the eigenvalues of $\textsf{L}_\m{ED}$ and by $d_\m{s} = \dim(\ker \textsf{L}_\m{ED})$ the dimension of its kernel.
Suppose there are $N$ scars, then the dimension of the kernel is bounded as $N \le d_\m{s} \le N^2$, where the lower bound corresponds to each scar constituting a disjoint decoherence-free subspace while the upper bound corresponds to the entire scar subspace being decoherence-free [cf.~Eq.~\eqref{eq:dfs-conditions}].
Let the eigenvalues be arranged by decreasing real part,
\begin{equation}
    0 = \m{Re}(\Lambda_0) = \cdots = \m{Re}(\Lambda_{d_\m{s} - 1}) > \m{Re}(\Lambda_{d_\m{s}}) \ge \m{Re}(\Lambda_{d_\m{s} + 1}) \ge \cdots.
\end{equation}
In general, we can identify the spectral gap 
\begin{equation}
    \eta_\m{ED} = \m{Re}(\Lambda_{d_\m{s}})
\end{equation}
as the second largest real part in the spectrum, which corresponds to the slowest mode and governs the asymptotic decay rate.
Note that, both non-normality and non-diagonalizability can independently cause anomalously large relaxation times that are not adequately captured by the simple estimate $\tau_\m{ED} \approx 1/\eta_\m{ED}$ \cite{Song_2019,Mori_2020,Lee_2023,Haga_2021,Znidaric_2023}.
Nevertheless, $\eta_\m{ED}$ does always determine the asymptotic convergence rate.

Along the Lindbladian dynamics Eq.~\eqref{eq:engineered-dissipation}, the probability to find the state in the scar subspace evolves according to the differential equation
\begin{equation}
  \dv{t} \tr[P_\cals \rho_t]
  = \sum_j \tr([L_j^\dag]_{\cals, \cals^\perp} [L_j]_{\cals, \cals^\perp} [\rho_t]_{\cals^\perp, \cals^\perp})^\dag \ge 0.
  \label{eq:S(t)}
\end{equation}
To arrive at the the above relation, we have used that the structural relations \cref{eq:invariance}, i.e., that $S$ is an invariant subspace.
The support $\tr[P_\cals \rho_t]$ increases monotonically \cite{Baumgartner_2008} as the engineered dissipation continuously takes probability from the decaying subspace and maps it into the scar subspace until $\tr[P_\cals \rho_t] = 1$ in the steady state.
Similarly, the support of an individual scar subspaces evolves as
\begin{equation}
  \dv{t} \tr[P_n \rho_t]
  = \sum_j \tr( [L_j]^\dag_{\cals_n, \cals^\perp} [L_j]_{\cals_n, \cals^\perp} [\rho_t]_{\cals_n^\perp, \cals_n^\perp} ) \ge 0.
  \label{eq:S_n(t)}
\end{equation}
As in Eq.~\eqref{eq:S(t)}, there is only inflow, no outflow of probability.
Each arrow in \cref{fig:scar-structure} corresponds to one of the rate equations \eqref{eq:S_n(t)}.

Because the asymptotic convergence towards $\cals$ is monotonic we define the relaxation time of the engineered dissipation for the scar subspace as the first-passage time to becoming $\epsilon$-close to reaching full support
\begin{equation}
    \tau_\m{ED}(\epsilon) = \inf\Big\{ t : \sum_n \bra{s_n} \rho_t \ket{s_n} > 1 - \epsilon \Big\},
\end{equation}
where the small parameter $\epsilon$ sets an error tolerance. 
Note that $\rho_t$ denotes here the solution to Eq.~\eqref{eq:engineered-dissipation}.
To provide a meaningful comparison to the full protocol where individual scars are selected, we accordingly define the individual relaxation times for each scar
\begin{equation}\label{eq:tau-n-ed}
    \tau_{\m{ED}, n}(\epsilon) = \inf\left\{ t : \frac{\bra{s_n} \rho_t \ket{s_n}}{p_{\m{ED},n}(\rho_0)} > 1 - \epsilon \right\},
\end{equation}
where $p_{\m{ED},n}(\rho_0) \coloneq \lim_{t \to \infty} \bra{s_n} P_n \exp(\textsf{L}_\m{ED} t) \rho_0 \ket{s_n}$ is the asymptotic support in the $n$-th scar of the engineered dissipation for a given initial state $\rho_0$.
In \cref{fig:charging-time}a, we show the evolution of the supports $\bra{s_n} \rho_t \ket{s_n}$ in a system of $N = 4$ spins for the initial state $\rho_0 = \mathds{1}/3^N$.
Although the Lindbladian is not a normal operator, the time to reach the stationary values seems, at least for this small system, to be well described by the spectral gap $\rat_{\m{ED}}$ (gray vertical dashed line in \cref{fig:charging-time}a).

\subsection{Scar selection time of the continuous monitoring}

A possible strategy to asymptotically select one of the scar states is to independently apply the two processes, engineered dissipation and continuous monitoring, in sequence.
Suppose the engineered dissipation has already driven the initial state into the scar manifold and only once full support has been reached ($\tr[P_\cals \rho_t] = 1$) it is turned off and the continuous monitoring is activated.
Then, the dynamics is given by only the monitoring
\begin{equation}
  \dd{\rho} 
  = -\mathrm{i} [H, \rho^\m{c}_t] \dd{t} 
    + \gamma \left( M \rho^\m{c}_t M^\dag - \frac{1}{2} \{M^\dag M, \rho^\m{c}_t\} \right) \dd{t} 
    + \sqrt{\gamma} [M \rho^\m{c}_t + \rho^\m{c}_t M^\dag - \tr(\rho^\m{c}_t M + M^\dag \rho^\m{c}_t) \rho^\m{c}_t] \dd{W_t}.
\end{equation}
The probability for the system to be found inside the scar subspace $\cals_n$ at time $t$ then evolves according to
\begin{equation}\label{eq:dsn}
  \tr[P_n \dd \rho^\m{c}_t]
  =
  \dd{(|\cals_n(t)|^2)} 
  = 2 \sqrt{\gamma} |\cals_n(t)|^2 \Big[\m{Re}(m_n) - \sum_i \m{Re}(m_i) |\cals_i(t)|^2 \Big] \dd{W_t},
\end{equation}
where $m_i$ are the eigenvalues of the operator $M$.
\Cref{eq:dsn} describes a state-dependent Brownian motion on a simplex, since total probability is conserved $\sum_i |\cals_i(t)|^2 = 1$.
In analogy to \cref{sec:stabilization} it can be shown that trajectories eventually converge to one of the scars, that is $|\cals_n(t)|^2 \to 1$ for some $n$.
Since convergence is asymptotically exponential \cite{Benoist_2014,Benoist_2025}, we again determine the time it takes for the state to get $\epsilon$-close to reaching a scar.
That is we define the stochastic first-passage time of the measurement-induced scar-selection process given that the (effective) initial state has full support on $\cals$, that is
\begin{equation}
    \tau^\m{c}_{\m{M}, n}(\epsilon | \rho_0) = \inf\{t : \bra{s_n} \rho^\m{c}_t \ket{s_n} > 1-\epsilon\}, \ \text{for} \ \rho_0 = P_\cals \rho_0 P_\cals.
\end{equation}
\Cref{eq:dsn} essentially describes stochastic transitions between decoherence-free subspaces.
For two subspaces the first-passage time distribution can be found explicitly \cite{Ladenburger_2025}.
For more than two subspaces, a pathwise upper bound is available.
Suppose that $\ket{s_n}$ is the selected scar ($|\cals_n(t)|^2 \to 1$).
It has been shown in Ref.~\cite{Benoist_2014} that the complementary subspace $\bar \cals^\perp_n = \cals \ominus \cals_n$ gets asymptotically drained exponentially fast with a time-independent convergence rate.
Concretely, we obtain the asymptotic behavior of $|\cals_n(t)|$ as \cite[Th.~5]{Benoist_2014}
\begin{align}
  1 - |\cals_n(t)|^2 \lesssim e^{- t [\alpha_n + o(1)]},
  \label{eq:x_n-asymptotic}
\end{align}
where $\alpha_n = 2 \gamma \min_{i} \{\m{Re}(m_n - m_i)^2\}$.
For the spin-$1$ $XY$ model, the eigenvalues of the total magnetization $M = S^z$ are spaced equidistantly with $m_{i+1} - m_i = 1$, and we thus have $\alpha_n = 2 \gamma, \ \forall n$.

\subsection{Ultra strong measurement limit}
\label{sec:strong-measurement}

Along the full charging protocol the support in the scar subspace evolves according to
\begin{equation}
  \begin{split}
  \dd{(|\cals(t)|^2)} 
  = \tr[P_\cals \textsf{L}(\rho^\m{c}_t)] \dd{t} 
     + \sqrt{\gamma}  \Big[ |\cals^\perp(t)|^2 \tr[\rho_{t}^{\m{c}} (P_\cals M + M^\dag P_\cals) ]
     - |\cals(t)|^2 \tr[\rho_{t}^{\m{c}} (P^\perp_\cals M + M^\dag P^\perp_\cals) ] \Big] \dd{W_t},
  \end{split}
  \label{eq:dS-full}
\end{equation}
with 
\begin{equation}
    \tr[P_\cals \textsf{L}(\rho^\m{c}_t)] = \sum_j \Gamma_j \tr( [L_j]^\dag_{\cals, \cals^\perp} [L_j]_{\cals, \cals^\perp} [\rho^\m{c}_t]_{\cals^\perp, \cals^\perp} ) \ge 0,
\end{equation}
and where $|\cals^\perp(t)|^2 = \tr[P^\perp_\cals \rho^\m{c}_t]$ is the instantaneous support of the state in the decaying subspace.

In the limit where the measurement is the dominant process, $\| \sqrt{\Gamma_j} L_j\| / \| \sqrt{\gamma} M\| \to 0$, the first term in Eq.~\eqref{eq:dS-full} is negligible and trajectories will hence become asymptotically supported on $\cals$ or $\cals^\perp$.
The probability to converge to the decaying subspace is then given by the initial support $p_{\cals^\perp} = \tr[P^\perp_\cals \rho_{0}]$, which is extensive for the maximally mixed state, $p_{\cals^\perp}(\mathds{1}/d) = 1 - (N+1)/d$, where $d$ denotes the dimension of the Hilbert space.
Measurement by itself is thus unable to charge the battery because the state is essentially always being mapped to the scar complement.
On the other hand, the engineered dissipation alone cannot single out individual scars.

\subsection{Strong dissipation}
\label{sec:strong-dissipation}

If the engineered dissipation is large with $\|\sqrt{\Gamma_j} L_j\| \gg \|\sqrt{\gamma} M\|$, the two processes become effectively independent.
Convergence towards the scar subspace takes place almost without being affected by the measurement and the mean time to converge to a scar decomposes into the individual contributions $\tau_n(\epsilon) \approx \tau_{\m{ED}, n}(\epsilon) + \tau_{\m{M}, n}(\epsilon | \rho_0)$.

\subsection{Convergence speed of the full protocol}

\begin{figure*}[t]
  \centering
  \begin{tikzpicture}
        \node (a) at (0,0) {\includegraphics{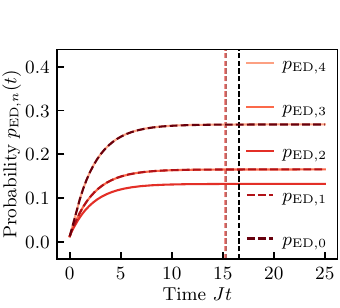}};
        \node (b) at (6.25,0) {\includegraphics{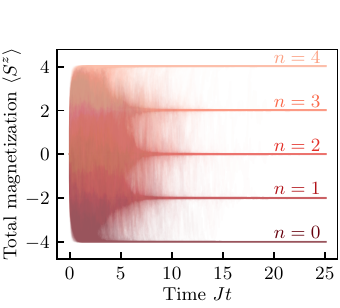}};
        \node (c) at (12.2,0) {\includegraphics{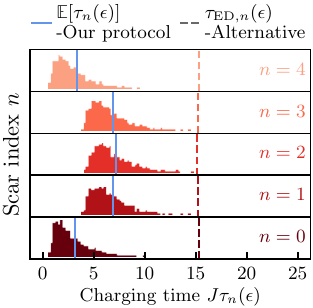}};

        \draw (-2.5,2.3) node {\bfseries a};
        \draw (3.45,2.3) node {\bfseries b};
        \draw (9.4, 2.3) node  {\bfseries c};

  \end{tikzpicture}
  \caption{Charging times of a $N = 4$ spin-$1$ $XY$ battery for the initial state $\rho(0) = \mathds{1}/3^N$. \textbf{a}, Time evolution of the support in each scar subspace $p^\m{ED}_n(t) = \tr[P_n \rho_\m{ED}(t)]$ [Eq.~\eqref{eq:engineered-dissipation}] due only to the engineered dissipation.
  The initial state is mapped onto $\cals$ as the individual probabilities converge to their stationary values. Vertical dashed lines denote the corresponding first-passage times $\tau_{n,\m{ED}}(\epsilon)$ [Eq.~\eqref{eq:tau-n-ed}] to become $\epsilon$-close (here: $\epsilon = \num{e-3}$). The black dashed line shows the spectral gap estimate $\tau_\m{ED}(\epsilon) \approx \ln(\epsilon)/\eta_{\m{ED}}$. \textbf{b}, Subensemble of $10^4$ quantum trajectories along the full charging protocol [Eq.~\eqref{eq:full-protocol}]. The total magnetization, $M = S^z$, witnesses which of the scar states is selected. \textbf{c}, First-passage times $\tau^\m{c}_n(\epsilon)$ of the full protocol for individual quantum trajectories to get $\epsilon$-close to a given scar [Eq.~\eqref{eq:tau-n}]. Blue lines indicate the mean first-passage time, red dashed lines show $\tau_{\m{ED},n}$ for reference, which corresponds to the fastest alternative consecutive protocol. Convergence towards extremal scars is faster than towards states in the bulk. Selection times are symmetric, resulting from the underlying symmetries of the operators $L_j$ and $M$ in this particular model. The parameters are $\Gamma_j / J = 2, \ \forall j$, $\gamma / J = 2$.}
  \label{fig:charging-time}
\end{figure*}

By performing engineered dissipation and continuous measurement concurrently a new, modified, convergence rate emerges, which consists of a contribution from the engineered dissipation and a contribution exclusively due to the monitoring.
This measurement-induced speedup is proportional to the distinguishability of the subspaces $\cals$ and $\cals^\perp$ and results from an effectively ongoing \enquote{which-subspace} measurement.
The relaxation rate of the engineered dissipation typically scales unfavorably with the system size and closes in the thermodynamic limit, leading to divergent relaxation times.
However, the subspace distinguishability does not suffer from such unfavorable scaling and we thus expect the relative enhancement to remain prominent as the number of particles increases. 
For the full protocol, we accordingly define the first-passage time [cf.~Eq.~\eqref{eq:first-passage}]
\begin{equation}\label{eq:tau-n}
    \tau^\m{c}_n(\epsilon) = \inf\{ t \; \colon \bra{s_n} \rho^\m{c}_t \ket{s_n} > 1-\epsilon\}.
\end{equation}
In \cref{eq:first-passage}c, we display the histograms of the first passage times for a $N = 4$ spin-1 battery.
Almost all trajectories get $\epsilon$-close to a scar in a time considerably less than the fastest possible alternative protocol which consists of engineered dissipation to reach the scar subspace, followed by an instantaneous projective measurement that then singles out one of the scars.
This ideal alternative procedure takes at least a time $\tau_{\m{ED},n}$ [cf.~\cref{fig:relaxation}].

We now show that convergence towards an individual scar along the universal charging protocol is exponential in time and analyze the asymptotic charging speed. 
For a given initial state $\rho_0$, we first identify the asymptotic convergence rate of the engineered dissipation by considering the outflow out of the scar complement
\begin{equation}\label{eq:sperp}
    \dd (|\cals^\perp(t)|^2)
    = -|\cals^\perp(t)|^2 \sum_j \Gamma_j \tr( [L_j]^\dag_{\cals, \cals^\perp} [L_j]_{\cals, \cals^\perp} [\tilde \rho_t]_{\cals^\perp, \cals^\perp} ) \dd t
\end{equation}
where $|\cals^\perp(t)|^2 = \tr[P^\perp_\cals \rho_t]$ is the support in the complement and
\begin{equation}
    [\tilde \rho_t]_{\cals^\perp, \cals^\perp} 
    = \frac{[\rho_t]_{\cals^\perp, \cals^\perp}}{|\cals^\perp(t)|^2}
\end{equation}
is the normalized state restricted to $\cals^\perp$.
The deterministic rate of outflow from $\cals^\perp$ is determined by the terms $[L_j]^\dag_{\cals, \cals^\perp} [L_j]_{\cals, \cals^\perp}$ which transfer probability from the complement into the scar subspace.
The formal solution to Eq.~\eqref{eq:sperp} is exponential and reads
\begin{equation}
    |\cals^\perp(t)|^2 = |\cals^\perp(0)|^2 \exp\Bigg(- \int_0^t \dd{s} \sum_j \Gamma_j \tr( [L_j]^\dag_{\cals, \cals^\perp} [L_j]_{\cals, \cals^\perp} [\tilde \rho_t]_{\cals^\perp, \cals^\perp}) \Bigg).
\end{equation}
We accordingly define the asymptotic rate as
\begin{equation}
    \rat_\m{ED} 
    \coloneq - \lim_{t \to \infty} \frac{1}{t} \ln(|\cals^\perp(t)|^2)
    = \lim_{t \to \infty} \frac{1}{t} \int_0^t \dd{s} \sum_j \Gamma_j \tr( [L_j]^\dag_{\cals, \cals^\perp} [L_j]_{\cals, \cals^\perp} [\tilde \rho_t]_{\cals^\perp, \cals^\perp}).
\end{equation}
It determines how fast the state $\rho_t$ looses support on $\cals^\perp$ and thus how fast it converges to $\cals$. 
The asymptotic rate $\rat_\m{ED}$ depends on the initial state $\rho_0$.
If $\rho_0$ has overlap with the slowest mode, it is formally lower bounded by the spectral gap of the corresponding restricted generator $[\textsf{L}]_{\cals^\perp,\cals^\perp}$ \cite[Eq.~(13)]{Benoist_2017}
\begin{equation}\label{eq:rED}
    \rat_\m{ED} \ge \min\{-\m{Re}(\Lambda) \, \vert \, \Lambda \in \spec([\textsf{L}]_{\cals^\perp,\cals^\perp})\},
\end{equation}
where 
\begin{equation}
    [\textsf{L}]_{\cals^\perp,\cals^\perp}\rho = -\m{i} [[H]_{\cals^\perp, \cals^\perp}, \rho] + \sum_j \Gamma_j \bigg( [L_j]_{\cals^\perp, \cals^\perp} \rho [L_j]^\dag_{\cals^\perp,\cals^\perp} - \frac{1}{2} \{[L_j]^\dag_{\cals,\cals^\perp} [L_j]_{\cals,\cals^\perp} + [L_j]^\dag_{\cals^\perp,\cals^\perp} [L_j]_{\cals^\perp,\cals^\perp}, \rho\} \bigg),
\end{equation}
with $\rho = P_\cals^\perp \rho P_\cals^\perp$. Note that $\rat_\m{ED}$ captures only the outflow out of $\cals^\perp$. Reaching the stationary state inside of $S$ may take additional time.

Benoist, Pellegrini and Ticozzi have shown in Ref.~\cite{Benoist_2017} that the exponential convergence rate of individual quantum trajectories from the decaying subspace ($\cald = \cals^\perp$) towards the recurrent subspace ($\calr = \cals$) is almost surely at least as fast as the slowest time scale of the ensemble averaged dynamics, the spectral gap.
In our universal charging protocol, a similar mechanism is at play.
In the $XY$ model we observe that trajectories are driven towards individual scars more rapidly than the Lindbladian dynamics of the engineered dissipation is able to drag the state into the scar subspace [cf.~\cref{fig:relaxation}].
As detailed below, we expect this to be a general feature of our protocol. 

Suppose that the charging protocol asymptotically selects the scar $\ket{s_n}$.
In analogy to Eq.~\eqref{eq:sperp} (recall that $M$ is block diagonal), leakage out of the scar complement $\cals^\perp_n$ is described by the differential equation
\begin{equation}\label{eq:sperp-full}
    \begin{split}
    \dd (|\cals^\perp_n(t)|^2)
    =& -|\cals_n^\perp(t)|^2 \sum_j \tr( [L_j]^\dag_{\cals_n, \cals_n^\perp} [L_j]_{\cals_n, \cals_n^\perp} [\tilde \rho^\m{c}_t]_{\cals_n^\perp, \cals_n^\perp} ) \dd t \\
     & + |\cals_n^\perp(t)|^2 \sqrt{\gamma} \left\{ \tr([\tilde \rho^\m{c}_t]_{\cals_n^\perp, \cals_n^\perp} ([M]_{\cals_n^\perp, \cals_n^\perp} + [M^\dag]_{\cals_n^\perp, \cals_n^\perp})) - \tr[\rho^\m{c}_t (M + M^\dag)] \right\} \dd W_t.
    \end{split}
\end{equation}
The formal solution is given by a Dol\'eans--Dade exponential
\begin{equation}
\begin{split}
    |\cals^\perp_n(t)|^2 
    =& |\cals^\perp_n(0)|^2 \exp\Bigg( - \int^t_0 \bigg\{\sum_j \Gamma_j \tr( [L_j]^\dag_{\cals_n, \cals_n^\perp} [L_j]_{\cals_n, \cals_n^\perp} [\tilde \rho^\m{c}_t]_{\cals_n^\perp, \cals_n^\perp} ) + \delta_n(\rho^\m{c}_t)\bigg\} \dd{s} \\
    &+ \sqrt{\gamma} \int^t_0  \bigg\{\tr([\tilde \rho^\m{c}_t]_{\cals_n^\perp, \cals_n^\perp} ([M]_{\cals_n^\perp, \cals_n^\perp} + [M^\dag]_{\cals_n^\perp, \cals_n^\perp})) - \tr[\rho^\m{c}_t (M + M^\dag)] \bigg\}\dd{W_s}\Bigg),
\end{split}
\end{equation}
where we have defined the \Ito correction 
\begin{equation}
    \delta_n(\rho^\m{c}_t) 
    = \frac{1}{2} \sqrt{\gamma} \abs{\tr([\tilde \rho^\m{c}_t]_{\cals^\perp_n, \cals^\perp_n} ([M]_{\cals^\perp_n, \cals^\perp_n} + [M^\dag]_{\cals^\perp_n, \cals^\perp_n})) - \tr[\rho^\m{c}_t (M + M^\dag)]}^2 \ge 0.
\end{equation}
At late times the state $\rho^\m{c}_t$ will be mostly supported on $\cals$, therefore $\delta_n(\rho^\m{c}_t)$ can be interpreted in terms of an instantaneous distinguishability between $\cals_n$ and $\cals^\perp_n$ induced by the operator $M$.
Since $\delta_n(\rho^\m{c}_t)$ is moreover non-negative, it can only ever increase the asymptotic convergence speed.
Note however that this measurement-induced speedup relies on the simultaneous action of continuous monitoring and engineered dissipation as each process by itself is unable to select an individual scar.
If the state is fully supported on $\cals$, this rate reduces to 
\begin{equation}\label{eq:sperp-full-solution}
     \delta_n(\rho^\m{c}_t) = \frac{1}{2} \left( 2\sqrt{\gamma} \frac{|\cals_n(t)|^2}{|\cals_n^\perp(t)|^2} \right)^2 \abs{ \left(\sum_i \m{Re}(m_i) |\cals_i(t)|^2 - \m{Re}(m_n) \right) }^2, 
     \quad \text{for } \rho^\m{c}_t = P_\cals \rho^\m{c}_t P_\cals, 
\end{equation}
which is consistent with Eq.~\eqref{eq:dsn} and recovers the corresponding asymptotic rate $\alpha_n = 2 \gamma \min_i\{\m{Re}(m_n - m_i)^2\}$ in Eq.~\eqref{eq:x_n-asymptotic}.

Note that for a general operator $M$ that is not block-diagonal, there would be the additional contribution $[M^\dag]_{\cals_n, \cals^\perp_n} [M]_{\cals_n, \cals^\perp_n}$ to the deterministic outflow which could even further increase the convergence speed.
It can be shown that the diffusive part in Eq.~\eqref{eq:sperp-full-solution} is a martingale \cite{Benoist_2017, Roldan_2023}
\begin{equation}
    M_W(t) \coloneq \int^t_0  \bigg\{\tr([\tilde \rho^\m{c}_t]_{\cals_n^\perp, \cals_n^\perp} ([M]_{\cals_n^\perp, \cals_n^\perp} + [M^\dag]_{\cals_n^\perp, \cals_n^\perp})) - \tr[\rho^\m{c}_t (M + M^\dag)] \bigg\}\dd{W_s}
\end{equation}
and that since it has no preferred direction
\begin{equation}
    \lim_{t \to \infty} \frac{1}{t} M_W(t) = 0.
\end{equation}
Therefore the asymptotic rate is determined only by the deterministic part
\begin{equation}
    \rat^\m{c} \coloneq -\lim_{t \to \infty} \frac{1}{t} \ln(|\cals^\perp_n(t)|^2) 
    = \lim_{t \to \infty} \frac{1}{t} \int^t_0 \bigg\{\sum_j \Gamma_j \tr( [L_j]^\dag_{\cals_n, \cals_n^\perp} [L_j]_{\cals_n, \cals_n^\perp} [\tilde \rho^\m{c}_t]_{\cals_n^\perp, \cals_n^\perp}) + \delta_n(\rho^\m{c}_s)\bigg\} \dd{s}
\end{equation}

We define the effective trajectory-dependent rate
\begin{equation}
    \bar \delta^\m{c}_n = \lim_{t \to \infty} \frac{1}{t} \int^t_0 \dd{s} \delta_n(\rho^\m{c}_s).
\end{equation}
which is formally bounded as 
\begin{equation}
    0 \le \bar \delta^\m{c}_n \le \sqrt{\gamma} \spr(M + M^\dag)^2 / 2,
\end{equation}
where $\spr(A)$ denotes the spectral spread of the operator $A$. 
For a typical trajectory that starts from a generic initial state we expect that $\bar \delta^\m{c}_n > 0$ leading to an enhanced asymptotic convergence rate.
In the spin-1 $XY$ model, the spectral spread of $S^z$ increases with the system size, thereby also increasing the potential to speed up the charging process as a whole.

\section{Scar certification time}

\begin{figure}[t]
  \centering
  \includegraphics{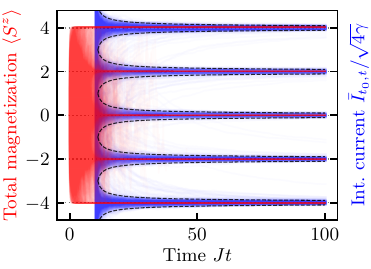}
  \caption{Scar certification for $N = 4$ spins [cf.~\cref{fig:charging-time}] for $\num{2e3}$ typical trajectories. Total magnetization $\langle S^z \rangle$ (red) and corresponding integrated measurement currents $\bar{I}_{t_0,t}$ (blue). To guarantee that most trajectories have already converged to a scar (as witnessed by $\langle S^z \rangle$) the start of the integration time is informed by the first-passage time distribution \cref{fig:charging-time} and here chosen to be $t_0 = 10$. The $\bar{I}_{t_0,t}$ are expected to converge in mean square to $2 \sqrt{\gamma} \m{Re}(m_n)$ with $\propto t^{-1}$. Black dashed lines indicate the theoretically predicted standard deviation $\propto (t-t_0)^{-1/2}$ (no fitting was performed).}
  \label{fig:integrated-current}
\end{figure}

Continuous monitoring yields a measurement current given by
\begin{equation}
    \dd{Y_t} = \sqrt{\gamma} \tr[(M+M^\dag) \rho^\m{c}_t] \dd{t} + \dd{W_t},
\end{equation}
where $\gamma$ is the measurement strength.
The current $\dd{Y_t} = I_t \dd{t}$ provides information about the observable $(M+M^\dag)$.
Here we show explicitly the convergence of the time integrated measurement current $\bar I_{t,t_0} = \frac{1}{t} \int^{t_{0} + t}_{t_{0}} \dd{Y_s}$ towards an eigenvalue of the operator $(M+M^\dag)$.
In \cref{sec:stabilization} we have shown that every quantum trajectory converges towards a scar $\ket{s_n}$ in time $\tau^\m{c}_n(\epsilon)$ [cf.~Eq.~\eqref{eq:charging}].
By construction of the charging protocol, each scar is moreover an eigenstate of $M$ with eigenvalue $m_n$. 
For $t \ge \tau^\m{c}_n(\epsilon)$ convergence (up to an error $\epsilon$) has occurred and we approximately have $\rho_{t}^{\m{c}} = \dyad{s_n}$ for some scar $\ket{s_n}$. 
The integrated current is hence (approximately) given by
\begin{equation}
    \bar I_{t,\tau^\m{c}_n(\epsilon)} = 2 \sqrt{\gamma} \m{Re}(m_n) + \frac{1}{t} \int^{\tau^\m{c}_n(\epsilon) + t}_{\tau^\m{c}_n(\epsilon)} \dd{W_s}.
\end{equation}
It then follows straightforwardly that the time integrated current converges in mean-square according to 
\begin{equation}
    \mathbb{E}[(\bar I_{t,\tau^\m{c}_n(\epsilon)} - 2 \sqrt{\gamma} \m{Re}(m_n))^2] = 1/t.
\end{equation}
\Cref{fig:integrated-current} shows both the total magnetization $M = S^z$ and $\bar I_{t,t_0}$ for an ensemble of trajectories in the spin-$1$ $XY$ model.
The start of the integration interval is chosen such that most trajectories have converged to a scar and the $t^{-1/2}$ convergence of the standard deviation is clearly visible.

\section{Spin-$1$ $XY$ chain}

\subsection{Shiraishi--Mori form}

Here we are going to show that the spin-$1$ $XY$ chain with Hamiltonian 
\begin{equation}
    H = J \sum_{j=1}^{N-1} S_{j}^{x}S_{j+1}^{x} + S_{j}^{y}S_{j+1}^{y} + \sum_{j= 1}^{N} \left[h S_{j}^{z} + D (S_{j}^{z})^{2} \right]
\end{equation}
conforms to the Shiraishi--Mori form. That is, it can be written as $H = \sum_{j} P_{[j]}^{\dagger} h_{[j]} P_{[j]} + H_{\mathrm{ZE}}$ where, in this case, $P_{[j]}$ are 2-local Hermitian projectors and $[H_{\mathrm{ZE}}, P_{[j]}] = 0$. Following Appendix F of Ref. \cite{Moudgalya_2024}, the Shiraishi--Mori projectors $P_{[j]}$ can be easily defined using the following two-site states:
\begin{alignat}{2}
    \ket{X_{1}}_{j,j+1} & = (\ket{1,-1} + \ket{-1,1}) / \sqrt{2} \,, \qquad && \ket{X_{4}}_{j,j+1} = \ket{0,1} \,, \\
    \ket{X_{2}}_{j,j+1} & = \ket{0,0} \,, \qquad && \ket{X_{5}}_{j,j+1} = \ket{-1,0} \,, \\
    \ket{X_{3}}_{j,j+1} & = \ket{1,0} \,, \qquad && \ket{X_{6}}_{j,j+1} = \ket{0,-1} \,.
\end{alignat}
First, note that
\begin{equation}
    S_{j}^{x}S_{j+1}^{x} + S_{j}^{y}S_{j+1}^{y} = \sqrt{2} \ketbra{X_{1}}{X_{2}}_{j,j+1} + \ketbra{X_{3}}{X_{4}}_{j,j+1} + \ketbra{X_{5}}{X_{6}}_{j,j+1} \, + \, \text{h.c.}  \,,
\end{equation}
and, therefore,
\begin{alignat}{2}
    (S_{j}^{x}S_{j+1}^{x} + S_{j}^{y}S_{j+1}^{y})\ket{X_{1}}_{j,j+1} & = \sqrt{2} \ket{X_{2}}_{j,j+1} \,, \qquad && (S_{j}^{x}S_{j+1}^{x} + S_{j}^{y}S_{j+1}^{y})\ket{X_{4}}_{j,j+1} = \ket{X_{3}}_{j,j+1} \,, \\
    (S_{j}^{x}S_{j+1}^{x} + S_{j}^{y}S_{j+1}^{y})\ket{X_{2}}_{j,j+1} & = \sqrt{2} \ket{X_{1}}_{j,j+1} \,, \qquad && (S_{j}^{x}S_{j+1}^{x} + S_{j}^{y}S_{j+1}^{y})\ket{X_{5}}_{j,j+1} = \ket{X_{6}}_{j,j+1} \,, \\
    (S_{j}^{x}S_{j+1}^{x} + S_{j}^{y}S_{j+1}^{y})\ket{X_{3}}_{j,j+1} & = \ket{X_{4}}_{j,j+1} \,, \qquad && (S_{j}^{x}S_{j+1}^{x} + S_{j}^{y}S_{j+1}^{y})\ket{X_{6}}_{j,j+1} = \ket{X_{5}}_{j,j+1} \,.
\end{alignat}
Thus, if we define the $2$-local Hermitian operator
\begin{equation}
    P_{[j]} = P_{j,j+1} = \sum_{k = 1}^{6} \dyad{X_{k}}_{j,j+1},
\end{equation}
we immediately get
\begin{equation}
    P_{j,j+1} (S_{j}^{x}S_{j+1}^{x} + S_{j}^{y}S_{j+1}^{y})  P_{j,j+1} = (S_{j}^{x}S_{j+1}^{x} + S_{j}^{y}S_{j+1}^{y}).
    \label{eq:SMPlocal}
\end{equation}
Since all the $\ket{X_{k}}$ are $2$-particle states and orthogonal to each other, $P_{j,j+1}$ is indeed a projector, and it trivially eliminates all scars $\ket{s_n}$. Additionally, it is straightforward to check
\begin{equation}
    \left[ P_{j,j+1} , \sum_{j= 1}^{N} \left[h S_{j}^{z} + D (S_{j}^{z})^{2} \right] \right] = 0.
\end{equation}
Therefore, we can write the spin-$1$ $XY$ chain Hamiltonian as 
\begin{equation}
    H = J \sum_{j=1}^{N-1} P_{j,j+1} (S_{j}^{x}S_{j+1}^{x} + S_{j}^{y}S_{j+1}^{y})  P_{j,j+1} + H_{\m{ZE}},
\end{equation}
where $H_{\mathrm{ZE}} = \sum_{j} \left[h S_{j}^{z} + D (S_{j}^{z})^{2} \right]$. As $S_{j}^{x}S_{j+1}^{x} + S_{j}^{y}S_{j+1}^{y}$ is also $2$-local, we can set $h_{[j]} = J (S_{j}^{x}S_{j+1}^{x} + S_{j}^{y}S_{j+1}^{y})$ and finally obtain the desired Shiraishi--Mori form.

\subsection{Lindblad operators for the engineered dissipation}

The protocol outlined in the main text suggests to use the Shiraishi--Mori projectors to construct the $L_{j}$ necessary to implement the engineered dissipation to steer any initial state inside $\cals$ ($L_{j} = O_{[j]} P_{[j]}$). Here, we are going to show that $\ker (P_{j,j+1})$ and $\ker (S_{j}^{x}S_{j+1}^{x} + S_{j}^{y}S_{j+1}^{y})$ coincide, so we can use $L_{j} \propto S_{j}^{x} (S_{j}^{x}S_{j+1}^{x} + S_{j}^{y}S_{j+1}^{y})$ instead. 

From Eq. \eqref{eq:SMPlocal} it is immediate that $\ker (P_{j,j+1}) \subseteq \ker (S_{j}^{x}S_{j+1}^{x} + S_{j}^{y}S_{j+1}^{y})$. The other inclusion is true if and only if, for any $\ket{\Psi} \in \Im (P_{j,j+1})$,
\begin{equation}
    (S_{j}^{x}S_{j+1}^{x} + S_{j}^{y}S_{j+1}^{y}) \ket{\Psi} \neq 0.
\end{equation}
Since $\Im (P_{j,j+1}) = \mathrm{span} \{ \ket{X_{k}}_{j, j+1} \}$, and we have already seen that $(S_{j}^{x}S_{j+1}^{x} + S_{j}^{y}S_{j+1}^{y}) \ket{X_{k}}_{j,j+1} \neq 0$, then $\ker (P_{j,j+1}) = \ker (S_{j}^{x}S_{j+1}^{x} + S_{j}^{y}S_{j+1}^{y})$ as desired.

\subsection{Parallel charge}

Here, for the spin-$1$ $XY$ chain example exhibited in the main text, we are going to find the parallel charge [Eq.~\eqref{parallel_erg_def}] and the optimal local unitaries $\check{u}_j$ extracting it.

First, let us show how to compute the reduced states of the density matrix corresponding to a single scar, $\dyad{s_n}$. The scars of the $N$-particle spin-$1$ $XY$ chain can be written as
\begin{equation}
    \ket{s_n} \coloneq \sqrt{\frac{(N - n)!}{ N! \, n!}} \, (Q^{+})^{n} \ket{s_{0}} \qquad \text{for} \qquad n \in \{0, \dots, N\},
\end{equation}
where the ladder-like operator is defined as $Q^{+} \coloneq \sum_{j = 1}^{N} (-1)^{j} (S_{j}^{x} + i S_{j}^{y})^{2} / 2$, and the lowest energy scar is the all spins down state, $\ket{s_{0}} \coloneq \ket{-1 , -1 , \dots , -1}$. Let us consider an arbitrary bipartition of the whole system into subsystems $A$ and $B$, with $N_{A}$ and $N_{B} = N - N_{A}$ spins each one, respectively. For this computation we are going to use subindices $N$, $N_{A}$ and $N_{B}$ in the operators to explicitly indicate which subsystem they act on and, additionally, the $n$-th scar state of the subsystem with $N_{i}$ spins will be denoted $\ket{s_n, N_{i}}$, where now $n \in \{ 0, \dots, N_{i} \}$. 

The ladder operator can always be decomposed as a sum $Q_n^{+} = Q_{N_{A}}^{+} + Q_{N_{B}}^{+}$ where both terms commute because they act on different subsystems, $[Q_{N_{A}}^{+} , Q_{N_{B}}^{+}] = 0$, and, therefore, the usual binomial formula can be used to write any power of it:
\begin{equation}
    (Q_n^{+})^{n} = (Q_{N_{A}}^{+} + Q_{N_{B}}^{+})^{n} = \sum_{k = 0}^{n} \binom{n}{k} (Q_{N_{A}}^{+})^{n-k} (Q_{N_{B}}^{+})^{k}. 
\end{equation}
Thus, the scar density matrix can be written as
\begin{equation}
    \dyad{s_n} = \frac{(N-n)!}{N! \, n!} \sum_{j = 0}^{n} \sum_{k = 0}^{n} \binom{n}{j} \binom{n}{k} (Q_{N_{A}}^{+})^{n-j}  \dyad{s_{0} , N_{A}} (Q_{N_{A}}^{-})^{n-k} \otimes (Q_{N_{B}}^{+})^{j} \dyad{s_{0} , N_{B}} (Q_{N_{B}}^{-})^{k},
\end{equation}
where $Q_{N_{i}}^{-}$ is the Hermitian adjoint of $Q_{N_{i}}^{+}$. The reduced density matrix of bipartition $A$ is therefore obtained by tracing over the $N_{B}$ part, 
\begin{equation}
    \rho^{A,n} =  \frac{(N-n)!}{N! \, n!} \sum_{j = 0}^{n} \sum_{k = 0}^{n} \binom{n}{j} \binom{n}{k} (Q_{N_{A}}^{+})^{n-j}  \dyad{s_{0} , N_{A}} (Q_{N_{A}}^{-})^{n-k} \, \bra{s_{0} , N_{B}} (Q_{N_{B}}^{-})^{k} (Q_{N_{B}}^{+})^{j} \ket{s_{0} , N_{B}},
    \label{eq:rhoAfirst}
\end{equation}
where we used that the states $(Q_{N_{B}}^{+})^{j} \ket{s_{0} , N_{B}}$ are orthogonal for different powers of $j$. Concretely, they are proportional to the scar states of $N_{B}$ spins and thus constitute a basis of the corresponding subspace:
\begin{equation}
    \bra{s_{0} , N_{B}} (Q_{N_{B}}^{-})^{k} (Q_{N_{B}}^{+})^{j} \ket{s_{0} , N_{B}} = \delta_{j,k} \, \frac{N_{B}! \, j!}{(N_{B} - j)!}.
\end{equation}
Replacing this in Eq.~\eqref{eq:rhoAfirst} we obtain the general result:
\begin{equation}
    \rho^{A,n} =  \binom{N}{n}^{-1} \sum_{j = 0}^{n} \binom{N-N_{A}}{j} \binom{N_{A}}{n-j}  \dyad{s_{n-j} , N_{A}} \qquad \text{for} \qquad n \in \{0, \dots, N_{A}\}.
 \end{equation}
The computation of $U$ below requires the single-spin and the two-spin reduced density matrices. That is, the cases $N_{A} = 1$ and $N_{A} = 2$. For the first one, since $N_{A} = 1$, there are only two possible scar states, $\ket{s_{0}}$ and $\ket{s_{i}}$. Therefore, the only two contributing terms in the sum are $n - j = 0$ and $n - j = 1$. The single-spin reduced state then is
\begin{equation}
    \rho^{1,n} = \frac{n}{N} \, \dyad{1} + \frac{N-n}{N} \, \dyad{-1}.
    \label{eq:scar1spin}
\end{equation}
If $N_{A} = 2$, there are three possible scars, $\ket{s_{0}}$, $\ket{s_{1}}$ and $\ket{s_{2}}$. Thus, we can only have $n - j = 0$, $n - j = 1$, or $n - j = 2$. Therefore, the two-spin reduced state is:
\begin{equation}
    \begin{aligned}
        \rho^{2,n} & = \frac{n(n-1)}{N(N-1)} \ketbra{1,1}{1,1} + \frac{(N-n) (N-n-1)}{N(N-1)} \ketbra{-1,-1}{-1,-1} \\
        & + \frac{2 (N-n)n}{N(N-1)} \left( \frac{\ket{1,-1} - \ket{-1,1}}{\sqrt{2}} \right) \left( \frac{\bra{1,-1} - \bra{-1,1}}{\sqrt{2}} \right).
    \end{aligned}
    \label{eq:scar2spin}
\end{equation}

We want to show the solution of the minimization problem
\begin{equation}
    \ergl (\dyad{s_n}) = \tr (\dyad{s_n} H) - \min_{\{ u_j\}} \tr\Big(\bigotimes_j u_j^{\phantom{\dagger}} \, \dyad{s_n} \, \bigotimes_j u_j^\dagger H\Big)
\end{equation}
has the form $u_{j} = U$ for all $j$. 
Let us first find the single-particle unitary operator $U$ that maximizes the work extracted from $\dyad{s_n}$ when applied to each member of the spin chain. That is, we want the $\tilde{u}$ delivering the minimum in
\begin{equation} \label{eq:ergLocalApp}
    \widetilde{\ergl}_\prll(\dyad{s_n}) \coloneq \tr [H \dyad{s_n}] - \min_{u} \tr [H u^{\otimes N} \dyad{s_n} u^{\dagger \, \otimes N}],
\end{equation}
where the tilde above $\ergl$ is to indicate that this is not the fully optimized $\ergl_\prll$ but rather a generally smaller quantity with a more constrained optimization. Contrary to the standard ergotropy $\erg$, there is no general procedure for obtaining a closed analytical solution for this type of minimization problem. Therefore, we are first going to proceed with an ansatz for $u$ and then numerically verify that it is indeed optimal. To derive the ansatz, let us restrict ourselves to unitary operators that rearrange the populations in the basis of the single-spin spin operator $S_z$. That is, of the form
\begin{equation}
    u = \ketbra{\curlyd{m}_{1}}{1} + \ketbra{\curlyd{m}_{0}}{0} + \ketbra{\curlyd{m}_{-1}}{-1},
\end{equation}
where $\{\ket{1}, \ket{0}, \ket{-1}\}$ is the basis of $S^{z}$, and $\{ \ket{\curlyd{m}_{1}}, \ket{\curlyd{m}_{0}}, \ket{\curlyd{m}_{-1}} \}$ is a permutation of that basis. The translation invariance of the $XY$ chain allows to rewrite the second term of Eq.~\eqref{eq:ergLocalApp} as
\begin{equation}
    \tr [H u^{\otimes N} \dyad{s_n} u^{\dagger \, \otimes N}] = \tr [ H^{(2)} \, u \otimes u \rho^{(2)} u^{\dagger} \otimes u^{\dagger}],
\end{equation}
where $\rho^{(2)}$ is the two-spin reduced state computed above, and
\begin{equation}
    H^{(2)} = \tilde{J} (S^{x} \otimes S^{x} + S^{y} \otimes S^{y}) + \tilde{h} (S^{z} \otimes \mathbb{1} + \mathbb{1} \otimes S^{z}) + \tilde{D} \left[ (S^{z})^{2} \otimes \mathbb{1} + \mathbb{1} \otimes (S^{z})^{2} \right],
\end{equation}
is a $2$-body effective Hamiltonian with rescaled parameters $\tilde{J} \coloneq J (N-1)$, $\tilde{h} \coloneq N h / 2$ and $\tilde{D} \coloneq N D / 2$, and with single-spin operators $S^{x}$ and $S^{y}$. Using the reduced density matrices in Eqs.~\eqref{eq:scar1spin} and~\eqref{eq:scar2spin}, a straightforward computation leads to
\begin{equation}
    \begin{aligned}
        \frac{1}{N} \widetilde{\ergl}_\prll (\dyad{s_n}) & = [1 - q_n \curlyd{m}_{1}^{2} - (1-q_n) \curlyd{m}_{-1}^{2} ] D + [q_n (1 - \curlyd{m}_{1}) - (1-q_n) (1 + \curlyd{m}_{1})] h \\
        & + 4 q_n ( 1 - q_n ) J (\delta_{\curlyd{m}_{1} , \curlyd{m}_{-1} + 1} + \delta_{\curlyd{m}_{1} , \curlyd{m}_{-1} - 1}),
    \end{aligned}
\end{equation}
where we defined $q_n \coloneq n / N$. Since only 6 combinations of the values $(\curlyd{m}_{1} , \curlyd{m}_{0}, \curlyd{m}_{-1})$ are possible, we can do an exhaustive search over all of them to find the optimal one. The different cases are
\begin{equation}
    \begin{aligned}
        (\curlyd{m}_{1}, \curlyd{m}_{0}, \curlyd{m}_{-1}) = (1, 0, -1) \qquad \Rightarrow \qquad \widetilde{\ergl}_\prll / N & = 0 \\
        (\curlyd{m}_{1}, \curlyd{m}_{0}, \curlyd{m}_{-1}) = (-1, 0, 1) \qquad \Rightarrow \qquad \widetilde{\ergl}_\prll / N & = 2 (2 q_n - 1) h \\
        (\curlyd{m}_{1}, \curlyd{m}_{0}, \curlyd{m}_{-1}) = (0, 1, -1) \qquad \Rightarrow \qquad \widetilde{\ergl}_\prll / N & = 4 q_n ( 1 - q_n ) J + q_n (D + h) \\
        (\curlyd{m}_{1}, \curlyd{m}_{0}, \curlyd{m}_{-1}) = (0, -1, 1) \qquad \Rightarrow \qquad \widetilde{\ergl}_\prll / N & = 4 q_n ( 1 - q_n ) J + q_n (D + h) - 2(1 - q_n) h \\
        (\curlyd{m}_{1}, \curlyd{m}_{0}, \curlyd{m}_{-1}) = (-1, 1, 0) \qquad \Rightarrow \qquad \widetilde{\ergl}_\prll / N & = 4 q_n ( 1 - q_n ) J + (1-q_n) (D + h) + 2(2 q_n - 1) h \\
        (\curlyd{m}_{1}, \curlyd{m}_{0}, \curlyd{m}_{-1}) = (1, -1, 0) \qquad \Rightarrow \qquad \widetilde{\ergl}_\prll / N & = 4 q_n ( 1 - q_n ) J + (1-q_n) (D + h) - 2 (1 - q_n) h
       \end{aligned}
\end{equation}
Note that for any value of $q_n$, the value of $\widetilde{\ergl}_\prll / N$ in the case $(\curlyd{m}_{1}, \curlyd{m}_{0}, \curlyd{m}_{-1}) = (-1, 1, 0)$ is larger than in the cases $(\curlyd{m}_{1}, \curlyd{m}_{0}, \curlyd{m}_{-1}) = (1, 0, -1)$, $(\curlyd{m}_{1}, \curlyd{m}_{0}, \curlyd{m}_{-1}) = (-1, 0, 1)$ and $(\curlyd{m}_{1}, \curlyd{m}_{0}, \curlyd{m}_{-1}) = (1, -1, 0)$. Also, the value in the case $(\curlyd{m}_{1}, \curlyd{m}_{0}, \curlyd{m}_{-1}) = (0, 1, -1)$ is larger than in $(\curlyd{m}_{1}, \curlyd{m}_{0}, \curlyd{m}_{-1}) = (0, -1, 1)$.  Therefore, if $q_n < 1/2$ we choose $(\curlyd{m}_{1}, \curlyd{m}_{0}, \curlyd{m}_{-1}) = (0, 1, -1)$ and if $q_n \geq 1/2$, $(\curlyd{m}_{1}, \curlyd{m}_{0}, \curlyd{m}_{-1}) = (-1, 1, 0)$. With this, we obtain for the parallel ergotropy
\begin{equation}
    \frac{1}{N} \widetilde{\ergl}_\prll (\dyad{s_n}) = 
    \begin{cases}
        q_n \left[ h + D + 4 (1 - q_n) J \right] & \qquad q_n < 1/2 \\
        q_n \left[ h + D + 4 (1 - q_n) J \right] + (2 q_n -1) (h - D) & \qquad q_n \geq 1/2
    \end{cases}
\end{equation}
delivered by
\begin{align} \label{optildu}
    \tilde{u} = \begin{pmatrix}
        0 & 1 & 0 \\ 1 & 0 & 0 \\ 0 & 0 & 1
    \end{pmatrix}, \quad \mathrm{for} \;\; q_n < 1/2, \qquad \mathrm{and} \qquad \tilde{u} = \begin{pmatrix}
        0 & 1 & 0 \\ 0 & 0 & 1 \\ 1 & 0 & 0
    \end{pmatrix}, \quad \mathrm{for} \;\; q_n \geq 1/2.
\end{align}
As mentioned above, upon performing a complete numerical optimization of $\widetilde{\ergl}_\prll$, we find that the optimal $u$ coincides with $\tilde{u}$ in Eq~\eqref{optildu}. To perform this optimization, we use a parametrization of the SU$(3)$ group introduced in Ref.~\cite{Bronzan_1988}. There, each $u \in \mathrm{SU}(3)$ is expressed via the $8$ angles $\theta_1$, $\theta_2$, $\theta_3$ $\in (0, \pi/2)$ and $\phi_1, \ldots, \phi_5 \, \in (0, 2 \pi)$ as
\begin{equation} \label{bronzano}
\begin{aligned}
    u_{11} &= \cos\theta_1 \cos\theta_2 e^{i\phi_1},
    \\
    u_{12} &= \sin\theta_1 e^{i\phi_3},
    \\
    u_{13} &= \cos\theta_1 \sin\theta_2 e^{i\phi_4},
    \\
    u_{21} &= \sin\theta_2 \sin\theta_3 e^{-i\phi_4-i\phi_5} -\sin\theta_1 \cos\theta_2 \cos\theta_3 e^{i\phi_1+i\phi_2-i\phi_3},
    \\
    u_{22} &= \cos\theta_1 \cos\theta_3 e^{i\phi_2},
    \\
    u_{23} &= -\cos\theta_2 \sin\theta_3 e^{-i\phi_1-i\phi_5} -\sin\theta_1 \sin\theta_2 \cos\theta_3 e^{i\phi_2-i\phi_3+i\phi_4},
    \\
    u_{31} &= -\sin\theta_1 \cos\theta_2 \sin\theta_3 e^{i\phi_1-i\phi_3+i\phi_5} -\sin\theta_2 \cos\theta_3 e^{-i\phi_2-i\phi_4},
    \\
    u_{32} &= \cos\theta_1 \sin\theta_3 e^{i\phi_5},
    \\
    u_{33} &= \cos\theta_2 \cos\theta_3 e^{-i\phi_1-i\phi_2} -\sin\theta_1 \sin\theta_2 \sin\theta_3 e^{-i\phi_3+i\phi_4+i\phi_5}.
\end{aligned}
\end{equation}
The rectangular shape of the parameter space greatly facilitates the numerical optimization, and in our tests, the basin-hopping algorithm \cite{Wales_1997} implemented in SciPy yielded consistently reliable results for this problem up to $N = 6$. 

\medskip

Importantly, the same $\tilde{u}$ delivers the single-site local ergotropy (defined in Eqs.~\eqref{locerg_E_k} and~\eqref{locergdef} and Ref.~\cite{Salvia_2023})
\begin{align}
    \ergl^{(j)}(\dyad{s_n}) \coloneq \tr(\dyad{s_n} H) - \min_{u_j} \tr \big( u_j \otimes \id_{\bar{j}} \dyad{s_n}  u_j^\dagger \otimes \id_{\bar{j}} \, H \big),
\end{align}
where $u_j$ acts on the $j$'th spin, and $\id_{\bar{j}}$ is the identity operator on all spins except the $j$'th. Note that $\tilde{u}$ is the optimal local-ergotropy-extracting unitary for all $j$.

\medskip

Next, for $N = 2, 3, 4$, we carried out the full optimization [Eq.~\eqref{parallel_erg_def}]
\begin{equation} \label{parallel_erg_def_app}
    \ergl_\prll (\dyad{s_n}) = \tr (\dyad{s_n} H) - \min_{\{ u_j\}} \tr\Big(\bigotimes_j u_j^{\phantom{\dagger}} \, \dyad{s_n} \, \bigotimes_j u_j^\dagger H\Big)
\end{equation}
to determine the set of $\optU_j$ delivering $\ergl_\prll$. To do so, we first reduce the number of parameters by utilizing the translation invariance of the problem. Namely, we formalize the structure of the Hamiltonian as $H = \sum_{j = 1}^{N} \mathfrak{H}^{(1)}_{j} + \sum_{j = 1}^{N-1} \mathfrak{H}^{(2)}_{j, j+1}$, where all $\mathfrak{H}^{(1)}_{j}$ are the same $\mathfrak{H}^{(1)}$ and all $\mathfrak{H}^{(2)}_{j, j+1}$ are the same $\mathfrak{H}^{(2)}$, and define $\mathfrak{h}_{j,j+1} = \mathfrak{H}^{(1)}_{j}/2 + \mathfrak{H}^{(1)}_{j+1}/2 + \mathfrak{H}^{(2)}_{j, j+1}$. Then,
\begin{equation}
    \begin{aligned}
        \tr\Big(\bigotimes_j u_j^{\phantom{\dagger}} \, \dyad{s_n} \, \bigotimes_j u_j^\dagger H\Big) = \sum_{j = 1}^{N-1} \tr (u_{j}^{\phantom{\dagger}} \otimes u_{j+1}^{\phantom{\dagger}} \rho^{2,n} u_{j}^{\dagger} \otimes u_{j+1}^{\dagger} \, \mathfrak{H}^{(2)}) + \frac{1}{2} \tr (u_{1}^{\phantom{\dagger}} \rho^{1,n} u_{1}^{\dagger} \, \mathfrak{H}^{(1)}) + \frac{1}{2} \tr (u_n^{\phantom{\dagger}} \rho^{1,n} u_n^{\dagger} \, \mathfrak{H}^{(1)})
    \end{aligned}
\end{equation}
This clearly demonstrates that the optimization problem is symmetric with respect to reflection at the center of the chain, so that $\optU_1 = \optU_N$, $\dots$, $\optU_j = \optU_{N-j}$, etc. Thus, we can run the optimization only over $u_1$, $\ldots$, $u_{\lceil N/2 \rceil}$, and set $u_{\lceil N/2 \rceil + k} = u_k$, $k = 1, \ldots, \lfloor N/2 \rfloor$. With the number of optimization parameters thereby essentially halved, the numerical optimization is carried out exactly as before: each independent $u_j$ is parametrized as in Eq.~\eqref{bronzano}, and then the basin-hopping algorithm is used. By trying a large number of combinations of $D$, $h$, and $J$, we find that almost always $\optU_1 = \cdots = \optU_N = \tilde{u}$, meaning that
\begin{align}
    \ergl_\prll(\dyad{s_n}) = \widetilde{\ergl}_\prll(\dyad{s_n}) = q_n \left[ h + D + 4 (1 - q_n) J \right] + \theta(2 q_n -1) (h - D),
\end{align}
where $\theta$ is the Heaviside step function. The only exception is the ultrastrong-coupling regime, $J \gg D, \, h$, where $\tilde{u}$ no longer attains the global optimum in Eq.~\eqref{parallel_erg_def_app}. Even in this regime, however, $\tilde{u}^{\otimes N}$ extracts most of $\ergl_\prll$. Conveniently, this regime is experimentally less relevant than the moderate- and strong-coupling regimes.

\medskip

Finally, to provide an independent numerical check of these findings, we employ the gradient-descent algorithm of Ref.~\cite{Cerisola_2026}, which is methodologically distinct from the basin-hopping approach used above. It yields the same optimal unitaries, up to small numerical artifacts. Overall, our extensive numerical analysis establishes with a high degree of confidence that $\tilde{u}$, given in Eq.~\eqref{optildu}, universally delivers the local and parallel ergotropies of the spin-$1$ $XY$ model throughout all experimentally relevant parameter regimes.

\subsection{Sequential work from scars in the $XY$ model}

\begin{figure*}[t]
  \centering
  \begin{tikzpicture}
  \node (a) at (0,0) {\includegraphics{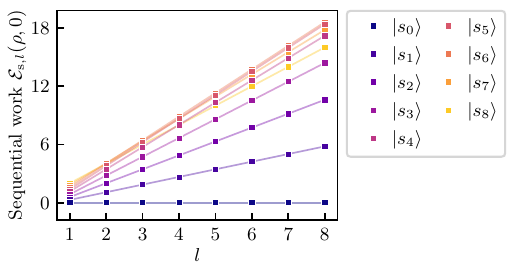}};
  \node (b) at (9,0) {\includegraphics{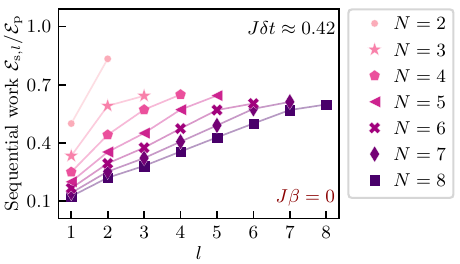}};

  \draw (-4, 2.3) node {\textbf{a}};
  \draw (5.5,2.3) node {\textbf{b}};
  \end{tikzpicture}
  \caption{Charge characterization of the spin-$1$ $XY$ battery. \textbf{a}, Charge of the battery that can be retrieved by locally discharging the cells sequentially $\ergl_{\m{s}, l}(\dyad{s_n}, \delta t = 0)$ [cf.~\cref{sequential_erg}] in the ideal scenario of vanishingly small waiting times $J \delta t = 0$ as a function of the cell index $l$ and for all scars of the $N = 8$ spin system. The linear scaling of the sequential work demonstrates that each cell holds a similar amount of charge. Depending on the Hamiltonian parameters $(h,J,D)$, the total parallel charge $\ergl_\m{p}(\dyad{s_n}) = \ergl_{\m{s}, N}(\dyad{s_n}, \delta t = 0)$ may not increase monotonically with $n$.
  \textbf{b}, Sequential work $\ergl(\dyad{s_N}, \delta t \approx 0.42)$ compared to the total parallel charge for a waiting time between sequential extractions of $J \delta t \approx 0.42$ [cf.~\cref{fig:battery}b] for the most energetic scar $\ket{s_N}$ as a function of $l$. The total work that can be extracted from each cell is diminished compared to the ideal case, however, the sequential work still increases monotonically. Parameters are the same as in \cref{fig:battery}.}
  \label{fig:cell-content}
\end{figure*}

Here we provide additional information on the local ergotropy held by the scars of the spin-$1$ battery analyzed in the main text.

One of the main features of our model is that once the battery is in its charged state (i.e., $\rho = \dyad{s_n}$), energy can be released in chunks in a controlled manner.
Instead of applying the energy extracting local unitary operations in parallel and therefore depleting the battery in its entirety by gaining the ergotropy $\ergl_\m{p}(\rho)$, work can be extracted intermittently, i.e., cell by cell using individual unitaries $\check{u}_j$, where $j = 1,2,\ldots,N$ as detailed in the main text [\cref{sequential_erg}].
In \cref{fig:cell-content} we show the sequential work (repeated here for convenience)
\begin{align} \label{si:sequential_erg}
    \ergl_{\seq,\ell}(\rho, \wt) \coloneq \tr(H \rho) - \tr\big( H \, U_{[1 \cdots \ell]}^{\phantom{\dagger}} \, \rho \, U_{[1 \cdots \ell]}^\dagger \big),
\end{align}
where $U_\fr = \exp(- \m{i} \wt H)$ generates the free evolution and $U_{[1 \cdots \ell]} = \check{u}_\ell \, U_\fr \, \cdots \, U_\fr \, \check{u}_1$.
Under ideal conditions that is, in the absence of disturbances between subsequent extractions $J \delta t = 0$, the sequential work scales linearly in the number $l$ of cells indicating that the total (parallel) charge is uniformly distributed over the individual spins.
Depending on the amount of energy or power output required to perform certain tasks one or multiple cells may thus be depleted.
Although \cref{fig:cell-content}a shows only the sequential work of $N = 8$ spins, we have confirmed that the behavior for system sizes $N < 8$ remains qualitatively similar.
If subsequent extractions are slow enough such that the free evolution becomes significant, only a certain percentage of the total parallel charge is available [cf.~\cref{fig:battery}b].
In \cref{fig:cell-content}b, we plot the sequential work in relation to $\erg_\m{p}(\rho)$ for the scar $\ket{s_N}$ and different system sizes.
At a waiting time of $J \delta t \approx 0.42$, each local unitary still extracts more work and $\erg_{\m{s},\ell}(\rho, \delta t)$ increases with $\ell$.
This monotonicity generally breaks down for longer waiting times but seems to remain increasingly robust for larger systems.

\subsection{Numerical simulation}

We here provide more details on the numerical simulations performed to obtain our results for the spin-1 $XY$ model.

We simulated our charging protocol by solving the stochastic differential equation \cref{eq:full-protocol} for the respective expectation values of interest using QuTiP's \texttt{smesolve} \cite{qutip5}.
The expectation value of the total magnetization $\langle S^z \rangle(t) = \tr[\rho^\m{c}_t S^z]$ in \cref{fig:relaxation}a and their corresponding measurement currents were obtained by simulating $\approx 100$ quantum trajectories and then postselecting a typical trajectory for each scar.
Basic numerical integration then yielded the integrated currents in \cref{fig:relaxation}c.
To obtain the stochastic charging times, we computed the expectation values $\tr[P_n \rho^\m{c}_t]$ for $10^4$ quantum trajectories for sufficiently long times.
Postprocessing of these data then yielded the stochastic charging times given a specified error tolerance $\epsilon$ [cf.~\cref{fig:charging-time}c].
From the distributions, we evaluated the respective sample means to get the mean first-passage times $\mathbb{E}[\tau_n(\epsilon)]$ shown in \cref{fig:relaxation}b.

To numerically compute the asymptotic scar distributions $p_n(\rho_0)$ shown in \cref{fig:battery}c, we solved the Lindblad equation using QuTip's \texttt{mesolve} to dynamically obtain $\tr[P_n \rho_t]$ starting from an initial state $\rho(0) = \exp(-\beta H) / Z(\beta)$ at inverse temperature $\beta$. 
For each system size, we confirmed that the integration time was sufficiently long to be in the steady state regime (see for instance \cref{fig:charging-time}a). We furthermore verified that $\sum_n p_n(\rho_0) = 1$ up to small errors.
The parallel ergotropy [\cref{fig:battery}a] was then obtained by directly evaluating \cref{eq:XY-local-ergotropy}.
To compare with the maximum possible ergotropy of the system $\ergl^\m{max} = E_\m{max} - E_\m{min}$ (the maximal capacity of the battery), we simply computed the largest and smallest eigenvalues of the Hamiltonian.

To obtain the average sequential ergotropy [\cref{sequential_erg} and \cref{fig:battery}b] we first computed the sequential ergotropy $\ergl_{\seq , N}^{\mathrm{av}} (\dyad{s_n}, \wt)$ for each scar individually.
This entails evolving $\dyad{s_n}$ unitarily with $U_\fr = \exp(-\m{i} H \wt)$, then, depending on the ratio $q_n = n/N$, applying the corresponding optimal extracting unitary $\tilde u$ of \cref{optildu} on the first spin and repeatedly iterating through the spin chain with $\tilde u$.
In \cref{fig:cell-content} we record the energy difference $\ergl_{\seq,\ell}(\dyad{s_n}, \wt) \coloneq \tr(H \rho) - \tr\big( H \, U_{[1 \cdots \ell]}^{\phantom{\dagger}} \, \rho \, U_{[1 \cdots \ell]}^\dagger \big)$ after the $\ell$-th step.
After all $N$ spins were addressed, we obtained the total sequential ergotropy $\ergl_{\seq,N}(\dyad{s_n}, \wt)$ for a given waiting time $\wt$.
This procedure was carried out for $20$ equidistantly spaced values between $\wt = 0$ and $\wt = 1$.
We finally constructed the average sequential ergotropy according to
\begin{equation}
    \ergl_{\seq,N}(\dyad{s_n}, \wt) = \sum_n p_n(\rho^\m{av}_0) \ergl_{\seq,N}(\dyad{s_n}, \wt).
\end{equation}

In the main panels of \cref{fig:battery,fig:relaxation}, we used the infinite temperature initial state $\rho(0) = \mathds{1} / 3^N$.
To investigate the performance of the battery for initial states states at lower temperature $J\beta \approx 1$, we employed the following strategy.
For each system size, we computed the arithmetic mean of the energy eigenvalues of the shifted Hamiltonian $\tilde H = H - \mathds{1} E_\m{min}$ which has only nonnegative eigenvalues
\begin{equation}
    \langle \tilde H \rangle = \tr[\tilde H] / 3^N
\end{equation}
and numerically found the unique temperature corresponding to $10 \%$ of $\langle \tilde H \rangle$, i.e., we solved 
\begin{equation}
    \tr[\exp(-\beta \tilde H) / Z(\beta) \tilde H] = 0.1 \langle \tilde H \rangle.
\end{equation}
The values of $J \beta$ in \cref{fig:battery} are listed in \cref{tab:betas} (rounded to the third decimal).

\begin{table}[t]
    \centering
    \begin{tabular}{c c}
        \toprule
         System size $N$ \, & \, Inverse temperature $J \beta$ \\
         \midrule 
         2 & 0.982\\
         3 & 0.959\\
         4 & 0.947\\
         5 & 0.924\\
         6 & 0.902\\
         7 & 0.884\\
         8 & 0.869 \\
         \bottomrule
    \end{tabular}
    \caption{Inverse temperatures for each system size corresponding to the insets displayed in \cref{fig:battery}.}
    \label{tab:betas}
\end{table}

\section{Average of the ergotropy and ergotropy of the average}

Let us consider an initial state $\rho_{0}$ and call the corresponding asymptotic state $\rho_{\infty} = \rho_{\infty} (\rho_{0})$. The ergotropy of this asymptotic state is simply given by $\ergl (\rho_{\infty}(\rho_{0}))$. If we now sample over different initial states and compute the sample average of the ergotropy, in the limit of infinitly many samples we would obtain
\begin{equation}
   \mathbb{E}[ \ergl ] = \lim_{K \to \infty} \frac{1}{K} \sum_{i = 1}^{K} \ergl (\rho_{\infty}(\rho_{0}^{i})).
\end{equation}
Under the dynamics induced by the protocol, each state will inevitably converge to a single scar, $\rho_{\infty}(\rho_{0}^{i}) = \dyad{s_n}$ for some $n$. Therefore, the ergotropy can only take $N+1$ different values, and will take each value $k_n = k_n (K)$ times, which is the number of initial states that converged to the $n$-th scar. Thus,
\begin{equation}
    \mathbb{E}[ \ergl ] = \lim_{K \to \infty} \frac{1}{K} \sum_{n = 0}^{N} \ergl (\dyad{s_n}) \, k_n (K) = \sum_{n = 0}^{N} \ergl (\dyad{s_n}) \left( \lim_{K \to \infty} \frac{k_n (K)}{K} \right).
\end{equation}
The quotient $k_n (K) / K$ is the ratio of states that converged to the scar $n$ and, in the limit $K \to \infty$, it will equal the probability of converging to that particular scar:
\begin{equation}
    \lim_{K \to \infty} \frac{k_n (K)}{K} = \prob_\mu (\ket{s_n}) = \int \dd \mu_{\rho_0} \, \tr (\textsf{P}^{\infty}_n \rho_{0}),
\end{equation}
where $\mu$ is the measure we are sampling initial states with respect to. Since the trace and the asymptotic projection $\textsf{P}^{\infty}_n$ are linear operations,
\begin{equation}
    \lim_{K \to \infty} \frac{k_n (K)}{K} = \tr (\textsf{P}^{\infty}_n  \mathbb{E}_{\mu} [ \rho_{0} ]).
\end{equation}
Therefore, we obtain the final result that the expected ergotropy is the ergotropy of the expected state
\begin{equation}
    \mathbb{E}[ \ergl ] = \sum_{n = 0}^{N} \ergl (\dyad{s_n}) \, \tr (\textsf{P}^{\infty}_n  \mathbb{E}_{\mu} [ \rho_{0} ]).
\end{equation}

\end{document}